\documentclass[aps,prd,10pt,notitlepage,nofootinbib]{revtex4-1}

\usepackage[utf8]{inputenc}
\usepackage{amsmath,amssymb,amsfonts}
\usepackage{newtxtext,newtxmath}

\usepackage{bbold}
\usepackage{bm}
\usepackage{graphicx}
\usepackage[usenames,dvipsnames]{xcolor}
\usepackage{color}
\usepackage[colorlinks=true,linkcolor=blue,urlcolor=blue,citecolor=blue]{hyperref}

\usepackage{slashed}
\usepackage[english]{babel}
\usepackage{dcolumn}
\usepackage{pifont}
\usepackage{dsfont,mathrsfs}
\usepackage{cancel}
\usepackage{bigints}
\usepackage{accents}
\usepackage{soul}
\usepackage{multirow}

\usepackage{amsmath, amssymb}
\usepackage{bbold}
\usepackage{makecell}
\usepackage{simpler-wick}
\usepackage{setspace}
\usepackage{scalerel}
\usepackage[normalem]{ulem}
\usepackage{hyperref} 
\usepackage{amsmath}

\usepackage{tikz}
\usetikzlibrary{svg.path}
\definecolor{orcidlogocol}{HTML}{A6CE39}
\tikzset{
	orcidlogo/.pic={
		\fill[orcidlogocol] svg{M256,128c0,70.7-57.3,128-128,128C57.3,256,0,198.7,0,128C0,57.3,57.3,0,128,0C198.7,0,256,57.3,256,128z};
		\fill[white] svg{M86.3,186.2H70.9V79.1h15.4v48.4V186.2z}
		svg{M108.9,79.1h41.6c39.6,0,57,28.3,57,53.6c0,27.5-21.5,53.6-56.8,53.6h-41.8V79.1z M124.3,172.4h24.5c34.9,0,42.9-26.5,42.9-39.7c0-21.5-13.7-39.7-43.7-39.7h-23.7V172.4z}
		svg{M88.7,56.8c0,5.5-4.5,10.1-10.1,10.1c-5.6,0-10.1-4.6-10.1-10.1c0-5.6,4.5-10.1,10.1-10.1C84.2,46.7,88.7,51.3,88.7,56.8z};
	}
}
\newcommand{\orcidlink}[1]{\href{https://orcid.org/#1}{\mbox{\scalerel*{\begin{tikzpicture}[yscale=-1,transform shape]\pic{orcidlogo};\end{tikzpicture}}{|}}}}

\newcommand{\nn}{\nonumber}

\newcommand{\FB}[1]{\left(#1\right)}

\newcommand{\SB}[1]{\left\{#1\right\}}
\newcommand{\TB}[1]{\left[#1\right]}
\newcommand{\AB}[1]{\left<#1\right>}

\newcommand{\munu}{{\mu\nu}}

\allowdisplaybreaks

\begin{document}
	\title{Speed of sound and isothermal compressibility in a magnetized quark matter with anomalous magnetic moment of quarks}
	
	\author{Rajkumar Mondal\orcidlink{0000-0002-1446-6560}$^{a,d}$}
	\email{rajkumarmondal.phy@gmail.com}

	\author{Sourav Duari \orcidlink{0009-0006-0795-5186}$^{a}$}
	\email{s.duari@vecc.gov.in}
	\email{sduari.vecc@gmail.com}	
	
	\author{Nilanjan Chaudhuri\orcidlink{0000-0002-7776-3503}$^{a,d}$}
	\email{nilanjan.vecc@gmail.com}
	\email{n.chaudhri@vecc.gov.in}
	
	\author{Sourav Sarkar\orcidlink{0000-0002-2952-3767}$^{a,d}$}
	\email{sourav.vecc@gmail.com}
	\email{sourav@vecc.gov.in}
	
	\author{Pradip Roy\orcidlink{0009-0002-7233-4408}$^{c,d}$}
	\email{pradipk.roy@saha.ac.in}

	\affiliation{$^a$Variable Energy Cyclotron Centre, 1/AF Bidhannagar, Kolkata - 700064, India}
	\affiliation{$^c$Saha Institute of Nuclear Physics, 1/AF Bidhannagar, Kolkata - 700064, India}
	\affiliation{$^d$Homi Bhabha National Institute, Training School Complex, Anushaktinagar, Mumbai - 400085, India} 
	

\begin{abstract}	
	We study the characteristics of quark matter under the influence of a background magnetic field with anomalous
	magnetic moment (AMM) of quarks at finite temperature and quark chemical potential in the framework of Polyakov loop extended Nambu Jona-Lasinio (PNJL) model. In presence of a magnetic field, the speed of sound and isothermal compressibility become anisotropic with respect to the direction of the background magnetic field, splitting into parallel and perpendicular directions with respect to the magnetic field. Though the qualitative nature of parallel and perpendicular components of squared speed of sound appear similar, they differ in magnitude at lower values of temperature. The parallel and perpendicular components of isothermal compressibility decrease with increasing temperature, indicating a trend towards increased incompressible strongly interacting matter. On inclusion of the AMM of quarks, the perpendicular component of isothermal compressibility becomes greater than the parallel component. Additionally, we investigate the quark number susceptibility normalized by its value at zero magnetic field, which may indicate the presence of magnetic fields in the system.
\end{abstract}
\maketitle

\section{Introduction}

In recent times, the study of properties of matter under extreme conditions such as high temperature and/or density  has been extended to include strong magnetic fields. Presently, it is well accepted that strong magnetic fields could exist in heavy ion collision (HIC) experiments at the Relativistic Heavy Ion Collider or Large Hadron Collider, in the cores of neutron stars, and during the early universe. The estimated value of the magnetic field in the HIC experiment is around $10^{15-18}$ Gauss~\cite{Kharzeev:2007jp,Skokov:2009qp,Tuchin:2013apa} generated by the rapid movement of electrically charged spectators in the early stages of the collision.  Although it experiences rapid decay within a few fm/c, the finite electrical conductivity ($\sim$ a few MeV) of the medium can possibly delay the decay process sufficiently allowing a nonzero magnetic field to persist even during the subsequent hadronic phase following the phase transition or crossover from the Quark-Gluon Plasma  (QGP)~\cite{Gursoy:2014aka,Inghirami:2016iru,Kalikotay:2020snc}. The magnitude of the magnetic field in astrophysical objects such as  the interior of neutron stars, magnetars  is of the order of $10^{15}$ Gauss. The strength of these magnetic fields is comparable to the typical energy scale of quantum chromodynamics (QCD), affecting various microscopic and macroscopic properties of strongly interacting matter (see Refs.~\cite{Friman:2011zz,Miransky:2015ava,Kharzeev:2013jha} for reviews). Furthermore, the presence of background magnetic field results in a large number of interesting phenomena such as chiral magnetic effect \cite{CMS:2012sap,Kharzeev:2007jp,Fukushima:2008xe,Kharzeev:2009pj}, magnetic catalysis (MC) \cite{Shovkovy:2012zn,Gusynin:1994re,Gusynin:1995nb,Gusynin:1999pq}, inverse magnetic catalysis (IMC) etc.~\cite{Preis:2010cq,Preis:2012fh}.

The spontaneous breakdown of chiral symmetry and color confinement are fundamental to the QCD vacuum at low energies. 
Due to the non-perturbative nature of QCD in this energy domain first principle calculations are untenable. Though lattice QCD (LQCD) simulations provide some useful insights at zero baryon density, extension to finite chemical potentials is plagued with complications~\cite{Muroya:2003qs,Aoki:2006br,Bazavov:2009zn,Cheng:2007jq}.
Consequently, phenomenological models are used to probe the thermodynamic and spectral properties of strongly interacting matter at low energies~\cite{Klevansky:1992qe,Vogl:1991qt,Buballa:2003qv}.
The Nambu–Jona-Lasinio (NJL) model \cite{Nambu:1961fr,Nambu:1961tp} is one such extensively used model which respects the global symmetries of QCD, especially chiral symmetry. It is non renormalizable due to the point like quark interactions, which arises from integrating out the gluonic degrees of freedom in this effective description~\cite{Klevansky:1992qe,Bijnens:1995ww}. Hence, a proper regularization scheme is needed to tame the divergent integrals. The model parameters are fixed by reproducing phenomenological quantities like the pion-decay constant, quark condensate, pion mass and so on. However, the NJL model lacks confinement. To address both chiral symmetry breaking and quark confinement, the Polyakov loop extended NJL (PNJL) model introduces interactions between quarks and a static, homogeneous gluon-like field \cite{Ratti:2005jh,Ratti:2006wg}.
Significant study has been found in literature where PNJL model is used to investigate the deconfinement transition and chiral symmetry restoration in strongly interacting magnetized matter \cite{Andersen:2014xxa,Kharzeev:2013jha,Chaudhuri:2020lga,Gatto:2010qs,Fukushima:2008wg,Mattos:2021alf,Mattos:2021tmz,Wang:2022xxp}.
The presence of a uniform background magnetic field tends to catalyze the chiral condensate. This leads to an increase in the transition temperature from chiral symmetry broken to the restored phase with increasing magnetic field, known as magnetic catalysis (MC)~\cite{Gusynin:1994re,Gusynin:1999pq}. 
As a result the transition temperature from chiral symmetry broken to the restored phase also increases.
Results from LQCD simulations at higher temperatures however show an opposite trend indicating inverse magnetic catalysis (IMC) of the transition temperature~\cite{Bali:2011qj,Bali:2012av}.  To address this discrepancy  in NJL-type models extensive efforts have been put in~\cite{Bandyopadhyay:2020zte,Ferreira:2014kpa,Farias:2016gmy,Avancini:2018svs,Sheng:2021evj,Mao:2016fha}.
An interesting alternative stems from the consideration of nonzero values of the anomalous magnetic moment (AMM) of the quarks which results in a decrease in chiral transition temperature implying IMC~\cite{Fayazbakhsh:2014mca,Chaudhuri:2019lbw,Chaudhuri:2020lga,Chaudhuri:2021skc,Chaudhuri:2021lui,Ghosh:2020xwp,Xu:2020yag,Mei:2020jzn,Aguirre:2021ljk,Ghosh:2021dlo,Farias:2021fci}.

In a magnetized medium, the spatial component of the  energy-momentum tensor in the presence of magnetic field becomes anisotropic due to the breaking of rotational symmetry~\cite{Ferrer:2010wz, Ferrer:2020tlz, Ferrer:2022afu}. This affects the equation of state (EoS) of strongly interacting matter. As a result, the pressure becomes direction-dependent, varying with the orientation of the magnetic field. The anisotropy in pressure is studied in magnetized quark matter with AMM in Ref.~\cite{Chaudhuri:2022oru}. In literature, numerous studies focus on investigating the EoS of compact stars such as neutron stars, quark stars, and hybrid stars, incorporating the anisotropic nature of pressure with substantial implications~\cite{Chatterjee:2014qsa, Canuto:1968apg, Martinez:2003dz, Noronha:2007wg, Huang:2009ue, Ferrer:2010wz, Strickland:2012vu,Dexheimer:2012mk,Sinha:2013dfa,PeresMenezes:2015ukv,Menezes:2015fla,Ferrer:2015wca}. However, the current generating magnetic field is not considered and the system is assumed as boundaryless in the above studies. 

The speed of sound reflects the propagation of disturbances through the medium. It is influenced by thermodynamic variables such as pressure, volume, temperature, and chemical potential, and therefore, it is directly correlated with the EoS of the system. Hence, speed of sound can be used as an ideal probe for studying strongly interacting matter during its space-time evolution.
It provides crucial insights, exhibiting a local minimum during a crossover and approaching zero at the critical point and along corresponding spinodal lines. Therefore, it is a suitable candidate for studying the phase structure of QCD. As discussed, the presence of a magnetic field induces anisotropy in pressure, which subsequently impacts the speed of sound, leading to different values depending on the direction of the magnetic field. Lattice QCD calculation demonstrates a minimum in the speed of sound at temperature $T_0=156.5\pm1.5$ MeV, indicating a crossover between hadron gas and QGP~\cite{HotQCD:2018pds}. The speed of sound in QCD matter has been determined using various methods including LQCD ~\cite{Borsanyi:2020fev,Philipsen:2012nu,Aoki:2006we,HotQCD:2014kol}, the (Polyakov–)Nambu–Jona-Lasinio [(P)NJL] model~\cite{Goswami:2023eol,He:2022kbc,Marty:2013ita,Peterson:2023bmr}, the quark-meson coupling model \cite{Abhishek:2017pkp,Schaefer:2009ui}, the hadron resonance gas (HRG) model \cite{Venugopalan:1992hy,Bluhm:2013yga}, the field correlator method (FCM)~\cite{Khaidukov:2018lor,Khaidukov:2019icg}, MIT bag model \cite{Pal:2023dlv,Pal:2023quk} and the quasiparticle model~\cite{Mykhaylova:2020pfk}. Additionally, the speed of sound holds particular significance for the studies of compact star properties. It has a substantial impact on the mass-radius relationship, cooling rate, the maximum possible mass of a neutron star~\cite{Ozel:2016oaf} and tidal deformability. Analysis of current neutron star data indicate a substantial increase in the speed of sound at densities beyond the nuclear saturation density~\cite{Bedaque:2014sqa,Tews:2018kmu,McLerran:2018hbz,Fujimoto:2019hxv}. Furthermore, as indicated in Ref.~\cite{Jaikumar:2021jbw}, the speed of sound has crucial impact on the frequencies of gravitational waves generated by the $g$-mode oscillation of a neutron star.
  
Isothermal compressibility is an important parameter whose value indicates the relative stiffness of EoS. It is also relevant in the analysis of the phase structure of the system. In the presence of magnetic field, pressure anisotropy affects isothermal compressibility and takes different values depending on the direction of magnetic fields. Ref.~\cite{Dolan:2011jm} investigates compressibility of rotating black holes. Refs.~\cite{Bhattacharyya:2011na,Iwasaki:2004nz,Yang:2021rdo} explore compressibility in quark matter, revealing insights into phase transitions and magnetic field effects.

The quark number susceptibility measures the response of the number density to an infinitesimal change of the quark chemical potential and is closely related to the fluctuations of conserved quantities, such as baryon number and electric charge. In recent times, there has been notable interest in investigating fluctuations of conserved charges within a magnetic field background~~\cite{Ding:2022uwj, Ding:2023bft,Marczenko:2024kko,Vovchenko:2024wbg}. It is observed that a finite magnetic field significantly affects these observables, which are of direct phenomenological interest and can be measured experimentally.

Recently, we have studied characteristics of nuclear matter in presence of a background magnetic field in Ref.~\cite{Mondal:2023baz}. In the present article, we will study the speed of sound and isothermal compressibility in magnetized quark matter using the PNJL model, illustrating the dependency of these observables on the AMM of quarks. Additionally, we investigate the quark number susceptibility, normalized by its zero field value, which may be an indicator for the existence of magnetic field in the system. 

The article is organized as follows: Section \ref{WM} gives a brief introduction to the general formalism of the PNJL model, while Section \ref{SpdSound} elaborates on the expressions for speed of sound in different thermodynamic situations and the isothermal compressibility. Section \ref{Numerical} discusses results, with a summary and conclusion provided in Section \ref{SC}. Further details are given in the appendix.
 
\section{PNJL Model}\label{WM} 
The PNJL model is an effective model of strongly interacting matter that combines chiral symmetry breaking with the effects of Polyakov loop which represents the deconfinement aspects of the matter. The Lagrangian density of the two-flavor PNJL model taking into account the AMM of free quarks in the presence of a constant background magnetic field can be expressed as follows~\cite{Chaudhuri:2020lga,Fayazbakhsh:2014mca,Chaudhuri:2019lbw,Chaudhuri:2022oru}:
\begin{eqnarray}
\mathcal{L}=\bar{q}(x)\FB{i\gamma^\mu D_\mu-m+\gamma^0\mu_q+\frac{1}{2}\hat{a}\sigma^{\munu}F_{\munu}}{q}(x)+G\SB{\FB{\bar{q}(x)q(x)}^2+\FB{\bar{q}(x)i\gamma_5\tau q(x)}^2}-U\FB{\Phi,\bar{\Phi},T}.\label{Lag_PNJL}
\end{eqnarray} 
Here, ${q}(x)$ represents the quark doublet fields where we have omitted the flavor  $(f=u,~d)$ and color $(c=r,~g,~b)$ indices from the quark field $(q^{fc}(x))$ for convenience, $m$ is the current quark mass with $m_u=m_d=m$ to ensure the isospin symmetry of the theory at vanishing magnetic field, $\mu_q$ represents quark chemical potential, $F_{\munu}=\FB{\partial_\mu A_\nu-\partial_\nu A_\mu}$ is the electromagnetic field strength tensor, and $\sigma^{\munu}=\frac{i}{2}\SB{\gamma^\mu\gamma^\nu-\gamma^\nu\gamma^\mu}$. The constituent quarks interact with the abelian gauge field $A_\mu$ and $\rm SU_c(3)$ gauge field $A_\mu^a$ via the covariant derivative:
\begin{eqnarray}
D_\mu=\partial_\mu-i\hat{Q}A_\mu-iA_\mu^a
\end{eqnarray}
where $A_\mu$ describes the vector potential of the external magnetic field along $\hat{z}$ direction ( for Landau gauge $A_\mu=(0,0,xB,0)$), the electric charge of quark is $\hat{Q}=\text{diag}(2e/3,-e/3)$, $\hat{a}=\hat{Q}\hat{\kappa}$ where $\hat{\kappa}=diag(\kappa_u,\kappa_d)$ is a matrix in flavor space containing AMM of the quarks. 
	In this work we adopt the logarithmic form which is given by~\cite{Roessner:2006xn,Roessner:2006xn} 
\begin{eqnarray}
	U\FB{\Phi,\bar{\Phi},T}&=&T^4\TB{-\frac{1}{2}A(T)\bar{\Phi}\Phi+B(T)\text{ln}\SB{1-6\bar{\Phi}\Phi+4\FB{\bar{\Phi}^3+\Phi^3}-3\FB{\bar{\Phi}\Phi}^2}}\label{U.potential}\\\rm{with}~~~A(T)&=&a_0+a_1\FB{\frac{T_0}{T}}+a_2\FB{\frac{T_0}{T}}^2,~~~B(T)=b_3\FB{\frac{T_0}{T}}^3~~.
\end{eqnarray}
Here the logarithmic divergence of the Polyakov loop potential as $\Phi, \bar{\Phi}\to1$ in the Haar measure formulation ensures that Polyakov loop $\Phi$  remains physically bounded and approaches its maximum value of 1 only at asymptotically high temperatures. However, as shown in Ref.~\cite{Chaudhuri:2020lga}, the results are qualitatively similar for different choices of Polyakov loop potential. The parameters of the model are provided in table~\ref{Table1}~\cite{Roessner:2006xn,Fukushima:2010fe,Chaudhuri:2022oru}. 
\begin{table*}[htb!]
	\caption{\label{Table1}Parameters in PNJL Model~\cite{Roessner:2006xn,Fukushima:2010fe,Chaudhuri:2022oru}}
	\begin{tabular}{ccccccccc}
		\\\hline\hline \\
		$T_0$~(GeV)~~~~~~~~~~~~~~&$a_0$~~~~~~~~~~~~~~&$a_1$~~~~~~~~~~~~~~&$a_2$~~~~~~~~~~~~~~&$a_3$~~~~~~~~~~~~~~&G\\
		\\\hline\\
		0.270~~~~~~~~~~~~~~&3.51~~~~~~~~~~~~~~&-2.47~~~~~~~~~~~~~~&15.2~~~~~~~~~~~~~~&-1.7~~~~~~~~~~~~~~&${2.2}/{\Lambda^2}$\\
		\hline\hline
	\end{tabular}   
\end{table*}
Now, employing mean field approximation on the Lagrangian provided in Eq.~\eqref{Lag_PNJL}, one can obtain the thermodynamic potential for 2-flavour PNJL model using imaginary time formalism of finite temperature field theory as~\cite{Kapusta:2006pm,Chaudhuri:2022oru,Chaudhuri:2020lga} :
\begin{eqnarray}
\Omega=\frac{B^2}{2}+\frac{\FB{M-m_0}^2}{4G}+U\FB{\Phi,\bar{\Phi},T}-3\sum_{nfs}^{}\frac{|e_fB|}{2\pi}\int^{+\infty}_{-\infty}\frac{dp_z}{2\pi}\omega_{nfs}f_\Lambda-T\sum_{nfs}\frac{|e_fB|}{2\pi}\int^{+\infty}_{-\infty}\frac{dp_z}{2\pi}\SB{\text{ln}g^++\text{ln}g^-}\label{Omega.Potential}
\end{eqnarray}
where  $n$ is Landau level, $s\in\FB{\pm 1}$ and 
\begin{eqnarray}
g^+&=&g^+(\Phi,\bar{\Phi},T,\mu_q)=1+3\FB{\Phi+\bar{\Phi}e^{-\frac{\omega_{nfs}-\mu_q}{T}}}e^{-\frac{\omega_{nfs}-\mu_q}{T}}+e^{-3\frac{\omega_{nfs}-\mu_q}{T}},\\
g^-&=&g^-(\Phi,\bar{\Phi},T,\mu_q)=g^+(\bar{\Phi},\Phi,T,-\mu_q)~~.
\end{eqnarray}
The PNJL model is known to be non-renormalizable due to the point-like interactions between the quarks~\cite{Klevansky:1992qe}. Consequently, a regularization scheme is to be chosen to eliminate divergences in the medium-independent term in Eq.~\eqref{Omega.Potential}. In a magnetic field, the sharp momentum cutoff causes artifacts due to the replacement of continuous momentum with discrete Landau levels. Here, we have opted for the smooth cutoff regularization procedure and introduced a multiplicative form factor ~\cite{Chaudhuri:2022oru,Fukushima:2010fe}, which is given by the expression
\begin{eqnarray}
f_\Lambda=\frac{\Lambda^{2N}}{\Lambda^{2N}+\SB{p_z^2+(2n+1-s)|e_fB|}^N}\label{F.Lambda}~,
\end{eqnarray}
in Eq.~\eqref{Omega.Potential} to tame the divergent medium-independent integral. The parameters $N$ and $\Lambda$ are chosen to reproduce the vacuum values of the pion decay constant $f_\pi$ and the pion mass $m_\pi$. The energy eigenvalues of the quarks in the presence of a background magnetic field with AMM of quarks is obtained as
\begin{eqnarray}
\omega_{nfs}=\sqrt{p_z^2+\SB{\sqrt{M^2+(2n+1-s)|e_fB|}-s\kappa_fe_fB}^2}.
\end{eqnarray}
The constituent quark mass $M$ and the expectation values of the Polyakov loops $\Phi$ and $\bar{\Phi}$ can be obtained self-consistently from Eq.\eqref{Lag_PNJL} by using the stationary conditions :
\begin{eqnarray}
\frac{\partial\Omega}{\partial M}=0;~~\frac{\partial\Omega}{\partial\Phi}=0;~
\frac{\partial\Omega}{\partial\bar{\Phi}}=0.\label{Sta.Omega}
\end{eqnarray}
Now, using Eq.~\eqref{Omega.Potential}, Eq.~\eqref{Sta.Omega} leads to the following coupled equations~\cite{Chaudhuri:2022oru,Chaudhuri:2020lga} 
\begin{eqnarray}
M&=&m+6G\sum_{nfs}\frac{|e_fB|}{2\pi}\int^{+\infty}_{-\infty}\frac{dp_z}{2\pi}\frac{1}{\omega_{nfs}}\frac{M}{M_{nfs}}\SB{M_{nfs}-s\kappa_fe_fB}\FB{1-f^+-f^-}\label{SCE1}\\
\frac{\partial U}{\partial\Phi}&=&3T\sum_{nfs}\frac{|e_fB|}{2\pi}\int^{+\infty}_{-\infty}\frac{dp_z}{2\pi}\SB{\frac{e^{-\frac{\omega_{nfs}-\mu_q}{T}}}{g^+}+\frac{e^{-2\frac{\omega_{nfs}+\mu_q}{T}}}{g^-}}\label{SCE2}\\
\frac{\partial U}{\partial\bar{\Phi}}&=&3T\sum_{nfs}\frac{|e_fB|}{2\pi}\int^{+\infty}_{-\infty}\frac{dp_z}{2\pi}\SB{\frac{e^{-2\frac{\omega_{nfs}-\mu_q}{T}}}{g^+}+\frac{e^{-\frac{\omega_{nfs}+\mu_q}{T}}}{g^-}}\label{SCE3}
\end{eqnarray}
where 
\begin{eqnarray}
M_{nfs}&=&\sqrt{M^2+(2n+1-s)|e_fB|},\\
f^+&=&f^+(\Phi,\bar{\Phi},T,\mu_q)=\frac{1}{g^+}\SB{\FB{\Phi+2\bar{\Phi}e^{-\frac{\omega_{nfs}-\mu_q}{T}}}e^{-\frac{\omega_{nfs}-\mu_q}{T}}+e^{-3\frac{\omega_{nfs}-\mu_q}{T}}},\label{fPlus}\\
f^-&=&f^-(\Phi,\bar{\Phi},T,\mu_q)=f^+(\bar{\Phi},\Phi,T,-\mu_q)~~\label{fMinus}.
\end{eqnarray}
$\frac{\partial U}{\partial\Phi}$ and $\frac{\partial U}{\partial\bar{\Phi}}$ are obtained from Eq.~\eqref{U.potential} and expressed as : 
\begin{eqnarray}
\frac{\partial U}{\partial\Phi}&=&	\frac{\partial U(\Phi,\bar{\Phi},T)}{\partial\Phi}=-T^4\SB{\frac{1}{2}A(T)\bar{\Phi}+6B(T)\frac{\bar{\Phi}-2\Phi^2+(\bar{\Phi}\Phi)\bar{\Phi}}{1-6\bar{\Phi}\Phi+4(\Phi^3+\bar{\Phi^3})-3(\bar{\Phi}\Phi)^2}}\\
\frac{\partial U}{\partial\bar{\Phi}}&=&\frac{\partial U(\Phi,\bar{\Phi},T)}{\partial\Phi}\FB{\Phi\to\bar{\Phi},\bar{\Phi}\to\Phi}
\end{eqnarray}
Furthermore, the expressions for quark the number density, entropy density and magnetization are obtained as follows~\cite{Chaudhuri:2022oru,Chaudhuri:2020lga}:
\begin{eqnarray}
n_q=-\frac{\partial\Omega}{\partial\mu_q}=3\sum_{nfs}^{}\frac{|e_fB|}{2\pi}\int^{+\infty}_{-\infty}\frac{dp_z}{2\pi}\SB{f^+-f^-}\label{nq}
\end{eqnarray}
\begin{eqnarray}
s=-\frac{\partial\Omega}{\partial T}=-\frac{\partial U}{\partial T}+\sum_{nfs}^{}\frac{|e_fB|}{2\pi}\int^{+\infty}_{-\infty}\frac{dp_z}{2\pi}\SB{\text{ln}g^++\text{ln}g^-}+3T\sum_{nfs}^{}\frac{|e_fB|}{2\pi}\int^{+\infty}_{-\infty}\frac{dp_z}{2\pi}\SB{\frac{\omega_{nfs}-\mu_q}{T^2}f^++\frac{\omega_{nfs}+\mu_q}{T^2}f^-}
\end{eqnarray}
\begin{eqnarray}\label{Magnetization}
\mathcal{M}&=&-\frac{\partial\Omega}{\partial B}=-B+3\sum_{nfs}^{}\frac{|{e_f}|}{2\pi}\int^{+\infty}_{-\infty}\frac{dp_z}{2\pi}\omega_{nfs}f_\Lambda+3\sum_{nfs}^{}\frac{|{e_fB}|}{2\pi}\int^{+\infty}_{-\infty}\frac{dp_z}{2\pi}\omega_{nfs}\frac{\partial f_\Lambda}{\partial  B}+T\sum_{nfs}^{}\frac{|e_f|}{2\pi}\int^{+\infty}_{-\infty}\frac{dp_z}{2\pi}\SB{\text{ln}g^++\text{ln}g^-}\nn\\&&+3\sum_{nfs}^{}\frac{|e_fB|}{2\pi}\int^{+\infty}_{-\infty}\frac{dp_z}{2\pi}\frac{1}{\omega_{nfs}}\SB{1-\frac{s\kappa_fe_fB}{M_{nfs}}}\SB{\frac{2n+1-s}{2}|e_f|-s\kappa_fe_fM_{nfs}}\SB{f_\Lambda-f^+-f^-}
\end{eqnarray}
where the expression for $\frac{\partial f_\Lambda}{\partial  B}$ can be found in Appendix~\ref{A1}. 
Note that the expressions for all the thermodynamic quantities in absence of AMM can be obtained by putting $ \kappa = 0 $ which agree with those in Ref.~\cite{Kharzeev:2013jha,Gatto:2010pt}. To arrive at the results for $eB=0$ i.e. for a thermal medium one has to make the replacement
\begin{equation}\label{key}
	\sum_{nfs}^{}\frac{|e_fB|}{2\pi}\int^{+\infty}_{-\infty}\frac{dp_z}{2\pi} \to \int \frac{d^3p}{(2\pi)^3}  ~~~~~\text{and}~~~~~ \omega_{nfs}\to \omega_p =\sqrt{\vec{p}^2 + M^2}
\end{equation}
and compare with Ref.~\cite{Sasaki:2006ww}.
 
\section{Speed of Sound And Isothermal Compressibility}\label{SpdSound}
The speed of sound is calculated as the square root of the ratio between a change in pressure ($p$) and the corresponding shift in energy density ($\epsilon$). Typically, the determination of sound speed generally involves specifying a constant parameter, denoted as $x$, where $x\equiv s/n_q$, $s$, $n_q$, $T$, $\mu_q$ etc. during the propagation of a compression wave through a medium. It is defined as 
\begin{eqnarray}
	c^2_x=\FB{\frac{\partial p}{\partial\epsilon}}_x.
\end{eqnarray}
When the speed of sound is  used to study the hydrodynamic evolution of the hot and dense matter created in heavy ion collisions, it is essential to consider the appropriate trajectory through the QCD phase diagram. This is usually  achieved by following isentropic curves such that the entropy per baryon (or entropy density per baryon density) $ s/ n_q$ is constant~\cite{Yao:2023yda}. 
	Additionally, it is also interesting to calculate the squared speed of sound with constant $T$ and $\mu_q$ . For example, $c^2_T$ is commonly used when there is a temperature reservoir or when the cooling timescale is fast compared to the sound wave period. Such a physical scenario may exist in the interstellar medium~\cite{Sorensen:2021zme}. In this article, we will study the sound speed in quark matter in a background magnetic field. The expressions for longitudinal and transverse components of the pressure and energy density are as follows~\cite{Ferrer:2010wz,Ferrer:2022afu}:
\begin{eqnarray}
\epsilon&=&\Omega+Ts+\mu_qn_q~~,\\
p^{\parallel}&=&-\Omega~~,~~~p^{\perp}=p^\parallel-B\mathcal{M}~~.
\end{eqnarray}
Correspondingly, the speed of sound becomes anisotropic due to the presence of magnetic field.
We will define the speed of sound using various thermodynamic relations expressed in terms of temperature $T$ and $\mu_q$ as ~\cite{Mondal:2023baz}:
\begin{eqnarray}
{c_x^2}(T,\mu_q)&=&{c_x^2}^{(\parallel)}(T,\mu_q)=\FB{\frac{\partial p^\parallel}{\partial \epsilon}}_x={\frac{\partial( p^\parallel,x)}{\partial( \epsilon,x)}}={\frac{\partial( p^\parallel,x)/\partial(T,\mu_q)}{\partial( \epsilon,x)/\partial(T,\mu_q)}}=\frac{\FB{\frac{\partial{p^\parallel}}{\partial T}}_{\mu_q}\FB{\frac{\partial x}{\partial\mu_q}}_T-\FB{\frac{\partial{p^\parallel}}{\partial\mu_q}}_T\FB{\frac{\partial x}{\partial T}}_{\mu_q}}{\FB{\frac{\partial\epsilon}{\partial T}}_{\mu_q}\FB{\frac{\partial x}{\partial\mu_q}}_T-\FB{\frac{\partial\epsilon}{\partial\mu_q}}_T\FB{\frac{\partial x}{\partial T}}_{\mu_q}}\label{C2xParallel}~~,
\end{eqnarray}
\begin{eqnarray}
{c_x^2}^{(\perp)}(T,\mu_q)&=&\FB{\frac{\partial p^\perp}{\partial \epsilon}}_x={c_x^2}^{(\parallel)}-B\FB{\frac{\partial\mathcal{M}}{\partial\epsilon}}_x
={c_x^2}^{(\parallel)}-B\frac{\FB{\frac{\partial\mathcal{M}}{\partial T}}_{\mu_q}\FB{\frac{\partial x}{\partial\mu_q}}_T-\FB{\frac{\partial\mathcal{M}}{\partial\mu_q}}_T\FB{\frac{\partial x}{\partial T}}_{\mu_q}}{\FB{\frac{\partial\epsilon}{\partial T}}_{\mu_q}\FB{\frac{\partial x}{\partial\mu_q}}_T-\FB{\frac{\partial\epsilon}{\partial\mu_q}}_T\FB{\frac{\partial x}{\partial T}}_{\mu_q}}\label{C2xPerpendicular}
\end{eqnarray}
where ${c_x}^{(\parallel)}$ and ${c_x}^{(\perp)}$ are the sound velocities along and perpendicular to the magnetic field direction respectively. Using the thermodynamic relations given in Appendix~\ref{A2}, we can further write down the sound velocity along the magnetic field direction as:
\begin{eqnarray}
	{c_{s/n_q}^{2(\parallel)}}&=&\frac{n_qs\FB{\frac{\partial s}{\partial \mu_q}}_T-s^2\FB{\frac{\partial n_q}{\partial \mu_q}}_T-n_q^2\FB{\frac{\partial s}{\partial T}}_{\mu_q}+sn_q\FB{\frac{\partial n_q}{\partial T}}_{\mu_q}}{\FB{sT+\mu_qn_q}\SB{\FB{\frac{\partial s}{\partial\mu_q}}_T\FB{\frac{\partial n_q}{\partial T}}_{\mu_q}-\FB{\frac{\partial s}{\partial T}}_{\mu_q}\FB{\frac{\partial n_q}{\partial \mu_q}}_T}}\label{C2sbynB}~~,
\end{eqnarray}
and sound velocity perpendicular to the magnetic field direction as:
\begin{eqnarray}
{c_{s/n_q}^{2(\perp)}}&=&{c_{s/n_q}^{2(\parallel)}}-B\frac{n_q\SB{\FB{\frac{\partial\mathcal{M}}{\partial T}}_{\mu_q}\FB{\frac{\partial s}{\partial\mu_q}}_T-\FB{\frac{\partial\mathcal{M}}{\partial\mu_q}}_T\FB{\frac{\partial s}{\partial T}}_{\mu_q}}-s\SB{\FB{\frac{\partial\mathcal{M}}{\partial T}}_{\mu_q}\FB{\frac{\partial n_q}{\partial\mu_q}}_T-\FB{\frac{\partial\mathcal{M}}{\partial\mu_q}}_T\FB{\frac{\partial n_q}{\partial T}}_{\mu_q}}}{\FB{sT+\mu_qn_q}\SB{\FB{\frac{\partial s}{\partial\mu_q}}_T\FB{\frac{\partial n_q}{\partial T}}_{\mu_q}-\FB{\frac{\partial s}{\partial T}}_{\mu_q}\FB{\frac{\partial n_q}{\partial \mu_q}}_T}}\label{C2sbynBP}~~.
\end{eqnarray}
The expressions for \(\left(\frac{\partial \mathcal{M}}{\partial T}\right)_{\mu_q}\), \(\left(\frac{\partial \mathcal{M}}{\partial \mu_q}\right)_T\), \(\left(\frac{\partial s}{\partial \mu_q}\right)_T\), \(\left(\frac{\partial s}{\partial T}\right)_{\mu_q}\), \(\left(\frac{\partial n_q}{\partial \mu_q}\right)_T\), and \(\left(\frac{\partial n_q}{\partial T}\right)_{\mu_q}\) in Eqs.~\eqref{C2sbynB} and \eqref{C2sbynBP} are provided in Appendix~\ref{A3}.

In thermodynamics, isothermal compressibility is a measure of change in volume of the system with increasing pressure. It is considered a sensitive quantity for indicating the fluctuation of the order parameter during a phase transition. Its smaller value reflects relatively stiffer EoS. Mathematically, it is defined in the absence of a magnetic field as:
\begin{eqnarray}
K_T=-\frac{1}{V}\FB{{\frac{\partial V}{\partial P}}}_T=\frac{1}{n_q^2}\FB{\frac{\partial n_q}{\partial\mu_q}}_T~.
\end{eqnarray}
Because of the anisotropy in pressure, the isothermal compressibility $K_T$ splits into parallel and perpendicular components with respect to the direction of the magnetic field. The expressions for $K_T^{\parallel}$ and $K_T^{\perp}$ are given by
\begin{eqnarray}
K_T^{(\parallel)}&=&\frac{1}{n_q^2}\FB{\frac{\partial n_q}{\partial\mu_q}}_T~~,\label{KT_para}\\
K_T^{(\perp)}&=&\frac{1}{n_q\SB{n_q-B\FB{\frac{\partial\mathcal{M}}{\partial\mu_q}}_T}}\FB{\frac{\partial n_q}{\partial\mu_q}}_T.\label{KT_per}
\end{eqnarray}

\section{Numerical Results}\label{Numerical}
In this section, we present results for thermodynamic quantities in magnetized quark matter using the PNJL model under various physical situations. 
The model parameters are determined by fitting the empirical values of the pion decay constant $f_\pi=92.4\rm~MeV$ and the pion mass $m_\pi=138\rm~MeV$ and chiral condensate $\FB{\AB{\bar{u}u}}^{1/3}=-245.7\rm~MeV$ at zero temperature and zero baryon density in the absence of a background magnetic field. Specifically, the model parameters are $\Lambda=620\rm~MeV$, $m=5.5\rm~MeV$. Since we are interested for the thermodynamic properties of magnetized quark matter, the choice of the representative values of background magnetic field $eB=0.05\rm~GeV^2$ and $eB=0.10\rm~GeV^2$ along with $eB=0$ which will provide us the opportunity to explore the interplay between the magnetic field and the thermal effects. The AMM of quarks are taken as $\kappa_u=0.29016\rm~GeV^{-1}$ and $\kappa_d=0.35986\rm~GeV^{-1}$~\cite{Chaudhuri:2022oru,Fayazbakhsh:2014mca}. In this section, we evaluate the results with a finite magnetic field, considering up to 2000 Landau levels for convergence.
 
\subsection{Constituent Quark Mass}\label{ConstituentQuarkMass}

\begin{figure}[h] 
	\includegraphics[angle = -90, scale=0.23]{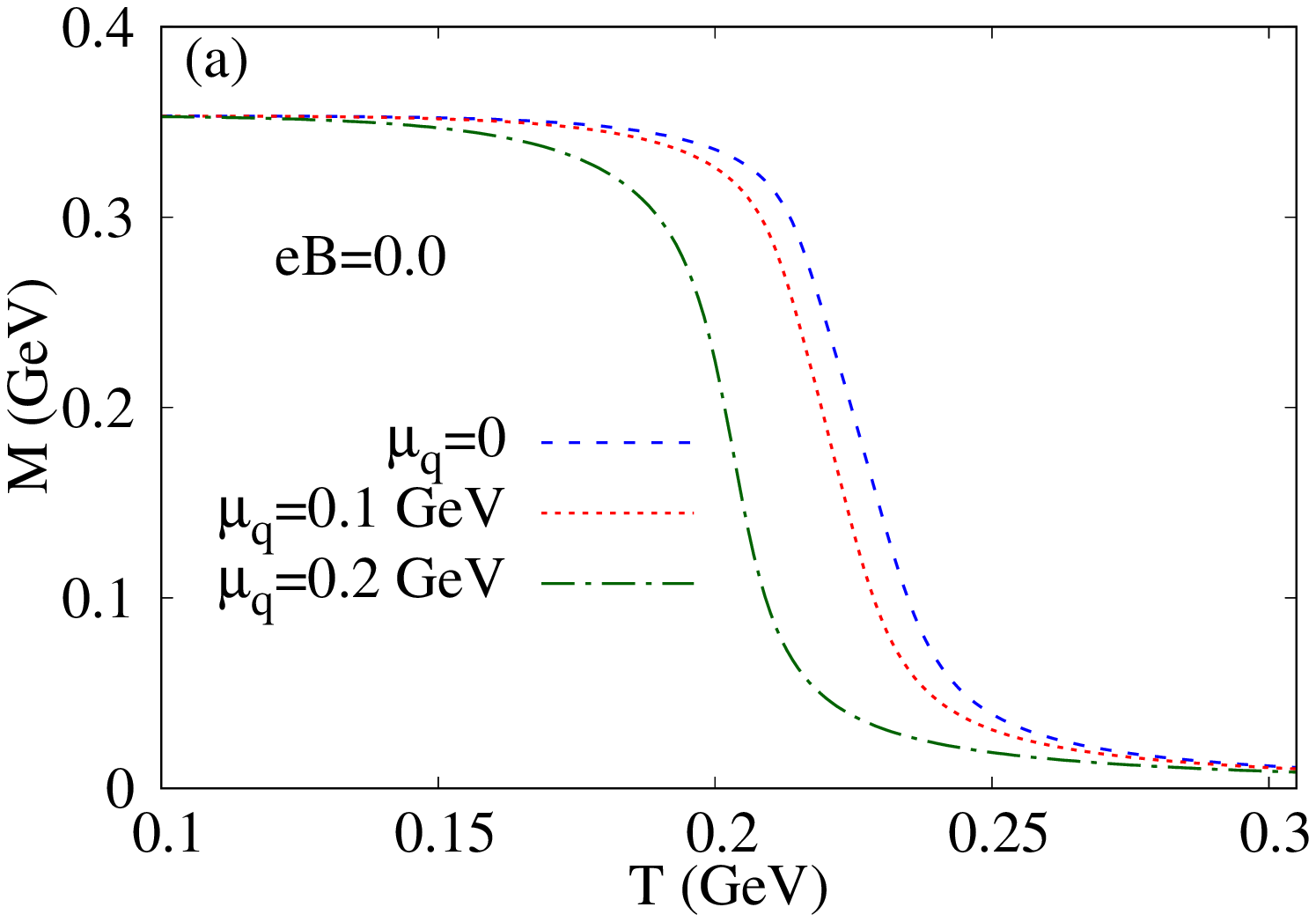}
	\includegraphics[angle = -90, scale=0.23]{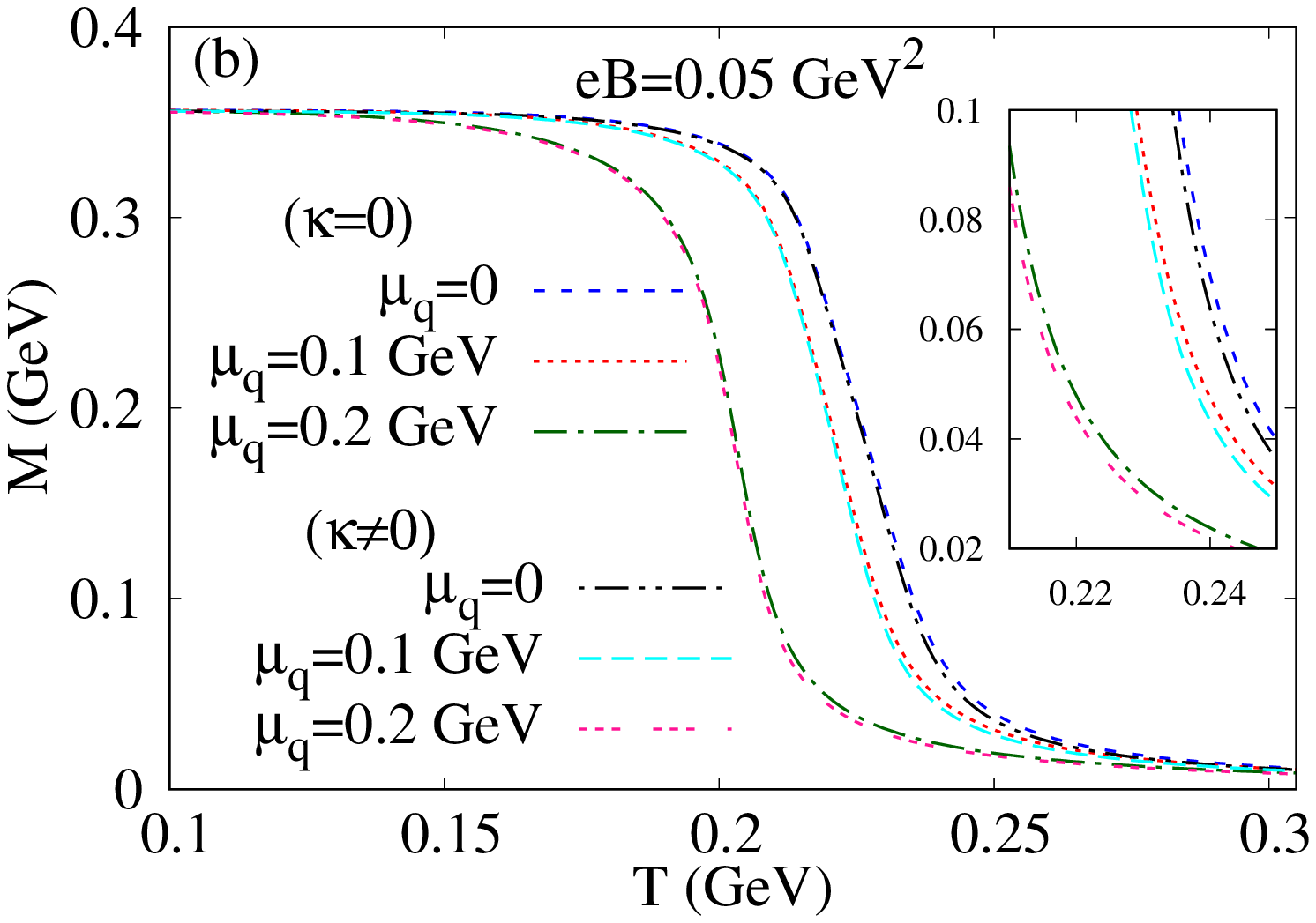} 
	\includegraphics[angle = -90, scale=0.23]{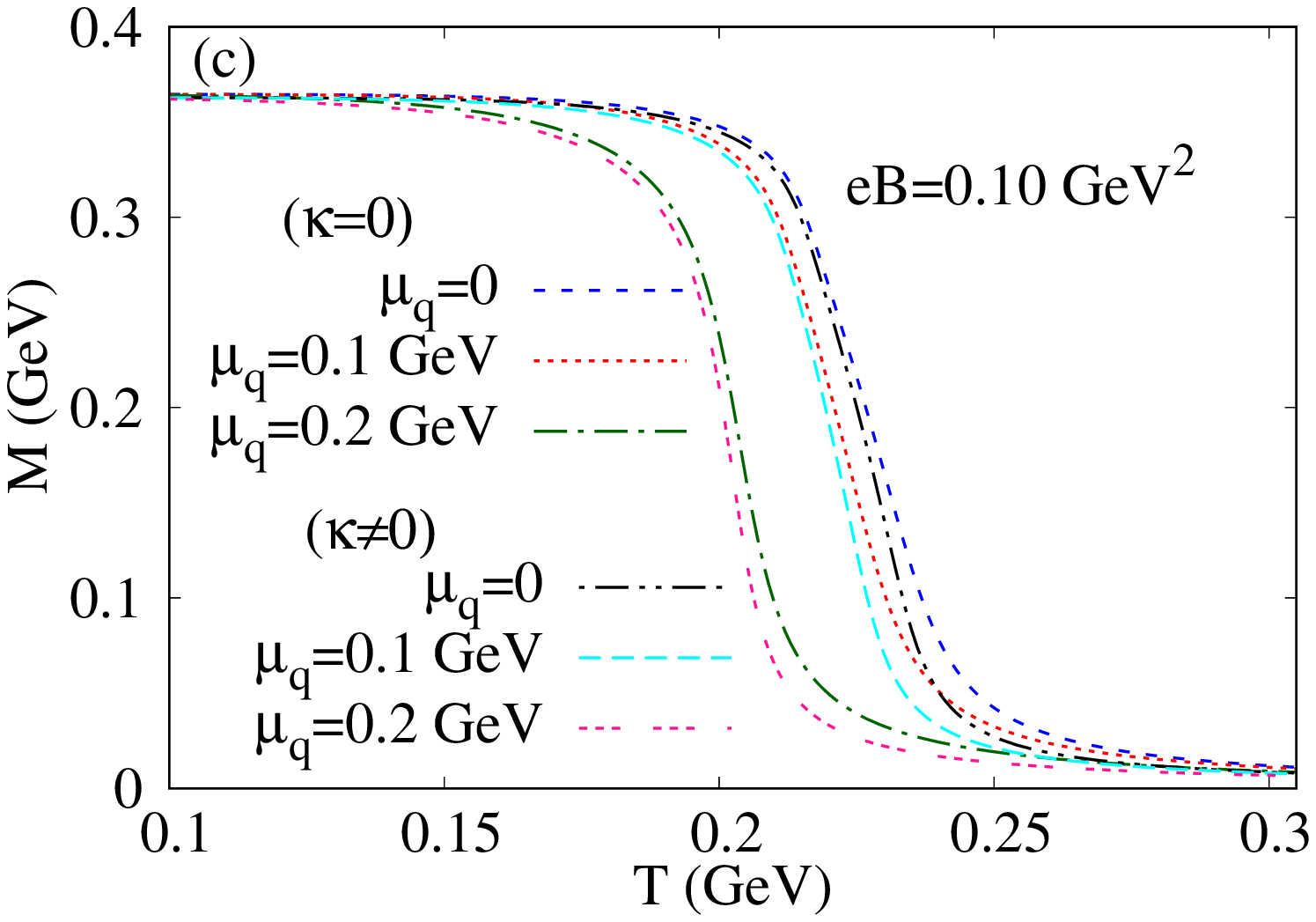}
	\caption{Constituent quark mass $(M)$ as function of temperature $(T)$ for different values of chemical potentials $\mu_q=0.0,~0.1,~0.2\rm~GeV$ at $(a)~eB=0$, $(b)~eB=0.05\rm~GeV^2$, $(c)~eB=0.10\rm~GeV^2$.}
	\label{TvsM}
\end{figure}

We start this section with the investigation of the constituent quark mass as a function of temperature, chemical potential and background magnetic field. This is achieved from the self-consistent solutions of Eqs.~\eqref{SCE1}-\eqref{SCE3}. Fig.~\ref{TvsM}(a) shows the variation of constituent quark mass with temperature for several values of $\mu_q=0.0,~0.1,~0.2\rm~MeV$ in the absence of magnetic field. The constituent quark mass starts at a high value and remains relatively stable at low temperatures indicating chirally broken phase. It then undergoes a rapid decrease within a narrow temperature range and eventually becomes nearly equal to the bare mass of quarks representing partial restoration of chiral symmetry. This transition occurs at $T^{\rm ch}\sim230 \rm~MeV$ in the absence of the magnetic field and zero chemical potential. As $\mu_q$ increases, Fig.~\ref{TvsM}(a) also depicts a decrease in the magnitude of $M$ and a shift of the transition temperature towards lower values.
In Figs.~\ref{TvsM}(b)-(c), the variation of constituent quark mass with temperature is shown for several values of $\mu_q$ at magnetic field strength $eB=0.05$ and $0.10\rm~GeV^2$ respectively considering both zero and non-zero values of AMM of quarks. It is observed that the constituent quark mass is greater in the scenario when quarks have zero AMM compared to when they have finite AMM (e.g. the cyan and red lines in the inset plot of Fig.~\ref{TvsM}(b) ). This difference becomes more evident with increasing magnetic field strength near the chiral phase transition (cyan and red lines in Figs.~\ref{TvsM}(b) and (c)). 
\begin{figure}[h] 
	\includegraphics[angle = -90, scale=0.23]{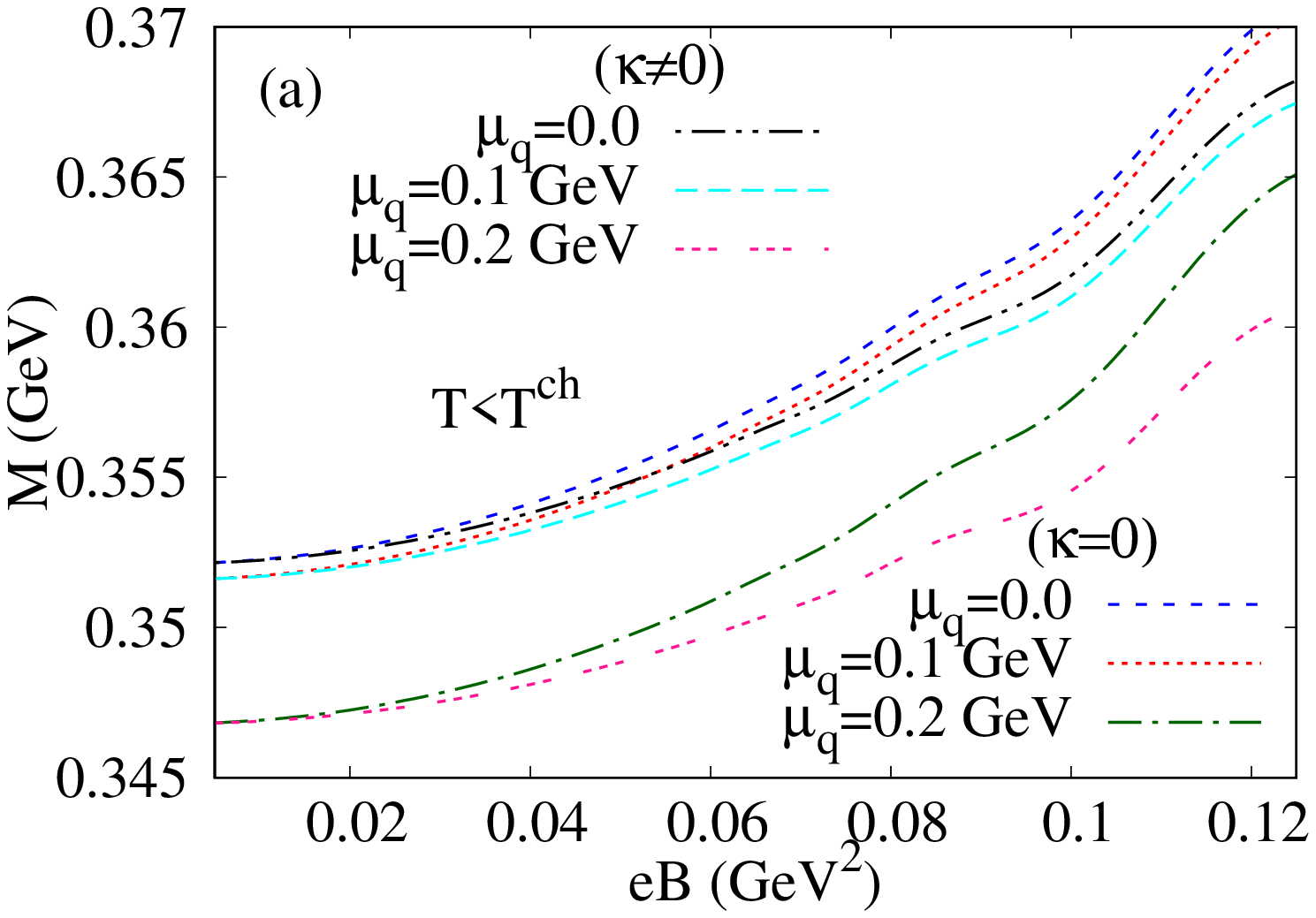}
	\includegraphics[angle = -90, scale=0.23]{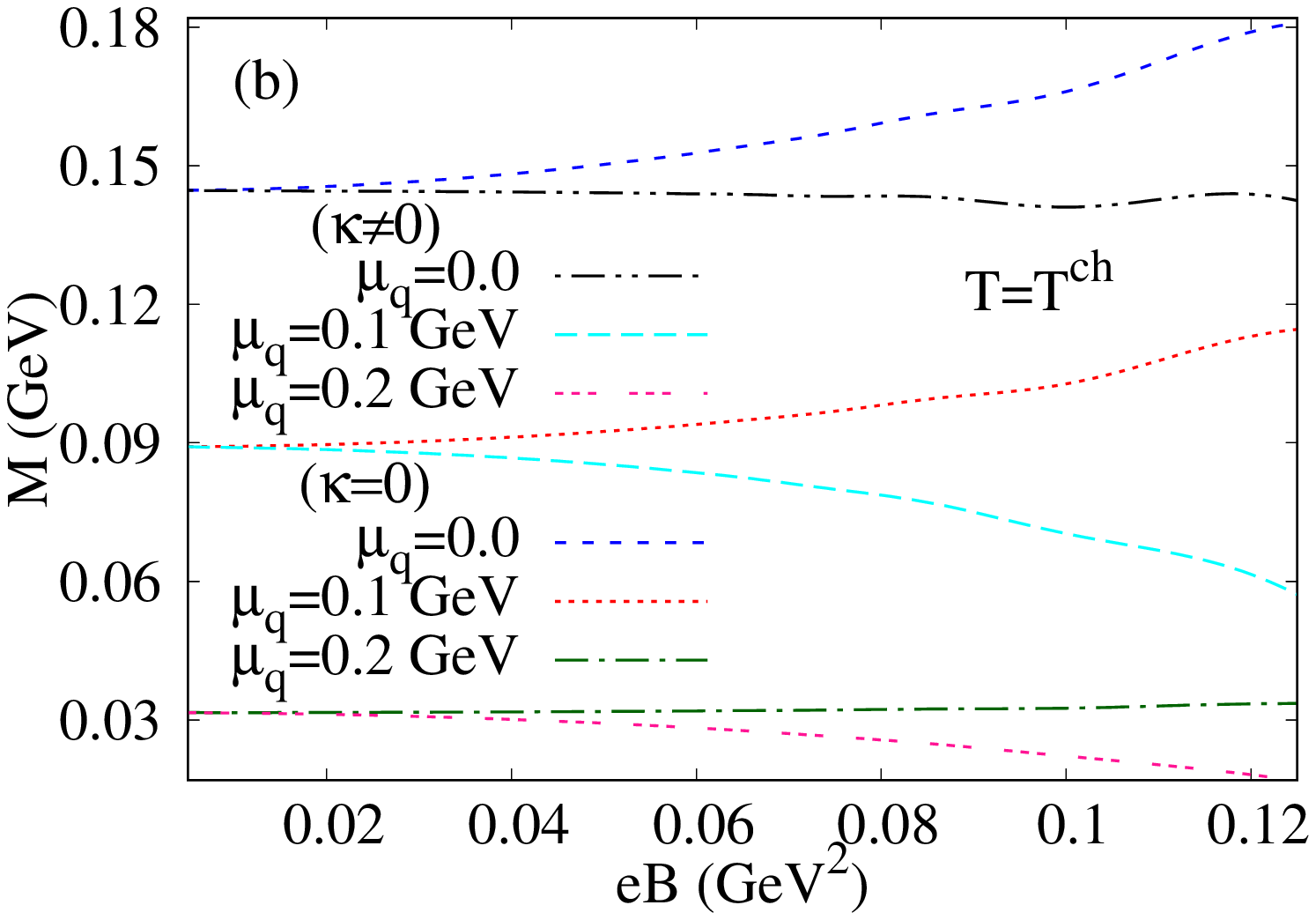} 
	\includegraphics[angle = -90, scale=0.23]{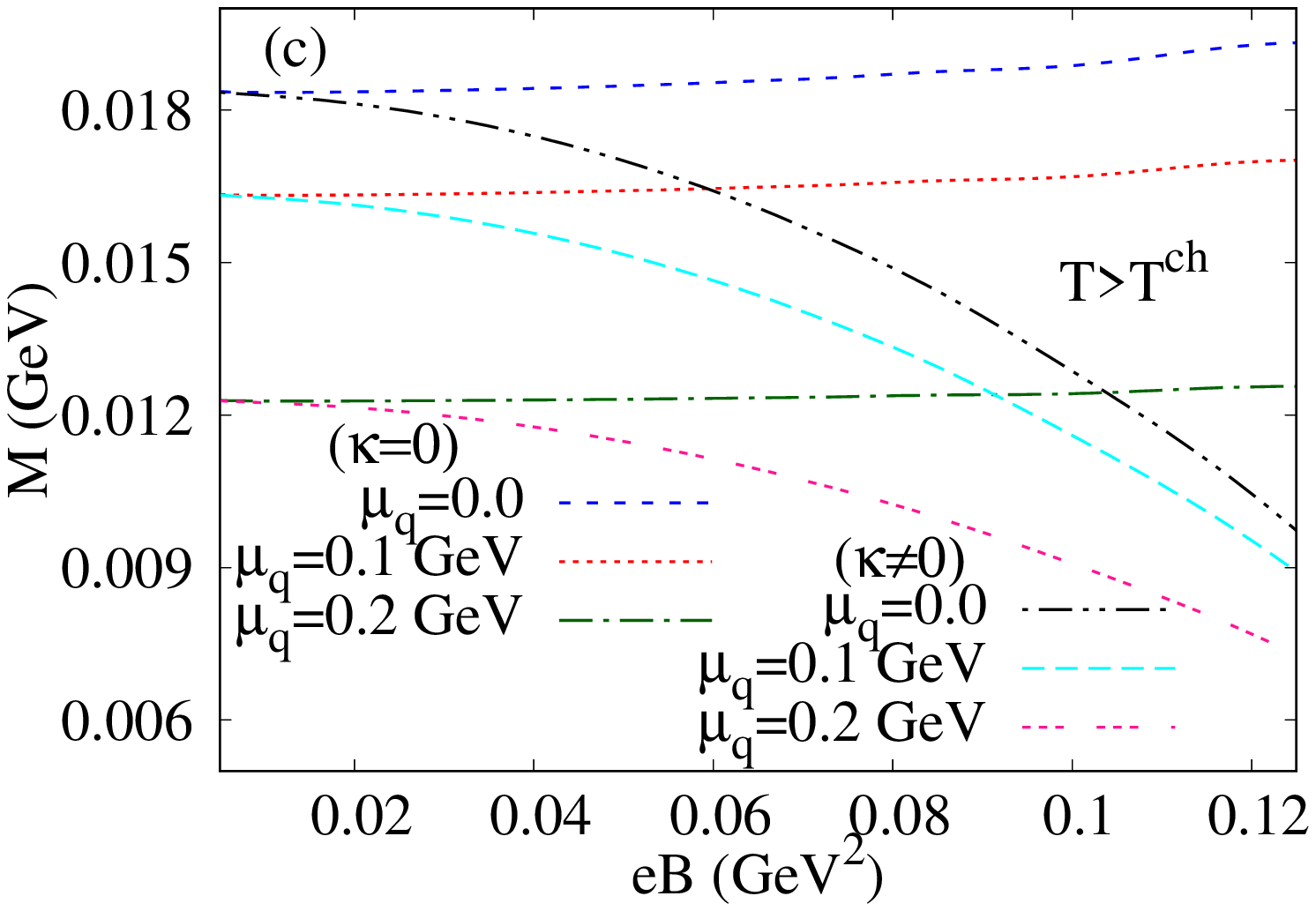}
	\caption{Constituent quark mass $(M)$ as function of background magnetic field $(eB)$ for different values of chemical potentials $\mu_q=0.0,~0.1,~0.2\rm~GeV$ at $(a)~T=150\rm~MeV$, $(b)~T=230\rm~MeV$, $(c)~T=275\rm~MeV$}
	\label{eBvsM}
\end{figure}

Next, we will investigate the magnetic field dependence of constituent quark mass for various values of $\mu_q$ both with and without AMM of quarks at three different temperatures representing various stages of the chiral phase. Analyzing Figs.~\ref{TvsM}(a)-(c), the different stages are as follows:

 (i) $T<T^{\rm ch}$ is indicating the broken chirally symmetry phase. We select $T=150\rm~MeV$ as representative temperature to investigate this phase.
 
 (ii) $T\sim T^{\rm ch}$ is representing region near chiral phase transition. We take $T=230\rm~MeV$ as representative temperature to investigate this phase.
 
 (iii) $T>T^{\rm ch}$ is signifying partial restoration of chiral symmetry phase. We choose $T=275\rm~MeV$ as representative temperature to investigate this phase.\\
Figs.~\ref{eBvsM}(a)-(c) illustrate the constituent quark mass as a function of background magnetic field for various
values of $\mu_q$ at three different temperatures representing the different phases as mentioned before considering both scenarios with and without AMM of quarks. In chirally broken phase as shown in Fig.~\ref{eBvsM}(a), $M$ increases with increasing $eB$ for all mentioned values of $\mu_q$ irrespective of whether the AMM of quarks are included or not, which is known as magnetic catalysis (MC). However, the increase in $M$ is slightly smaller with increasing $eB$ at each values of $\mu_q$ when AMM of quarks is included (e.g., blue and black lines in Figs.~\ref{eBvsM}(a)). Notably, the magnitude of $M$ consistently appears higher for a smaller value of $\mu_q$ across all values of $eB$ both with and without AMM of quarks. Near chiral phase transition as shown in Fig.~\ref{eBvsM}(b), $M$ increases with increasing $eB$ in the absence of AMM of quarks. Conversely, an opposite trend is observed when AMM of quarks is present and this is known as inverse magnetic catalysis (IMC). As the temperature increases partial chiral symmetry has been restored as shown in Fig.~\ref{eBvsM}(c). Here, the qualitative behaviour of the plots are similar as observed in Fig.~\ref{eBvsM}(b). Finally, in all the stages of phase transition the constituent quark mass difference between zero and non-zero AMM of quarks increases with increase $eB$ (e.g. black and blue lines in any of the subplots in Fig.~\ref{eBvsM}(a)-(c)), which is also evident from Figs.~\ref{TvsM}(a)-(c).
\subsection{Quark number density and quark number susceptibility}
\begin{figure}[h] 
	\includegraphics[angle = -90, scale=0.34]{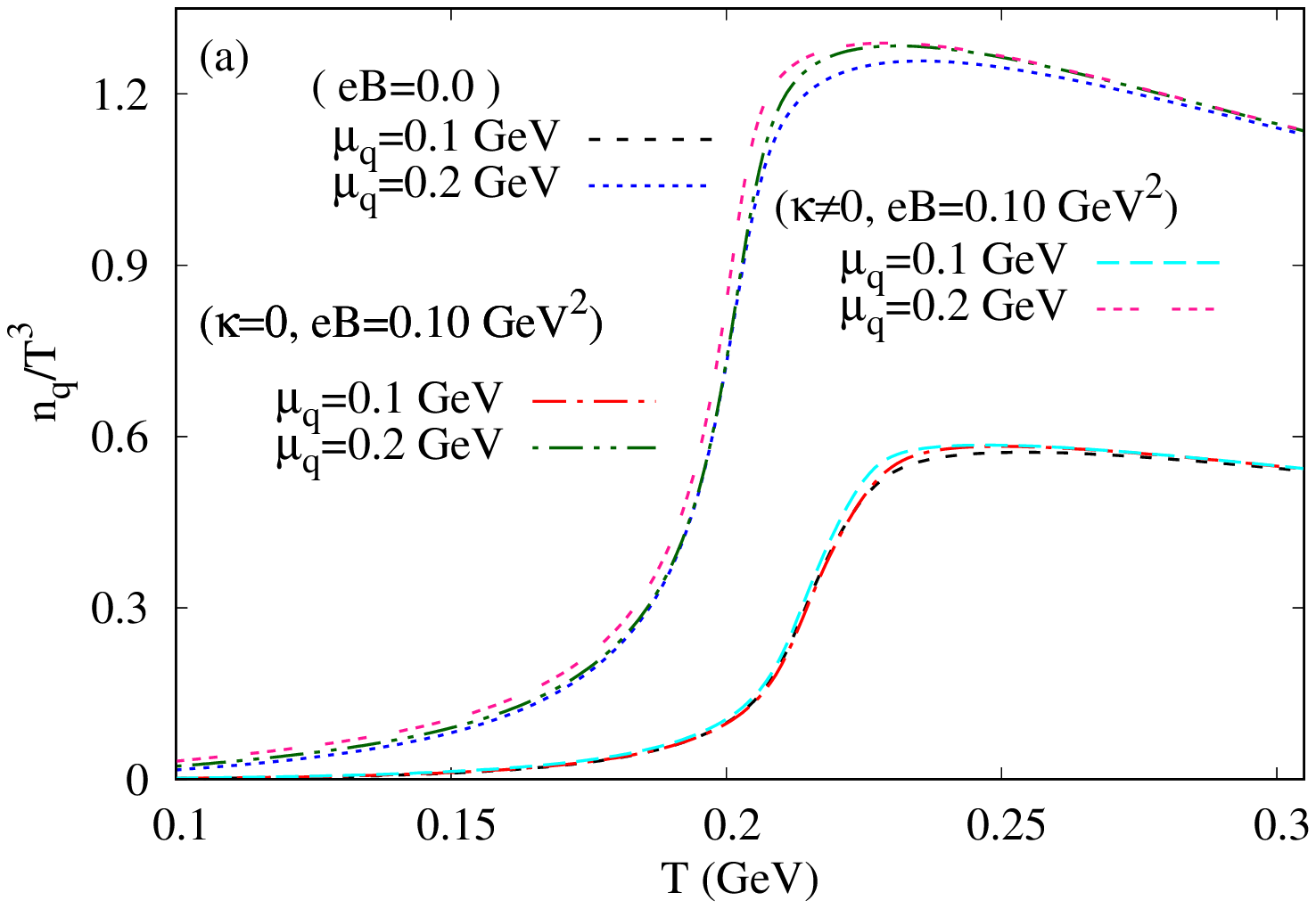}
	\includegraphics[angle = -90, scale=0.34]{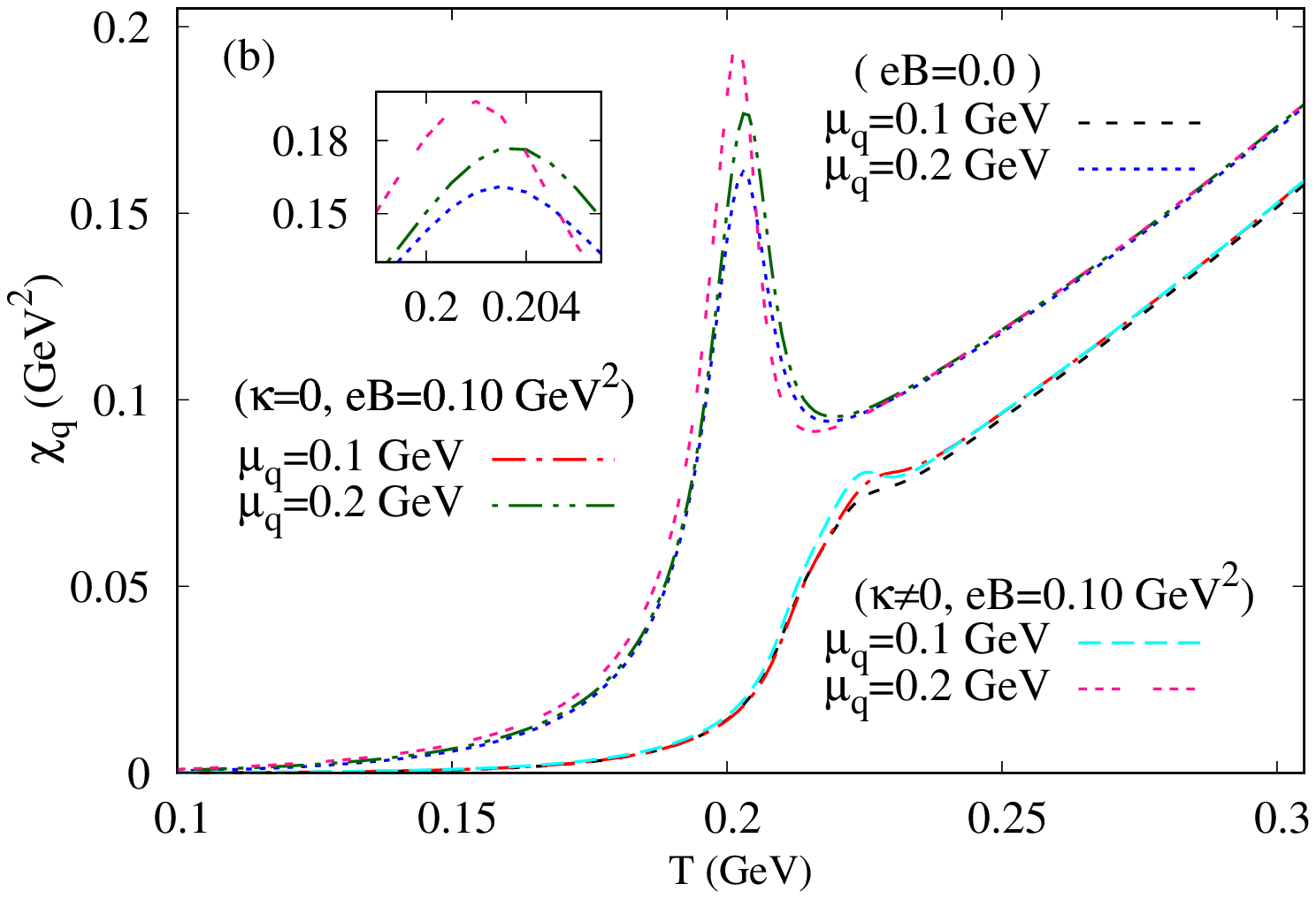}
	\caption{Variation of (a) scaled quark number density $(n_q/T^3)$ and (b) $\chi_q$=$\frac{dn_q}{d\mu_q}$ as function temperature $(T)$ for different values of $\mu_q$ and $eB$.}
	\label{TvsnqbyT3sbyT3}
\end{figure}

The scaled quark number density is defined as $n_q/T^3$. The behaviour of the quark number density is investigated following Refs.~\cite{Ratti:2005jh,Chaudhuri:2020lga}. The results for $n_q/T^3$ as a function of temperature are presented in Fig.~\ref{TvsnqbyT3sbyT3}(a) at $\mu_q=0.1$ and $0.2$ GeV considering the three cases: (i) $eB=0.0$, (ii) $eB=0.10~\rm GeV^2$, $\kappa=0$, (iii) $eB=0.10~\rm GeV^2$, $\kappa\ne0$. For case (i), the interactions of an effective gluon field suppress contributions from both one-quark and two-quark states to the density below the transition temperature. As a result, the three-quark state becomes more dominant. Thus, we observe a strong suppression in the density of quarks below the transition. However, this suppression becomes less effective for temperatures above the transition (e.g. blue line in Fig.~\ref{TvsnqbyT3sbyT3}(a)). This can be understood from Eqs.~\eqref{fPlus}, \eqref{fMinus} and \eqref{nq}~(see Ref.~\cite{Chaudhuri:2022oru} for details). Moreover, it is evident that the scaled quark density rises as $\mu_q$ increases. The same qualitative behaviour is also observed in the presence of background magnetic fields in case of (ii). This similarity arises because the change in $M$ is almost negligible compared to the zero field case at high temperature and the contributions from one-quark and two-quark states remain strongly superessed in the low temperature region below the transition as in the zero field case. Furthermore, when we include the AMM of quarks, i.e, in case of (iii), the value of $n_q/T^3$ increases (see e.g. magenta, blue and green lines Fig.~\ref{TvsnqbyT3sbyT3}(a)) compared to cases of (i) and (ii).    

Next, we focus on susceptibilities. For example, the chiral susceptibility carries signals of phase transitions and can be considered an order parameter for chiral transitions. The expression for the quark number susceptibility is given in Eq.~\ref{Appendix.dnqbydNu}. Fig.~\ref{TvsnqbyT3sbyT3}(b) shows the result for quark number susceptibilities ($\chi_q$) varying with temperature for $\mu$ = 0.1 and 0.2 GeV in the following three cases: (i) $eB=0.0$, (ii) $eB=0.10~\rm GeV^2$, $\kappa=0$, (iii) $eB=0.10~\rm GeV^2$, $\kappa\ne0$. In the presence of background magnetic field, the transition temperature ($T^{\chi_q}_c$) shifts towards higher values (blue and green lines in the inset plot of Fig.~\ref{TvsnqbyT3sbyT3}(b)) indicating magnetic catalysis (MC). On the other hand, $T^{\chi_q}_c$ moves towards lower values of temperature when AMM of quarks are included implying inverse magnetic catalysis (IMC) (blue and magenta lines in the inset plot of Fig.~\ref{TvsnqbyT3sbyT3}(b)). This can be observed from the inset plot in Fig.~\ref{TvsnqbyT3sbyT3}(b) at $\mu_q$ = 0.2 GeV. As the quark chemical potential decreases, the discontinuity in the curves vanishes indicating a crossover (black, red and cyan lines in Fig.~\ref{TvsnqbyT3sbyT3}(b)). 
\begin{figure}[h] 
	\includegraphics[angle = -90, scale=0.23]{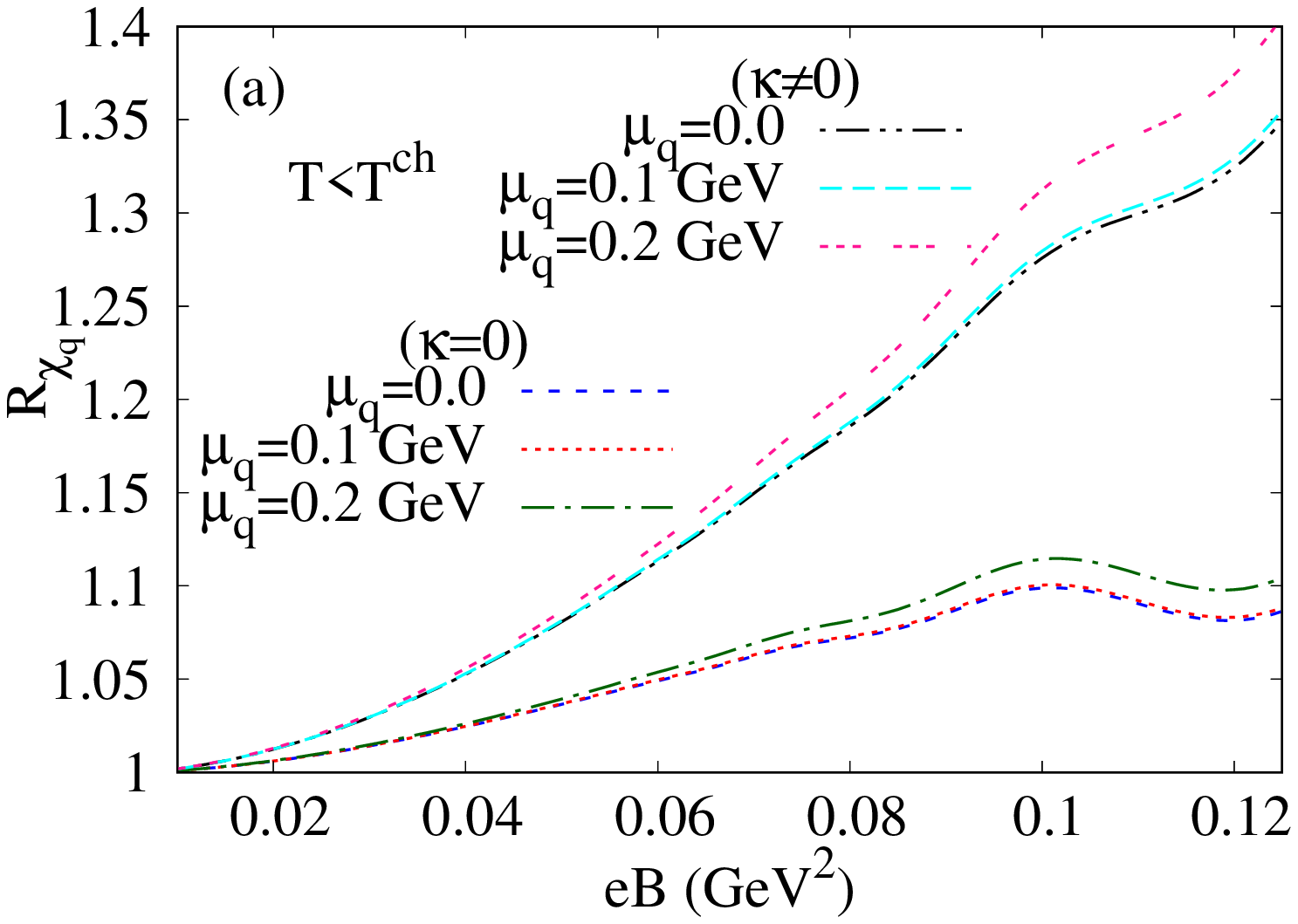}
	\includegraphics[angle = -90, scale=0.23]{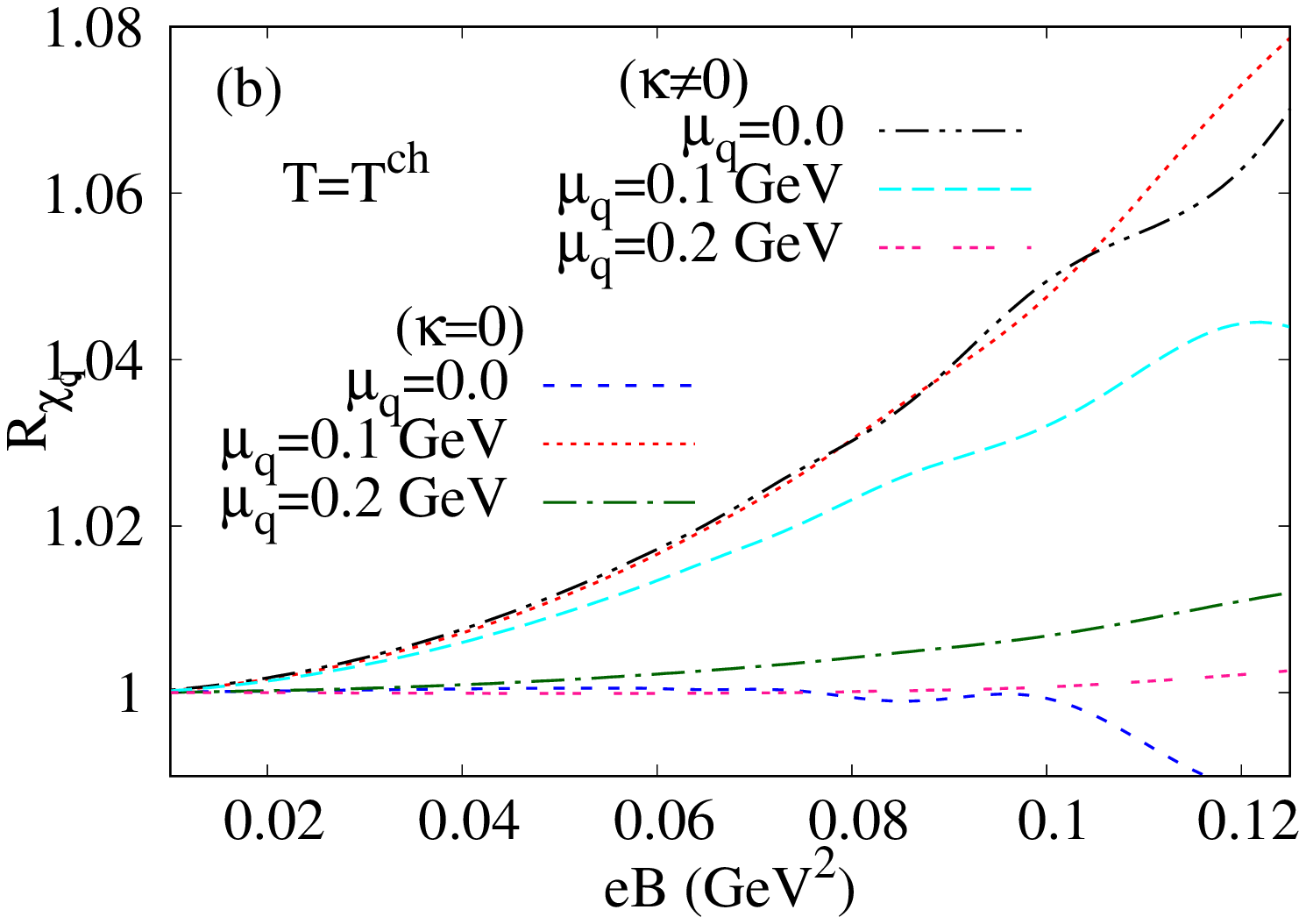}
	\includegraphics[angle = -90, scale=0.23]{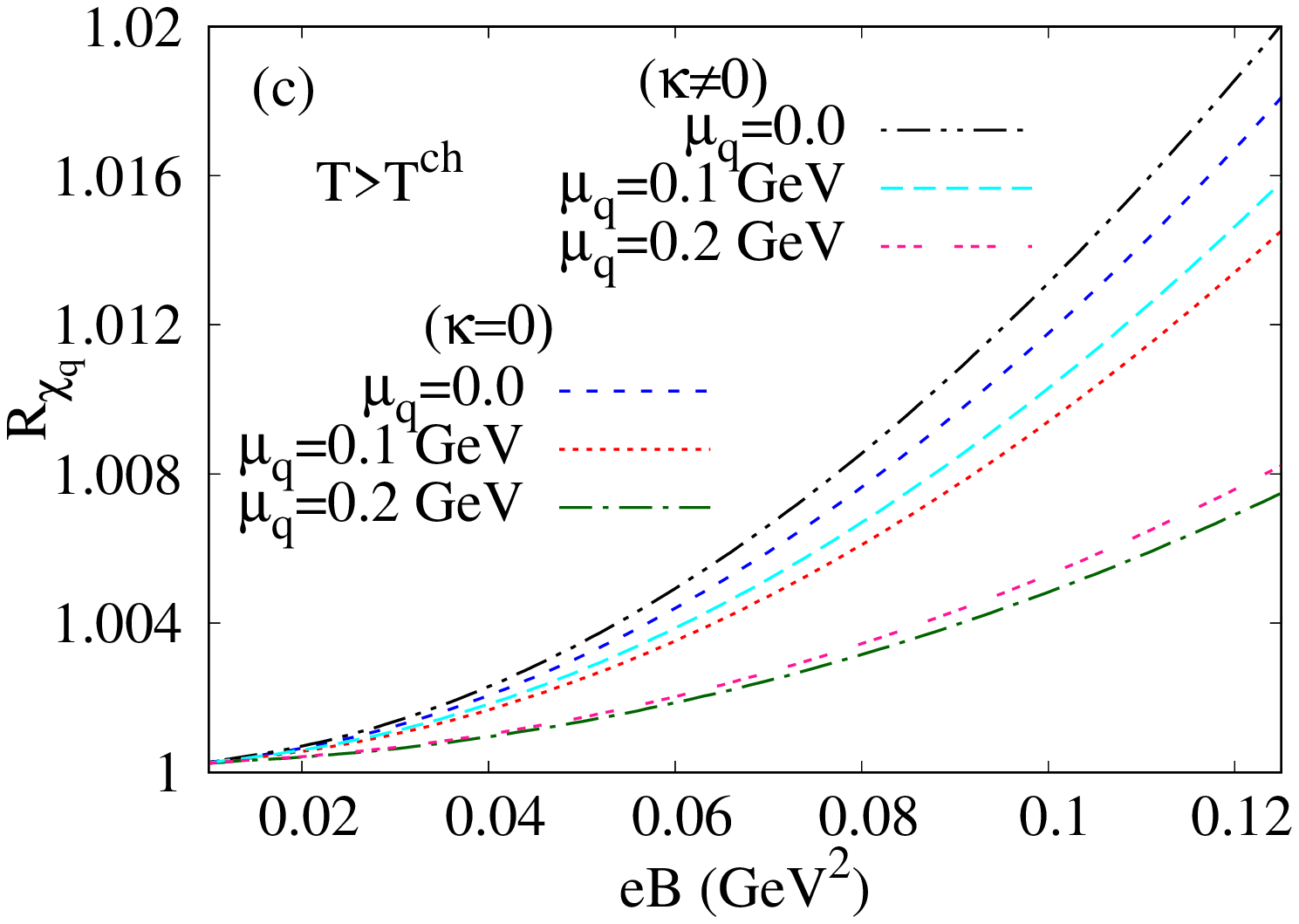}
	\caption{$R_{\chi_q}$ as a function of $eB$ for various values of $\mu_q$ with and without AMM at (a) chiral broken phase, (b) in the vicinity of chiral phase transition for $\mu_q=0,~eB=0$ (c) partial chiral restored phase.}
	\label{eBvsRxq}
\end{figure}
 We now study the normalized susceptibility at three different temperatures where it is defined as the ratio of susceptibilities obtained in non-central collisions (strong magnetic fields) to that of central collisions (weak or vanishing magnetic fields) following Refs.~\cite{Ding:2022uwj, Ding:2023bft}. It is defined as
\begin{eqnarray}
	R_{\chi_q}=\frac{\chi_q(eB\ne0)}{\chi_q(eB=0)}.
\end{eqnarray}
Therefore, by definition, $R_{\chi_q}$ will be unity in central collisions or vanishing (weak) magnetic fields. So, $R_{\chi_q}$ can possibly be considered as a quantity sensitive to the existence of magnetic field in the system. Here, we will present $R_{\chi_q}$ as a function of $eB$ for various values of $\mu_q$ with and without AMM of quarks at different stages of chiral phase transition as mentioned previously in subsection \ref{ConstituentQuarkMass}. Figs.~\ref{eBvsRxq}(a)-(c) represent the stages as $T<T^{\rm ch}$, $T=T^{\rm ch}$ and $T>T^{\rm ch}$ respectively. In chirally broken phase, Fig.~\ref{eBvsRxq}(a) shows that $R_{\chi_q}$ has a rising nature with increasing magnetic fields and $R_{\chi_q}$ also increases with higher values of $\mu_q$ (see e.g. red and green lines in Figs.~\ref{eBvsRxq}(a)). Furthermore, the overall magnitude of $R_{\chi_q}$ is larger with AMM of quarks (black and blue lines in Figs.~\ref{eBvsRxq}(a)) but the behavioural nature is the same. In the vicinity of chiral phase transition Fig.~\ref{eBvsRxq}(b) also shows that $R_{\chi_q}$ increases in presence of AMM of quarks. Fig.~\ref{eBvsRxq}(c)  represents the partial chirally restored phase. Here $R_{\chi_q}$ exhibits a rising nature with increasing $eB$ and  its magnitude further increases with the inclusion of the AMM of quarks. However, $R_{\chi_q}$ decreases with higher values of $\mu_q$ with or without AMM of quarks. It is also evident from the Figs.~\ref{eBvsRxq}(a)-(c) that  $R_{\chi_q}$ decreases with increasing temperature. Therefore, it can be concluded that magnetic field effect is small in the partial chirally restored phase compared to the chirally broken and nearly restored phase.  

\subsection{Magnetization of The Medium}
The expression for the magnetization is given in Eq.~\eqref{Magnetization}. We define, for convenience, the scaled magnetization ($\mathcal{M}_{\rm Scaled}$) as:
\begin{eqnarray}
	e\mathcal{M}_{\rm Scaled}=\mathcal{M}(T, \mu_q, eB)-\mathcal{M}(T=0, \mu_q=0, eB).
\end{eqnarray}
\begin{figure}[h] 
	\includegraphics[angle = -90, scale=0.23]{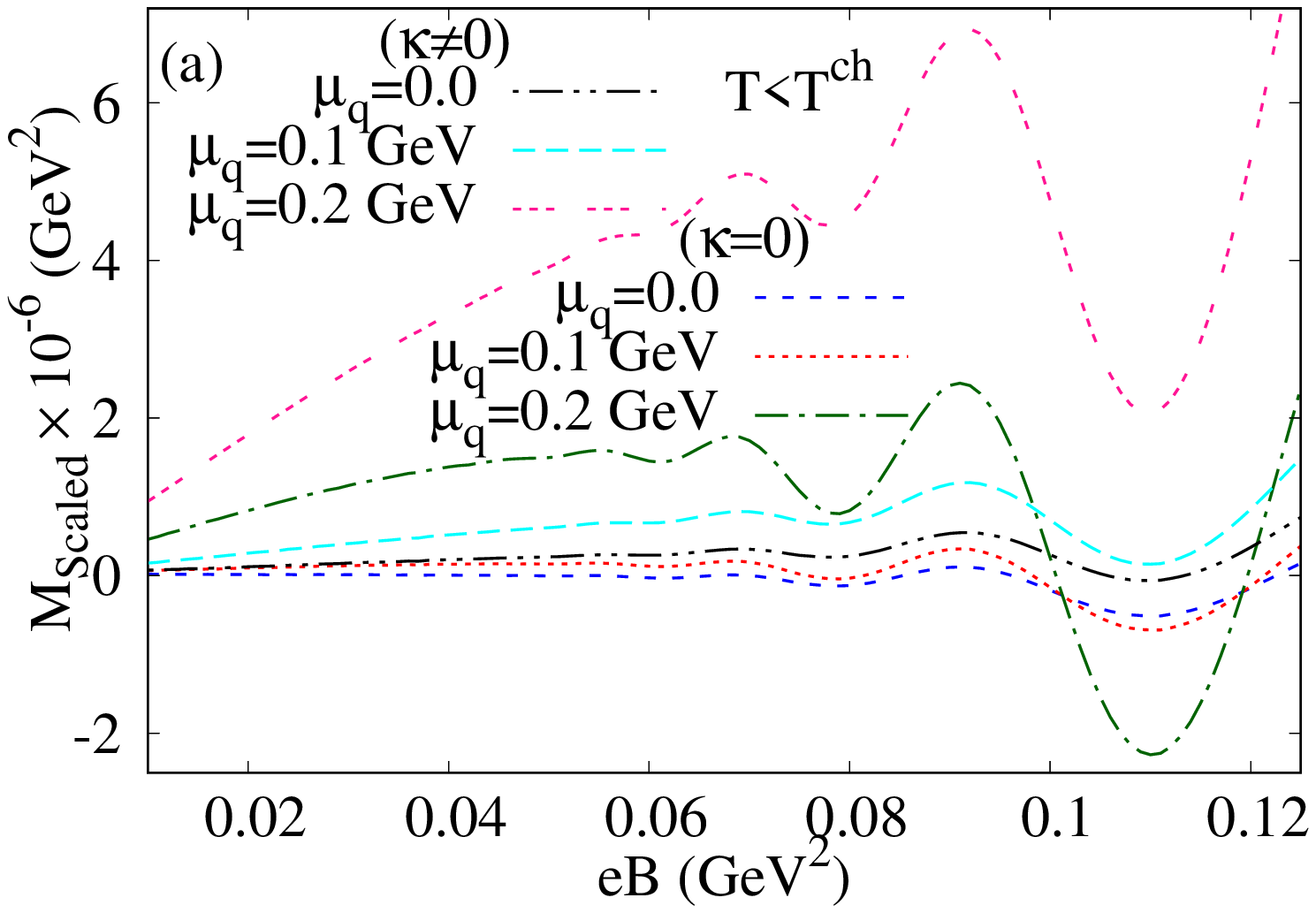}
	\includegraphics[angle = -90, scale=0.23]{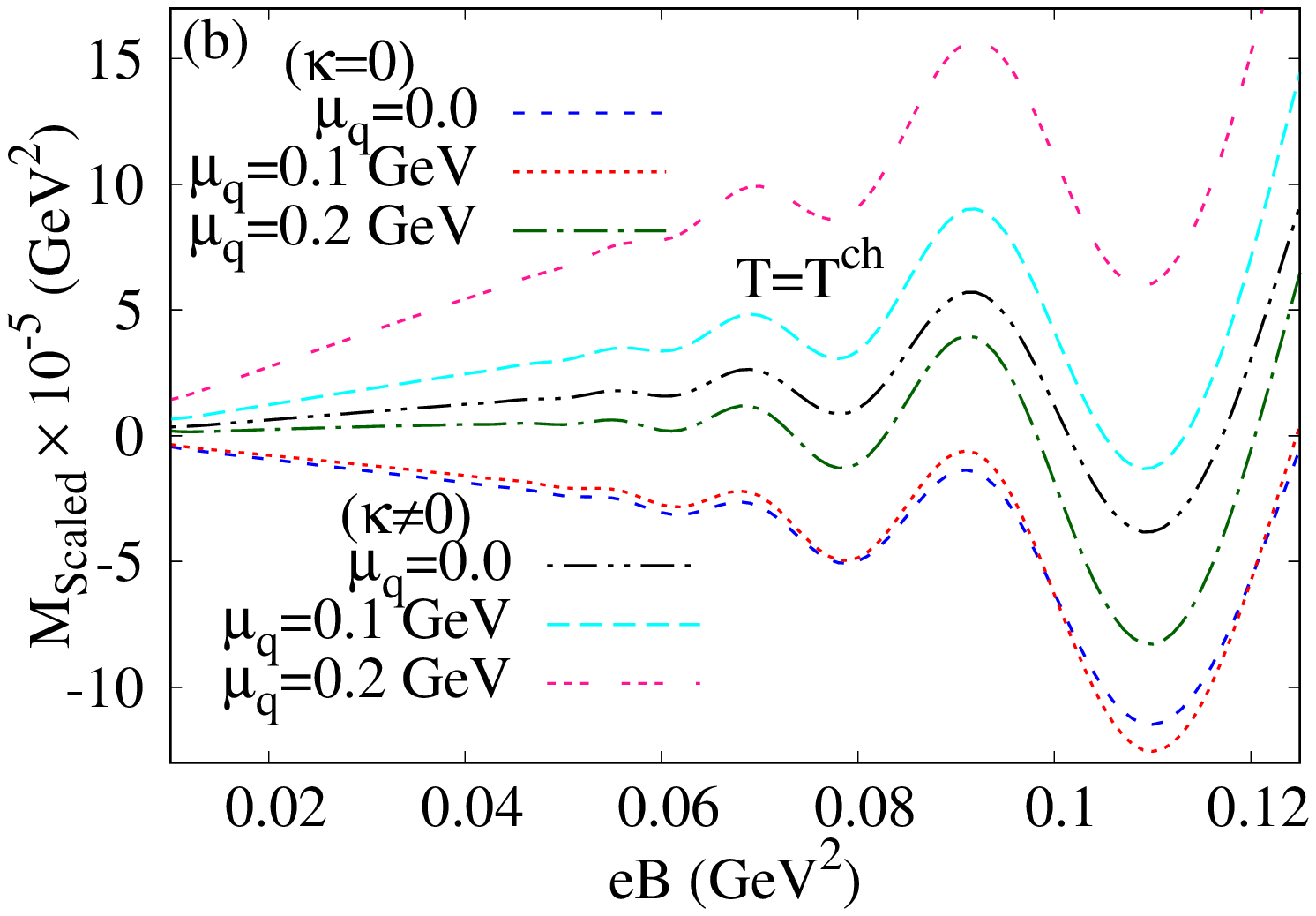}
	\includegraphics[angle = -90, scale=0.23]{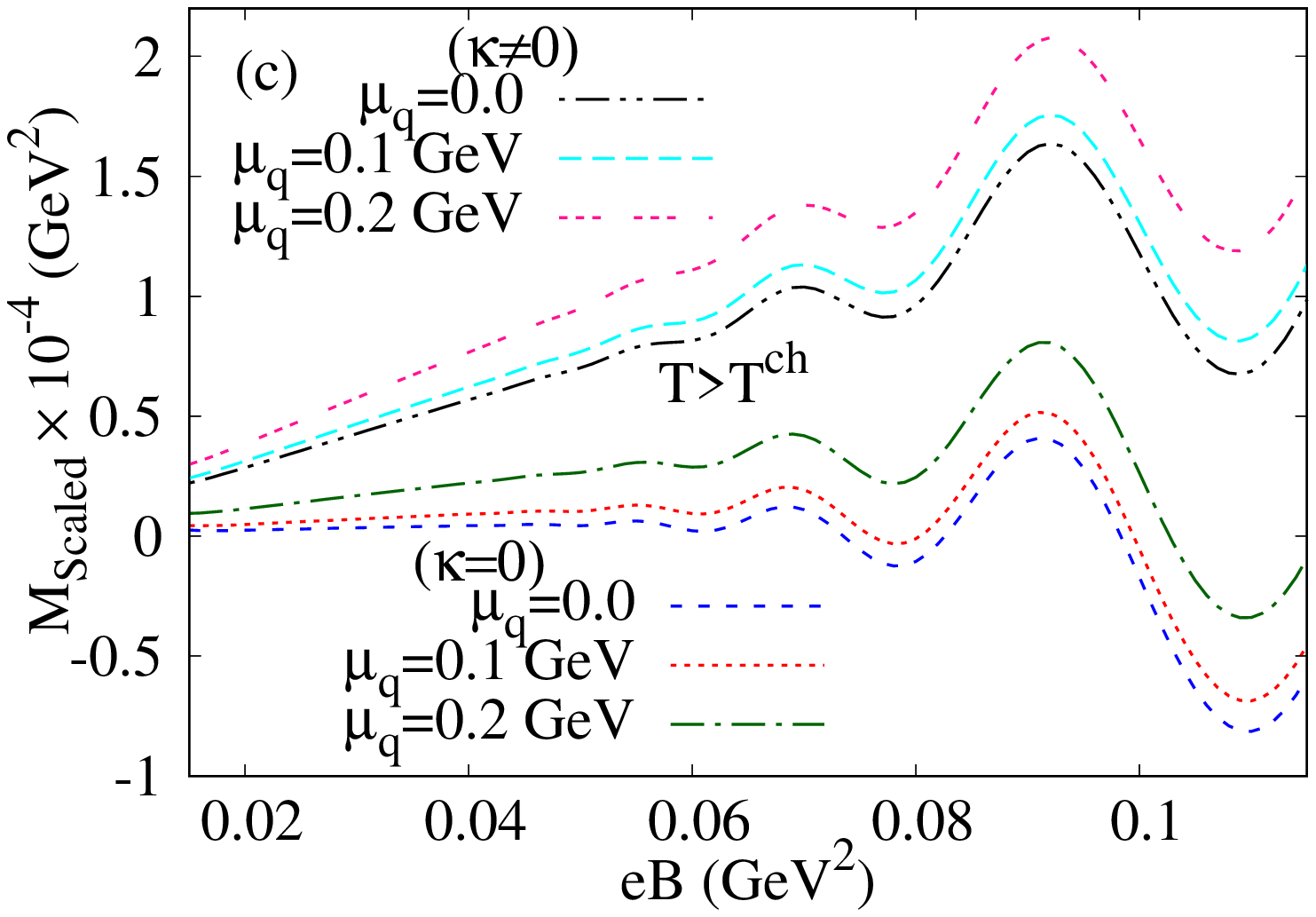}
	\caption{Scaled magnetization ($\mathcal{M}_{\rm Scaled}$) as a function of $eB$ for various values of $\mu_q$ at (a) chiral broken phase, (b) in the vicinity of chiral phase transition (c) partial chiral restored phase.}
	\label{eBMagnetization}
\end{figure}
In Figs.~\ref{eBMagnetization}(a)-(c), we illustrate the variation of scaled magnetization ($\mathcal{M}_{\rm scaled}$) as a function of the background magnetic field for various values of $\mu_q$ both with and without AMM of the quarks at three different temperatures representing chiral symmetry breaking, near chiral phase transition and its restoration respectively as discussed in subsection \ref{ConstituentQuarkMass}. In chirally broken phase, it is evident from Fig.~\ref{eBMagnetization}(a) that $\mathcal{M}_{\rm Scaled}$ shows an oscillating trend as the magnetic field $eB$ increases.
The oscillating trend and magnitude of $\mathcal{M}_{\rm Scaled}$ increase as the temperature increases as evident in Figs.~\ref{eBMagnetization}(a)-(c). Finally, it is observed from each subplot that $\mathcal{M}_{\rm Scaled}$ is large when the AMM of quarks is considered (black and blue lines in any of the subplots of Figs.~\ref{eBMagnetization}(a)-(c)).

\subsection{Speed of Sound at Constant $s/n_B$}
In this subsection, we study the variation of speed of sound with temperature in presence of a background magnetic field in quark matter. As discussed in section \ref{SpdSound}, ${c_{s/n_q}^2}$ splits into ${c_{s/n_q}^{2(\parallel)}}$ and ${c_{s/n_q}^{2(\perp)}}$ along and perpendicular to the direction of the background magnetic field respectively. We use Eqs.~\eqref{C2sbynB} and \eqref{C2sbynBP} to determine the speed of sound.

\begin{figure}[h] 
	\includegraphics[angle = -90, scale=0.23]{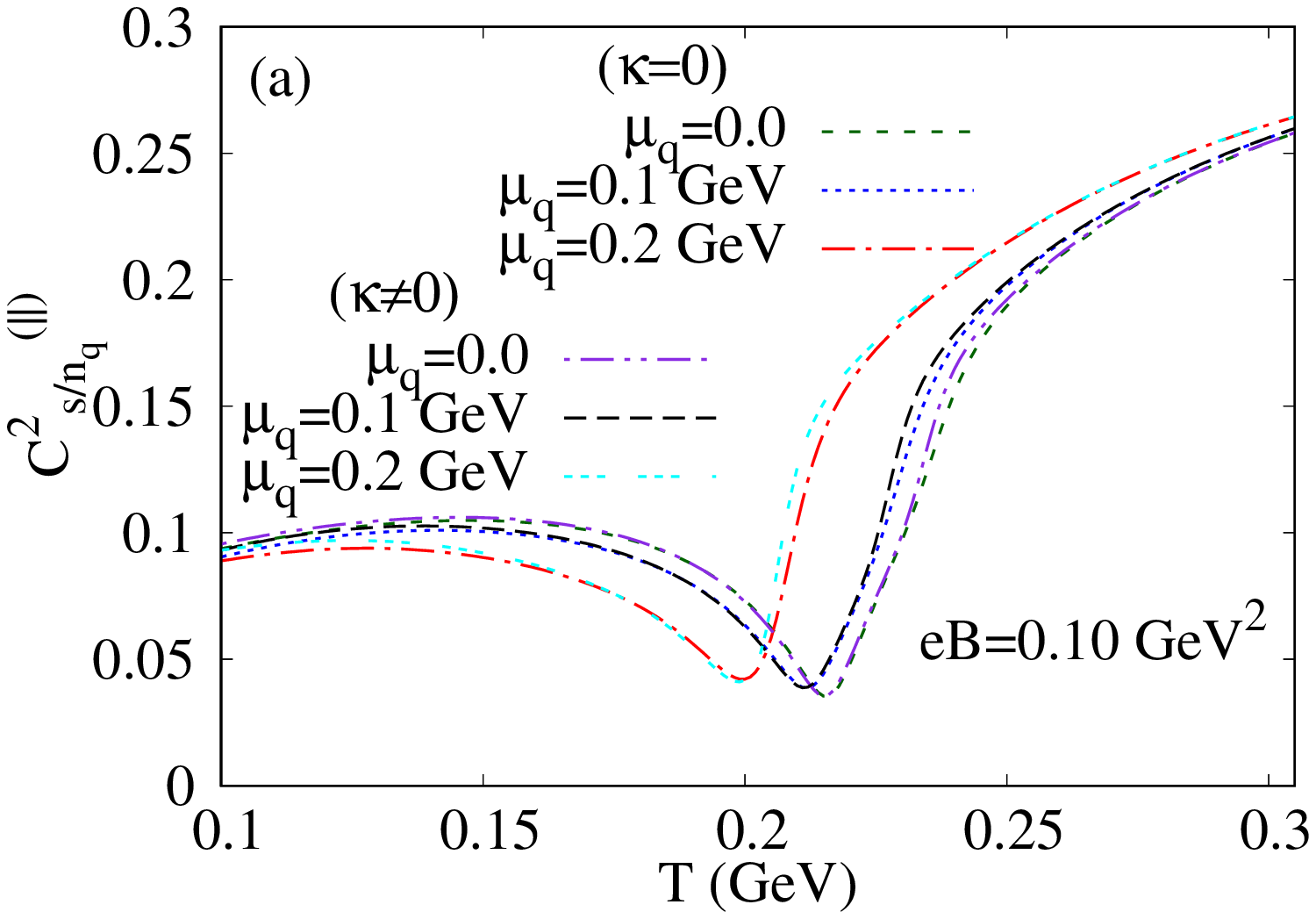}
	\includegraphics[angle = -90, scale=0.23]{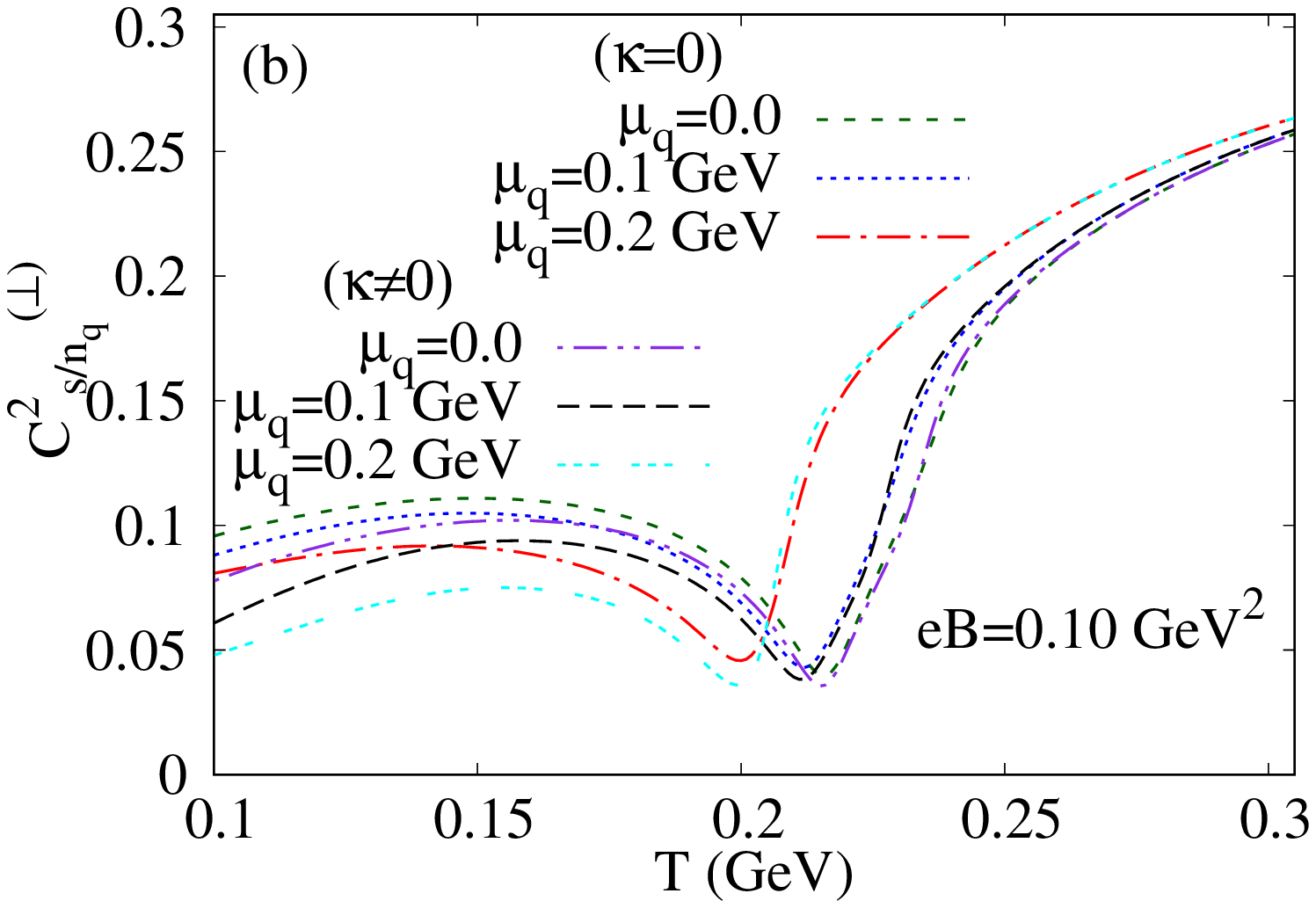}
	\includegraphics[angle = -90, scale=0.23]{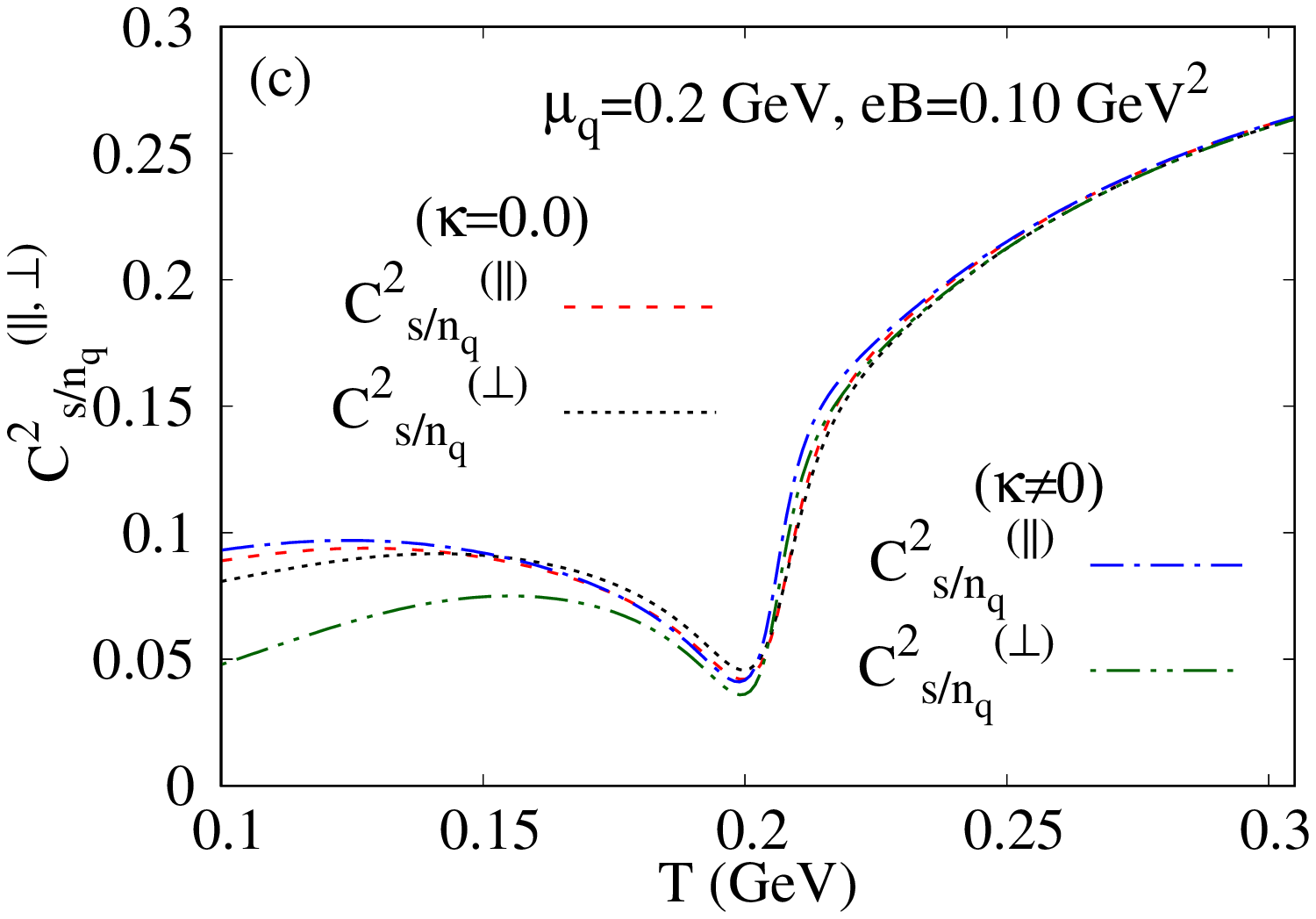}
	\caption{(a) $C_{s/n_B}^{2(\parallel)}$ and (b) $C_{s/n_B}^{2(\perp)}$ as a function of $T$ for various values of $\mu_q=0,~0.1,~0.2\rm~GeV$ at $eB=0.10\rm~GeV^2$, (c) $C_{s/n_B}^{2(\parallel,\perp)}$ as a function of $T$ at $\mu_q=0.2\rm~GeV$ and $eB=0.10\rm~GeV^2$.}\label{Fig.TC2sbynB}
\end{figure}

Fig.~\ref{Fig.TC2sbynB}(a) illustrates the variation of ${c_{s/n_q}^{2(\parallel)}}$ as a function of temperature for various values of $\mu_q$ at $eB=0.10~\rm GeV^2$ both with and without AMM of quarks. In all the plots, ${c_{s/n_q}^{2(\parallel)}}$ decreases initially with temperature, reaches its lowest value, then sharply increases over a short temperature range before saturating slightly below the ideal gas value. This minimum value, known as the softest point, may be an important indicator of the transitions observed in heavy-ion collisions \cite{Hung:1994eq}. Following a transition or crossover, the release of new degrees of freedom leads to a rapid increase in the speed of sound. In all plots, this transition point shifts towards lower temperatures with increasing $\mu_q$ (green, blue and red lines in Fig.~\ref{Fig.TC2sbynB}(a)). Conversely, the magnitude of ${c_{s/n_q}^{2(\parallel)}}$ decreases with the increase of $\mu_q$ below the transition temperature. However, ${c_{s/n_q}^{2(\parallel)}}$ increases with the increase of $\mu_q$ above the transition temperature. The figure also shows that near the transition region the plots shift towards lower $T$ values with the inclusion of AMM of quarks (red and cyan lines in Fig.~\ref{Fig.TC2sbynB}(a)) but the behaviour of the speed of sound remains same as observed without AMM of quarks. 
\begin{figure}[h] 
	\includegraphics[angle = -90, scale=0.23]{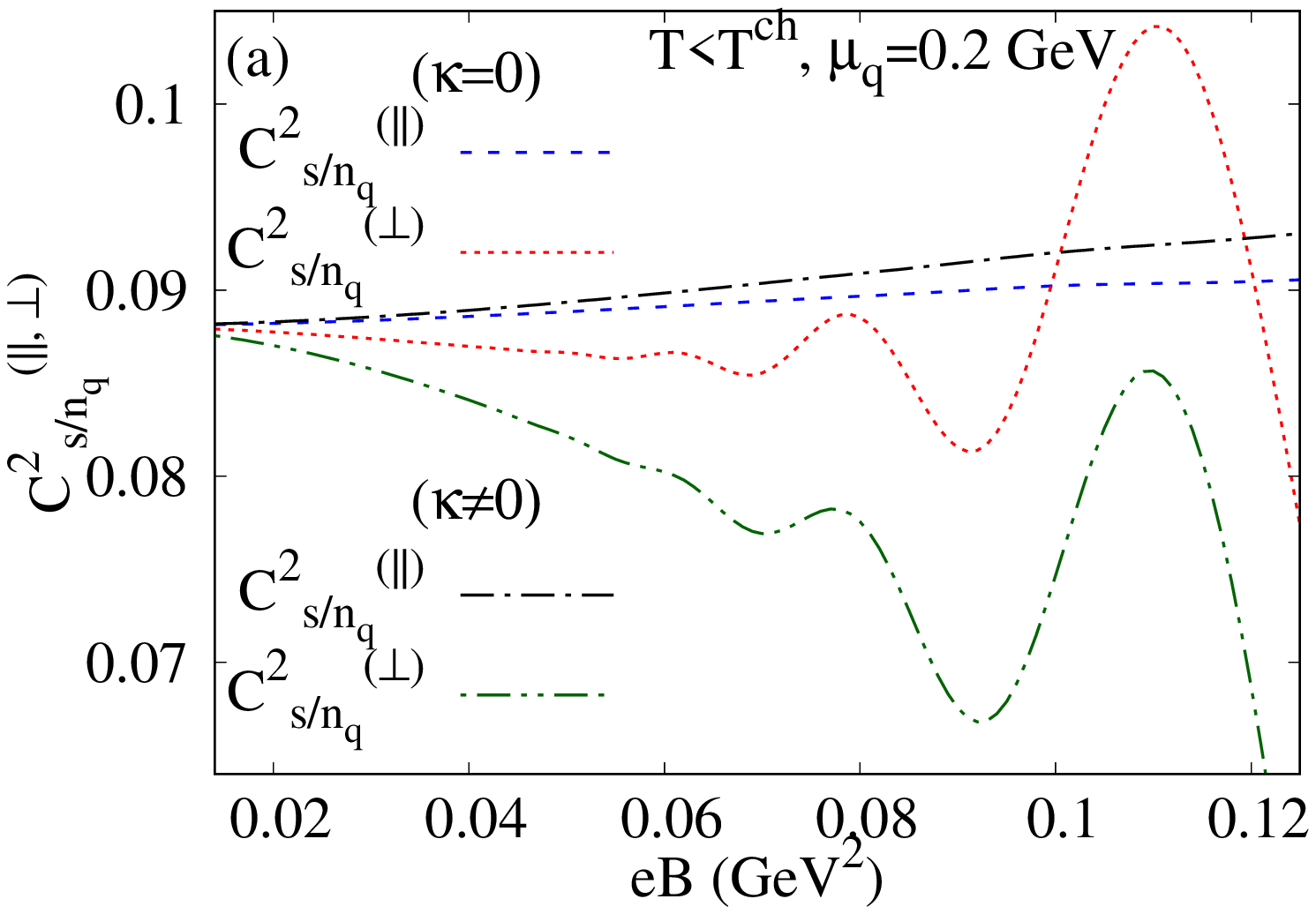}
	\includegraphics[angle = -90, scale=0.23]{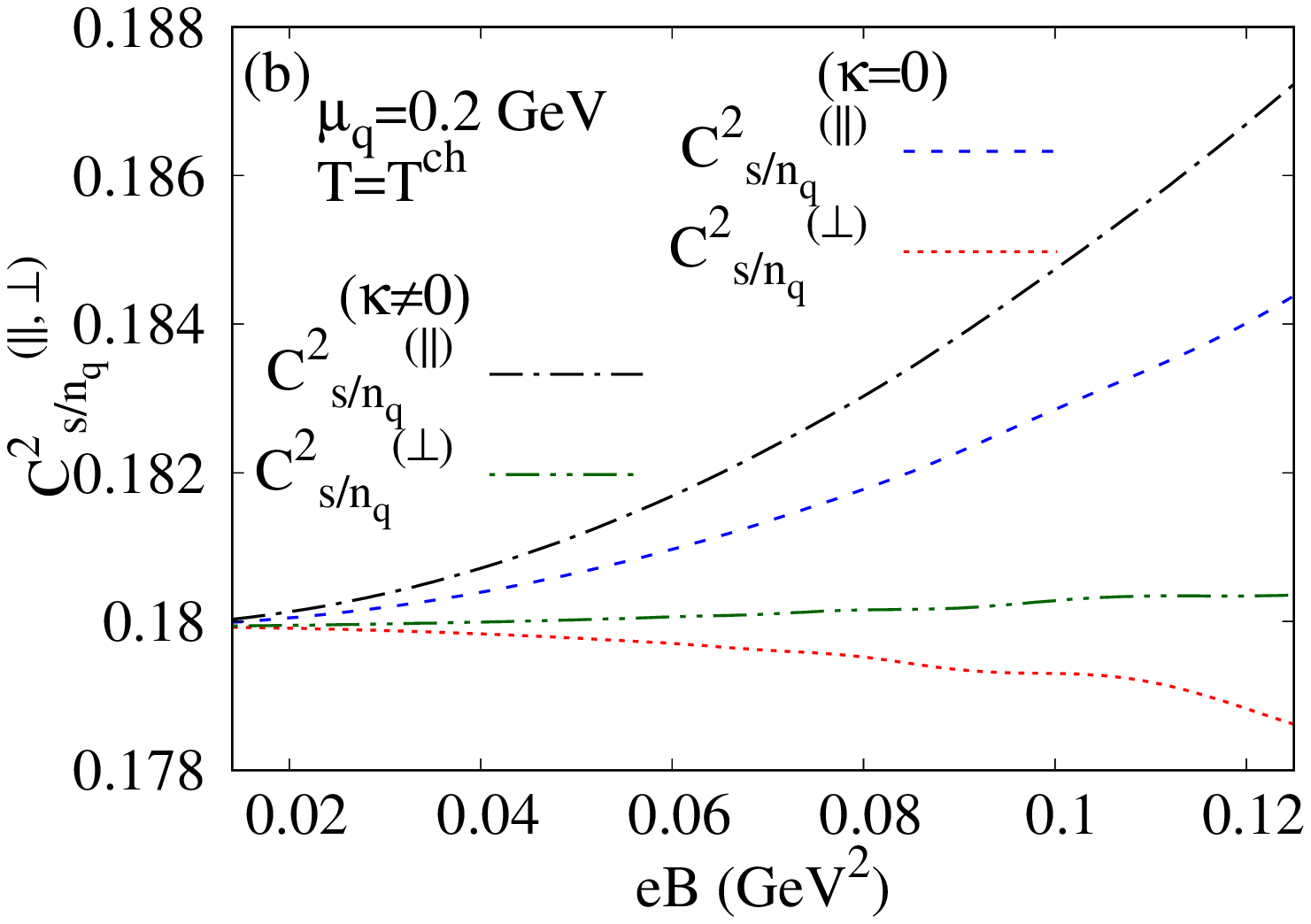}
	\includegraphics[angle = -90, scale=0.23]{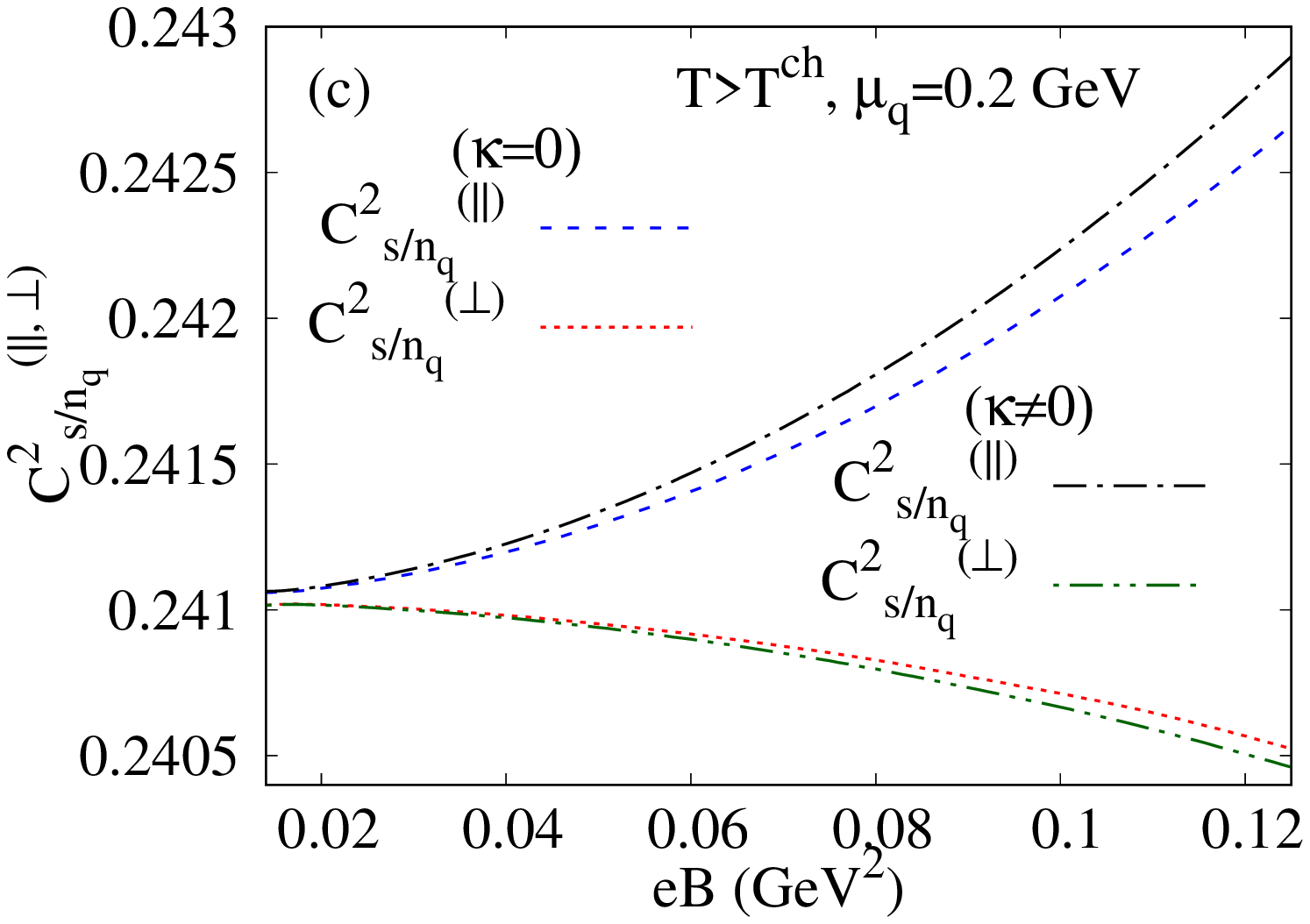}
	\caption{${c_{s/n_q}^{2(\parallel,\perp)}}$ as a function of $eB$ for $\mu_q=0.2\rm~GeV$ with and without AMM of quark at (a) chiral broken phase, (b) in the vicinity of chiral phase transition (c) partial chiral restored phase.}
	\label{Fig.eBC2sbynB}
\end{figure}
Similarly, Fig.~\ref{Fig.TC2sbynB}(b) presents the variation of ${c_{s/n_q}^{2(\perp)}}$ as a function of temperature for various values of $\mu_q$ at $eB=0.10~\rm GeV^2$ considering both with and without AMM of quarks. The qualitative behaviour of ${c_{s/n_q}^{2(\perp)}}$ is same as ${c_{s/n_q}^{2(\parallel)}}$. With the inclusion of AMM of quarks Fig.~\ref{Fig.TC2sbynB}(b) shows that the magnitude of ${c_{s/n_q}^{2(\perp)}}$ decreases more for a given value of $\mu_q$ at lower temperature region (black and blue lines in Fig.~\ref{Fig.TC2sbynB}(b)) which is opposite to the behaviour of ${c_{s/n_q}^{2(\parallel)}}$ observed in Fig.~\ref{Fig.TC2sbynB}(a). Additionally, the lowest value of ${c_{s/n_q}^{2(\perp)}}$ decreases due to AMM of quarks (red and cyan lines in Fig.~\ref{Fig.TC2sbynB}(b)). The comparison between the parallel and perpendicular components of ${c_{s/n_q}^{2}}$ as a function of temperature at $eB=0.1\rm~GeV^2$ and $\mu_q=0.2\rm GeV$ is presented in Fig.~\ref{Fig.TC2sbynB}(c). Notably, there is a significant difference between the components. In the absence of AMM of quarks, ${c_{s/n_q}^{2(\perp)}}$ follows a similar trend. However, in presence of AMM of quarks, ${c_{s/n_q}^{2(\parallel)}}$ is greater than ${c_{s/n_q}^{2(\perp)}}$ in the symmetry broken phase (blue and green lines in Fig.~\ref{Fig.TC2sbynB}(c)).

Next, we will discuss the variation of ${c_{s/n_q}^{2(\parallel,\perp)}}$ as a function of the background magnetic field for different values of $\mu_q$ both with and without AMM of quarks at different stages of chiral phase transition as mentioned previously in subsection \ref{ConstituentQuarkMass}. Figs.~\ref{Fig.eBC2sbynB}(a)-(c) represent speed of sound in the chirally broken phase ($T<T^{\rm ch}$), near chiral phase transition ($T\simeq T^{\rm ch}$) and partial chirally restored phase ($T>T^{\rm ch}$) respectively. For both with and without AMM of quarks, Fig.~\ref{Fig.eBC2sbynB}(a) shows that ${c_{s/n_q}^{2(\parallel)}}$ increases while ${c_{s/n_q}^{2(\perp)}}$ oscillates with increasing background magnetic field. Around the chiral phase transition, Figs.~\ref{Fig.eBC2sbynB}(b) shows that the oscillations in ${c_{s/n_q}^{2(\perp)}}$ has disappeared. However, in the  partially chiral restored phase as shown in Figs.~\ref{Fig.eBC2sbynB}(c), ${c_{s/n_q}^{2(\parallel)}}$ increases and ${c_{s/n_q}^{2(\perp)}}$ decreases smoothly with increasing $eB$. {Finally, in both the chirally broken and partially chiral restored phase, ${c_{s/n_q}^{2(\parallel)}}$ increases more and ${c_{s/n_q}^{2(\perp)}}$ decreases more with the increase of the background magnetic field when AMM of quarks is included.}   
\subsection{Isothermal Compressibility}
\begin{figure}[h]
	\includegraphics[angle = -90, scale=0.30]{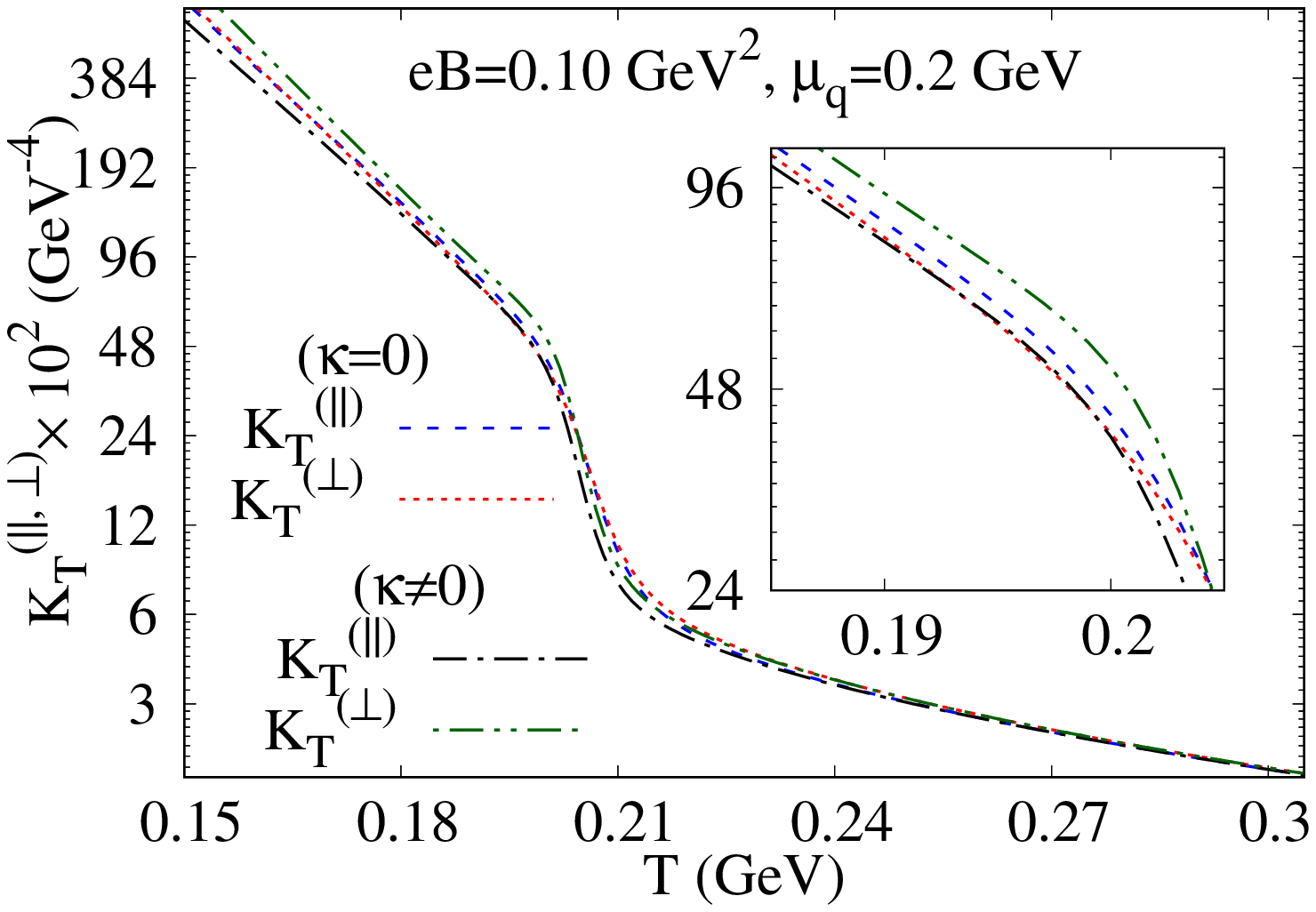}
	\caption{$K_T^{(\parallel,\perp)}$ as a function of $T$ at $\mu_q=0.2\rm~GeV$ and $eB=0.10~GeV^2$ in both with and without AMM of quarks.}
	\label{Fig.TeBKT}
\end{figure}

The isothermal compressibility ($K_T$) remains isotropic in all directions when there is no magnetic field present. Consequently, the equation of state (EoS) is the same across all directions. However, $K_T$ exhibits anisotropic behavior and splits into $K_T^{\parallel,\perp}$ along and perpendicular to the direction of background magnetic field respectively. We use Eqs.~\eqref{KT_para}-\eqref{KT_per} to calculate isothermal compressibility. Fig.~\ref{Fig.TeBKT}(a) illustrates $K_T^{\parallel,\perp}$ as a function of temperature at $eB=0.10\rm~GeV^2$ and $\mu_q=0.2\rm~GeV$ considering both scenarios with and without the AMM of quarks. In both cases, $K_T^{\parallel,\perp}$ decrease as temperature increases. This suggests that QCD matter becomes highly incompressible with increasing temperature irrespective of directions. With the inclusion of the AMM of quarks, $K_T^{\perp}$ is greater than $K_T^{\parallel}$ (green and black lines in inset plot of  Fig.~\ref{Fig.TeBKT}(a)) and the difference between $K_T^{\parallel}$ and $K_T^{\perp}$ decreases with increasing temperature. Therefore, the EoS becomes less stiff along the field direction compared to other directions when the AMM of quarks is considered and $K_T^{\parallel,\perp}$ becomes nearly isotropic at very high temperature. Conversely, the difference between $K_T^{\parallel}$ and  $K_T^{\perp}$ is very small without AMM of quarks (blue and red lines in inset plot of  Fig.~\ref{Fig.TeBKT}(a)). Moreover, plots in the figure clearly indicate a crossover or transition occurring around $T\simeq205~\rm MeV$ which can be connected to the observation in Fig.~\ref{TvsnqbyT3sbyT3}(b). 

\begin{figure}[h]
	\includegraphics[angle = -90, scale=0.23]{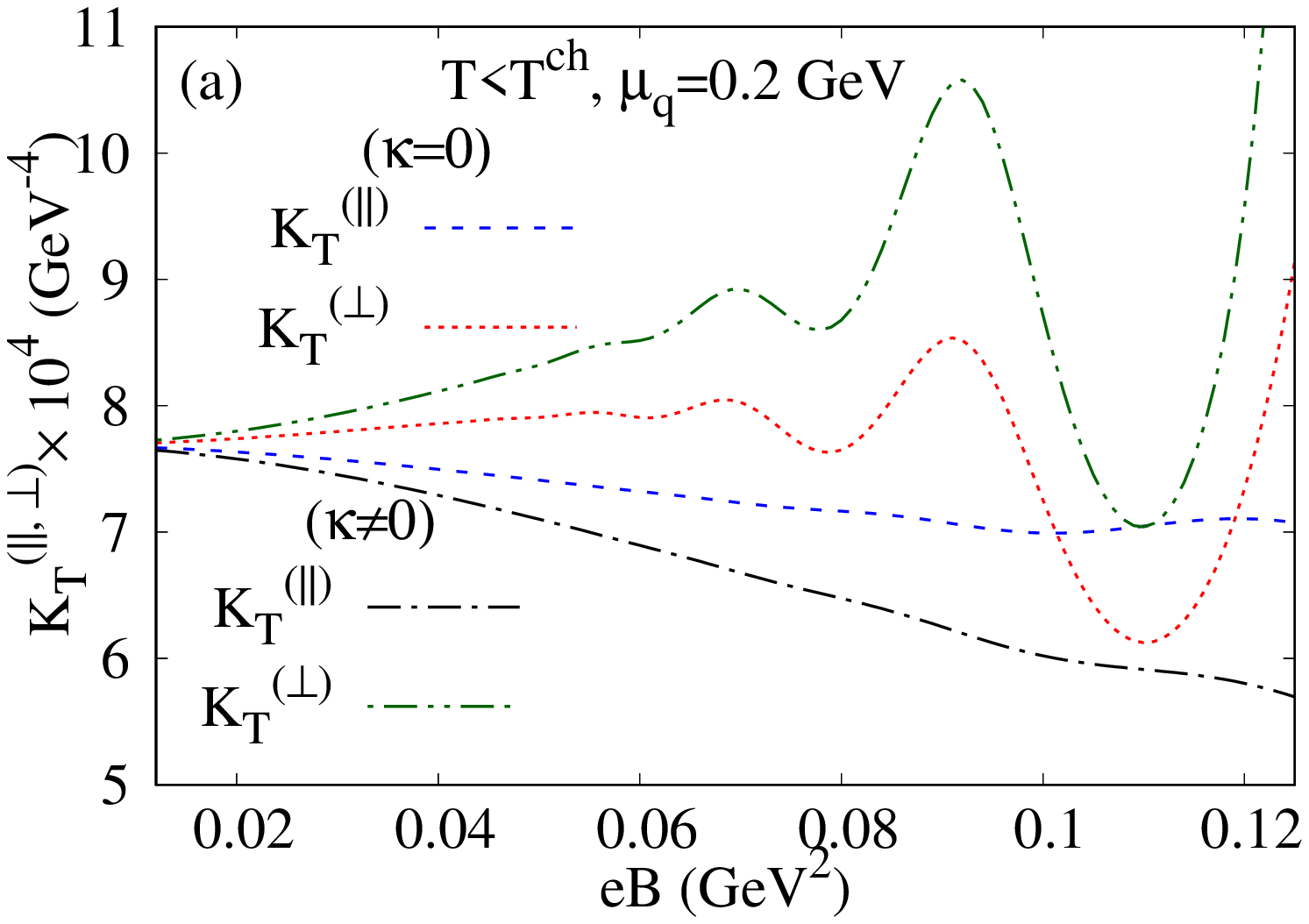}
	\includegraphics[angle = -90, scale=0.23]{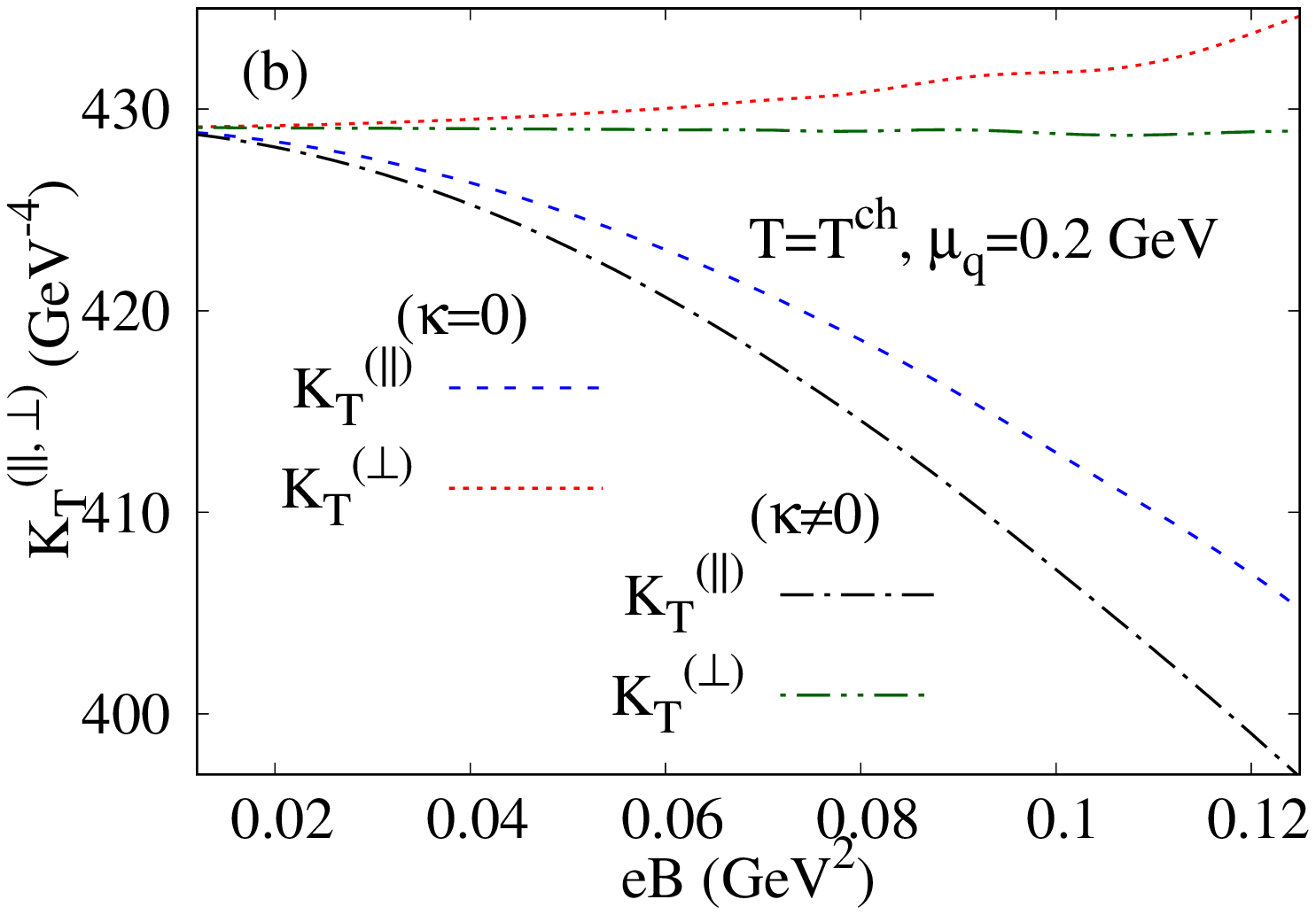}
	\includegraphics[angle = -90, scale=0.23]{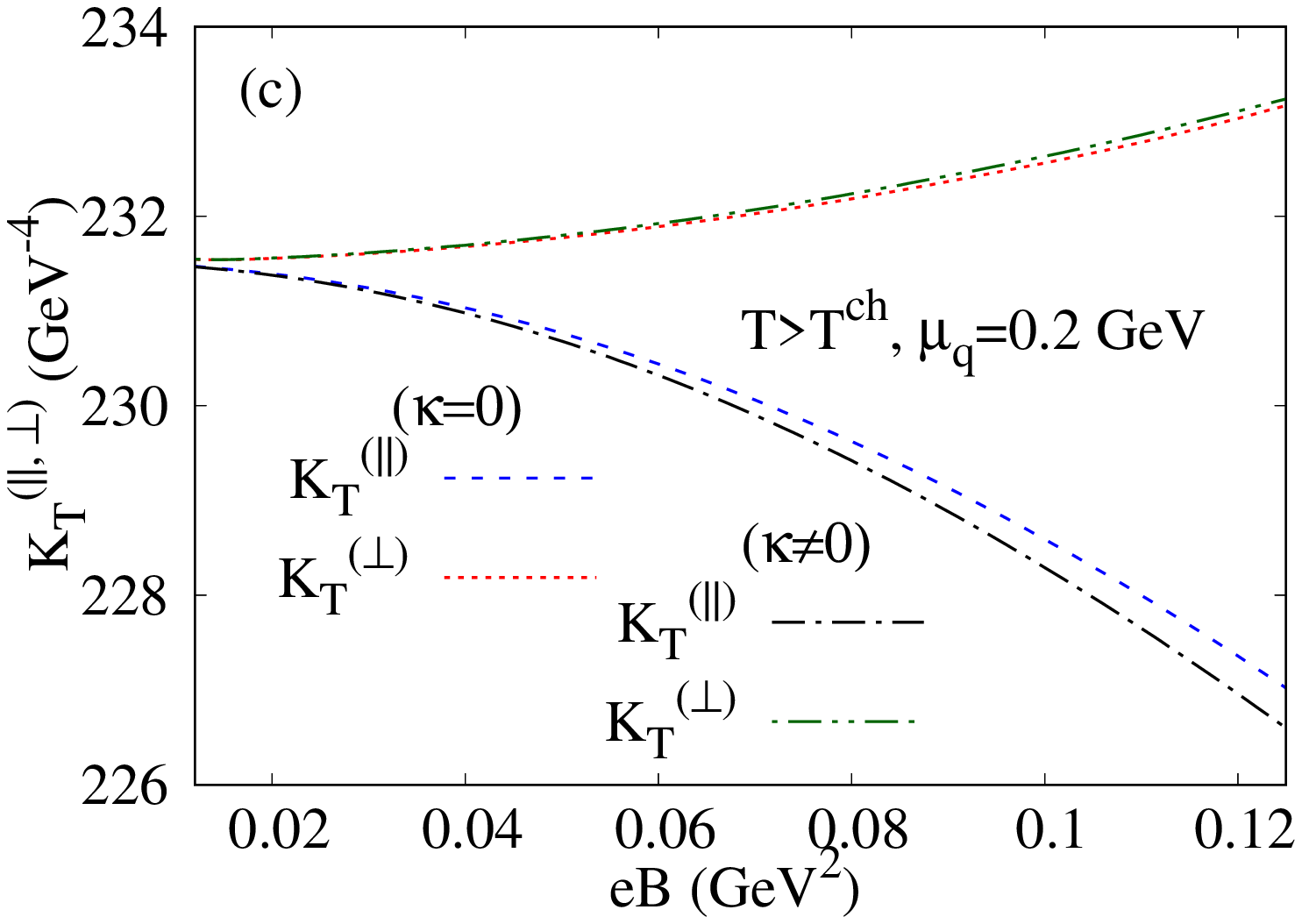}
	\caption{$K_T^{(\parallel,\perp)}$ as a function of $eB$ for $\mu_q=0.2\rm~GeV$ in both with and without AMM of quark at (a) chiral broken phase, (b) near chiral phase	transition (c) partial chiral restored phase.}
	\label{Fig.eBKT}
\end{figure}

Next, we present $K_T^{\parallel,\perp}$ as a function $eB$ for $\mu_q=0.2\rm~GeV$ both with and without the AMM of quarks at various stages of chiral phase transition as mentioned previously in subsection \ref{ConstituentQuarkMass}. Figs.~\ref{Fig.eBKT}(a)-(c) represent chirally broken phase, near the chiral phase transition and partially restored phase respectively. In Fig.~\ref{Fig.eBKT}(a), $K_T^{\perp}$ exhibits oscillations with increasing background magnetic fields and its values are higher indicating  more compressible QCD matter when considering the AMM of quark (red and green lines in Fig.~\ref{Fig.eBKT}(a)). Conversely, $K_T^{\parallel}$ decreases with increasing background magnetic fields (blue and black lines in Fig.~\ref{Fig.eBKT}(a)) reflecting a lower compressibility in the presence of the AMM of quarks. Fig.~\ref{Fig.eBKT}(b) reveals that without the AMM of quarks, $K_T^{\perp}$ increases more with increasing $eB$ unlike the scenario with the inclusion of the AMM of quarks (red and green lines in Fig.~\ref{Fig.eBKT}(b)). On the other hand, $K_T^{\parallel}$ decreases with increasing background magnetic field and its decreasing nature is larger in presence of AMM of quarks (blue and black lines in Fig.~\ref{Fig.eBKT}(a)). Moreover, Fig.~\ref{Fig.eBKT}(c) indicates an increase in $K_T^{\perp}$ (red line in Fig.~\ref{Fig.eBKT}(c)) and a decrease in $K_T^{\parallel}$ (blue line in Fig.~\ref{Fig.eBKT}(c)) with rising magnetic fields. These changes are more prominent when the AMM of quarks is taken into account (black and blue lines in Fig.~\ref{Fig.eBKT}(c)). 
%

\section{Summary \& Conclusion}\label{SC}
In summary, we have investigated several properties of quark matter subjected to a background magnetic field  both with and without the AMM of quarks at finite temperature and chemical potential within the framework of the PNJL model. 
The study involves the constituent quark mass, quark number density, and quark number susceptibility (normalized by its value at zero or weak magnetic field), as well as the speed of sound and isothermal compressibility.

The constituent quark mass initially starts at a high value and remains relatively stable at lower temperatures indicating a chirally broken phase. However, it decreases rapidly within a narrow temperature range and eventually approaches the bare mass of quarks indicating partial restoration of chiral symmetry. This transition typically occurs around $T^{\rm ch}\sim$230 MeV in a scenario without the magnetic field and zero chemical potential. In the chirally broken phase, we observe an increasing trend in the quark mass as the magnetic field strength rises, a phenomenon known as MC. Conversely, when the AMM of quarks is considered, we observe an opposite trend referred to as IMC.

Additionally, we have observed a suppression in the quark number density at low temperatures attributed to the effective interactions of the gluon field. These interactions restrict contributions from both the one-quark and two-quark states. However, this suppression becomes less pronounced at higher temperatures.

It is observed that the presence of a background magnetic field tends to shift the transition temperature ($T^{\chi_q}_c$) to higher values indicating MC. On the contrary, inclusion of AMM of quarks decreases the transition temperature confirming that IMC observed by analysing the results of gap equations.
The susceptibility normalized by its value at zero or weak magnetic field is an important probe for studying magnetic fields observed in non-central HICs.
 
Our study reveals that the presence of a magnetic field induces anisotropy in the speed of sound leading to variations between the ${c^2_{s/n_q}}^{(\parallel, \perp)}$ components. Notably, the qualitative behaviour of ${c_{s/n_q}^{2(\perp)}}$ is similar to that of ${c_{s/n_q}^{2(\parallel)}}$. When AMM of quarks is included, we find that the magnitude of ${c_{s/n_q}^{2(\perp)}}$ decreases more for a given value of $\mu_q$ in the lower temperature region, opposite to the behavior of ${c_{s/n_q}^{2(\parallel)}}$. 

In the presence of the magnetic field, the isothermal compressibility also becomes anisotropic dividing into $K_T^{\parallel,\perp}$ along and perpendicular to the direction of magnetic field. The $K_T^{\parallel,\perp}$ decrease as temperature rises indicating a trend towards increased incompressibility of QCD matter regardless of its orientation with respect to the direction of magnetic field. Upon considering AMM of quarks, $K_T^{\perp}$ is greater than $K_T^{\parallel}$ and the difference between $K_T^{\parallel,\perp}$ decreases with increasing temperature. Therefore, the EoS becomes more stiff along the field direction compared to other directions and becomes nearly isotropic at very high temperature. On the other hand, $K_T^{\parallel}$ decreases and $K_T^{\perp}$ increases with increasing background magnetic field. Hence, the system is less compressible along the magnetic field direction with increasing background magnetic field.

\appendix
\section{Mathematical Details}\label{A1}
\begin{eqnarray}
	\frac{\partial f_\Lambda}{\partial B}&=&-\frac{N(2n+1-s)\Lambda^{2N}P^{2(N-1)}_{nfs}}{\FB{\Lambda^{2N}+P^{2N}_{nfs}}^2}|e_f|,~~P_{nfs}=\sqrt{p_z^2+(2n+1-s)|e_fB|}\\
	\frac{\partial \omega_{nfs}}{\partial M}&=&\frac{M}{\omega_{nfs}}\FB{1-\frac{sk_fe_fB}{M_{nfs}}},~~\frac{\partial M_{nfs}}{\partial M}=\frac{M}{M_{nfs}}
\end{eqnarray}
\begin{eqnarray}
	\frac{\partial {\text{ln}g^+}}{\partial T}&=&\FB{\frac{\partial {\text{ln}g^+}}{\partial M}}\frac{\partial M}{\partial T}+\FB{\frac{\partial {\text{ln}g^+}}{\partial\Phi}}\frac{\partial \Phi}{\partial T}+\FB{\frac{\partial {\text{ln}g^+}}{\partial\bar{\Phi}}}\frac{\partial \bar{\Phi}}{\partial T}+\FB{\frac{\partial {\text{ln}g^+}}{\partial{T}}}_{M,\Phi,\bar{\Phi}}\\
	\frac{\partial {\text{ln}g^+}}{\partial\mu_q}&=&\FB{\frac{\partial {\text{ln}g^+}}{\partial M}}\frac{\partial M}{\partial\mu_q}+\FB{\frac{\partial {\text{ln}g^+}}{\partial\Phi}}\frac{\partial \Phi}{\partial\mu_q}+\FB{\frac{\partial {\text{ln}g^+}}{\partial\bar{\Phi}}}\frac{\partial \bar{\Phi}}{\partial\mu_q}+\FB{\frac{\partial {\text{ln}g^+}}{\partial{\mu_q}}}_{M,\Phi,\bar{\Phi}}
\end{eqnarray}
\begin{eqnarray}
	&&\frac{\partial\text{ln}g^+}{\partial M}=-3\beta\frac{\partial\omega_{nfs}}{\partial M}f^+,~~\frac{\partial\text{ln}g^+}{\partial \Phi}=\frac{3e^{-\beta(\omega_{nfs}-\mu_q)}}{g^+},~~\frac{\partial\text{ln}g^+}{\partial\bar{\Phi}}=\frac{3e^{-2\beta(\omega_{nfs}-\mu_q)}}{g^+},~~\FB{\frac{\partial\text{ln}g^+}{\partial{\mu_q}}}_{M,\Phi,\bar{\Phi}}=3\beta f^+,\\
	&&\FB{\frac{\partial\text{ln}g^+}{\partial{T}}}_{M,\Phi,\bar{\Phi}}=\frac{\omega_{nfs}-\mu_q}{T^2}3f^+.
\end{eqnarray}
\begin{eqnarray}
	\frac{\partial {\text{ln}g^-}}{\partial T}&=&\FB{\frac{\partial {\text{ln}g^-}}{\partial M}}\frac{\partial M}{\partial T}+\FB{\frac{\partial {\text{ln}g^-}}{\partial\Phi}}\frac{\partial \Phi}{\partial T}+\FB{\frac{\partial {\text{ln}g^-}}{\partial\bar{\Phi}}}\frac{\partial \bar{\Phi}}{\partial T}+\FB{\frac{\partial {\text{ln}g^-}}{\partial{T}}}_{M,\Phi,\bar{\Phi}}\\
	\frac{\partial {\text{ln}g^-}}{\partial\mu_q}&=&\FB{\frac{\partial {\text{ln}g^-}}{\partial M}}\frac{\partial M}{\partial\mu_q}+\FB{\frac{\partial {\text{ln}g^-}}{\partial\Phi}}\frac{\partial \Phi}{\partial\mu_q}+\FB{\frac{\partial {\text{ln}g^-}}{\partial\bar{\Phi}}}\frac{\partial \bar{\Phi}}{\partial\mu_q}+\FB{\frac{\partial {\text{ln}g^-}}{\partial{\mu_q}}}_{M,\Phi,\bar{\Phi}}
\end{eqnarray}
\begin{eqnarray}	
	&&\frac{\partial\text{ln}g^-}{\partial M}=-3\beta\frac{\partial\omega_{nfs}}{\partial M}f^-,~~\frac{\partial\text{ln}g^-}{\partial \Phi}=\frac{3e^{-2\beta(\omega_{nfs}+\mu_q)}}{g^-},~~	\frac{\partial\text{ln}g^-}{\partial \bar{\Phi}}=\frac{3e^{-\beta(\omega_{nfs}+\mu_q)}}{g^-},~~\FB{\frac{\partial\text{ln}g^-}{\partial{\mu_q}}}_{M,\Phi,\bar{\Phi}}=-3\beta f^-,\\
	&&\FB{\frac{\partial\text{ln}g^-}{\partial{T}}}_{M,\Phi,\bar{\Phi}}=\frac{\omega_{nfs}+\mu_q}{T^2}3f^-.
\end{eqnarray}
\begin{eqnarray}
	\frac{\partial f^+}{\partial T}&=&\FB{\frac{\partial f^+}{\partial M}}\frac{\partial M}{\partial T}+\FB{\frac{\partial f^+}{\partial\Phi}}\frac{\partial \Phi}{\partial T}+\FB{\frac{\partial f^+}{\partial\bar{\Phi}}}\frac{\partial \bar{\Phi}}{\partial T}+\FB{\frac{\partial f^+}{\partial{T}}}_{M,\Phi,\bar{\Phi}}\\
	\frac{\partial f^+}{\partial\mu_q}&=&\FB{\frac{\partial f^+}{\partial M}}\frac{\partial M}{\partial\mu_q}+\FB{\frac{\partial f^+}{\partial\Phi}}\frac{\partial \Phi}{\partial\mu_q}+\FB{\frac{\partial f^+}{\partial\bar{\Phi}}}\frac{\partial \bar{\Phi}}{\partial\mu_q}+\FB{\frac{\partial f^+}{\partial{\mu_q}}}_{M,\Phi,\bar{\Phi}}
\end{eqnarray}
\begin{eqnarray}
	&&\frac{\partial f^+}{\partial M}=\beta\frac{\partial \omega_{nfs}}{\partial M}\TB{3{f^+}^2-\frac{1}{g^+}\TB{\SB{\Phi+4\bar{\Phi}e^{-\beta(\omega_{nfs}-\mu_q)}}e^{-\beta(\omega_{nfs}-\mu_q)}+3e^{-3\beta(\omega_{nfs}-\mu_q)}}}\\
	&&\frac{\partial f^+}{\partial\Phi}=\frac{1-3f^+}{g^+}e^{-\beta(\omega_{nfs}-\mu_q)},~~\frac{\partial f^+}{\partial\bar{\Phi}}=\frac{2-3f^+}{g^+}e^{-2\beta(\omega_{nfs}-\mu_q)}\\
	&&\FB{\frac{\partial{f^+}}{\partial{\mu_q}}}_{M,\Phi,\bar{\Phi}}=\beta\TB{\frac{1}{g^+}\SB{\FB{\Phi+4\bar{\Phi}e^{-\beta(\omega_{nfs}-\mu_q)}}e^{-\beta(\omega_{nfs}-\mu_q)}+3e^{-3\beta(\omega_{nfs}-\mu_q)}}-3{f^+}^2}\\
	&&\FB{\frac{\partial{f^+}}{\partial{T}}}_{M,\Phi,\bar{\Phi}}=\frac{\omega_{nfs}-\mu_q}{T^2}\TB{\frac{1}{g^+}\SB{\FB{\Phi+4\bar{\Phi}e^{-\beta(\omega_{nfs}-\mu_q)}}e^{-\beta(\omega_{nfs}-\mu_q)}+3e^{-3\beta(\omega_{nfs}-\mu_q)}}-3{f^+}^2}
\end{eqnarray}
\begin{eqnarray}
	\frac{\partial f^-}{\partial T}&=&\FB{\frac{\partial f^-}{\partial M}}\frac{\partial M}{\partial T}+\FB{\frac{\partial f^-}{\partial\Phi}}\frac{\partial \Phi}{\partial T}+\FB{\frac{\partial f^-}{\partial\bar{\Phi}}}\frac{\partial \bar{\Phi}}{\partial T}+\FB{\frac{\partial f^-}{\partial{T}}}_{M,\Phi,\bar{\Phi}}\\
	\frac{\partial f^-}{\partial\mu_q}&=&\FB{\frac{\partial f^-}{\partial M}}\frac{\partial M}{\partial\mu_q}+\FB{\frac{\partial f^-}{\partial\Phi}}\frac{\partial \Phi}{\partial\mu_q}+\FB{\frac{\partial f^-}{\partial\bar{\Phi}}}\frac{\partial \bar{\Phi}}{\partial\mu_q}+\FB{\frac{\partial f^-}{\partial{\mu_q}}}_{M,\Phi,\bar{\Phi}}
\end{eqnarray}
\begin{eqnarray}
	&&\frac{\partial f^-}{\partial M}=\frac{\partial f^+}{\partial M}\FB{\Phi\to\bar{\Phi},\mu_q\to-\mu_q},~~\frac{\partial f^-}{\partial{\Phi}}=\frac{2-3f^-}{g^-}e^{-2\beta(\omega_{nfs}+\mu_q)},~~\frac{\partial f^-}{\partial\bar{\Phi}}=\frac{1-3f^-}{g^-}e^{-\beta(\omega_{nfs}+\mu_q)}\\
	&&\FB{\frac{\partial{f^-}}{\partial{\mu_q}}}_{M,\Phi,\bar{\Phi}}=-\beta\TB{\frac{1}{g^-}\SB{\FB{\bar{\Phi}+4{\Phi}e^{-\beta(\omega_{nfs}+\mu_q)}}e^{-\beta(\omega_{nfs}+\mu_q)}+3e^{-3\beta(\omega_{nfs}+\mu_q)}}-3\FB{f^-}^2}\\
	&&\FB{\frac{\partial{f^-}}{\partial{T}}}_{M,\Phi,\bar{\Phi}}=\FB{\frac{\partial f^+}{\partial\mu_q}}_{M,\Phi,\bar{\Phi}}\FB{\Phi\to\bar{\Phi},\mu_q\to-\mu_q}
\end{eqnarray}
\section{Thermodynamical Relations}\label{A2}
\begin{eqnarray}
	\FB{\frac{\partial(s/n_q)}{\partial\mu_q}}_{T}=\frac{1}{n_q}\FB{\frac{\partial s}{\partial\mu_q}}_T-\frac{s}{n_q^2}\FB{\frac{\partial n_q}{\partial\mu_q}}_T&,&~~\FB{\frac{\partial(s/n_q)}{\partial T}}_{\mu_q}=\frac{1}{n_q}\FB{\frac{\partial s}{\partial T}}_T-\frac{s}{n_q^2}\FB{\frac{\partial n_q}{\partial T}}_{\mu_q}~~\\
	\FB{\frac{\partial\epsilon}{\partial T}}_{\mu_q}=T\FB{\frac{\partial s}{\partial T}}_{\mu_q}+\mu_q\FB{\frac{\partial n_q}{\partial T}}_{\mu_q}&,&~~
	\FB{\frac{\partial\epsilon}{\partial\mu_q}}_T=T\FB{\frac{\partial s}{\partial\mu_q}}_T+\mu_q\FB{\frac{\partial n_q}{\partial\mu_q}}_T.
\end{eqnarray}
\section{Susceptibilies}\label{A3}
In section \ref{SpdSound}, we have observed that the speed of sound, isothermal compressibility in Eqs.~\eqref{C2xParallel}-\eqref{KT_per} involve the partial derivatives \(\left(\frac{\partial \mathcal{M}}{\partial T}\right)_{\mu_q}\), \(\left(\frac{\partial \mathcal{M}}{\partial \mu_q}\right)_T\), \(\left(\frac{\partial s}{\partial \mu_q}\right)_T\), \(\left(\frac{\partial s}{\partial T}\right)_{\mu_q}\), \(\left(\frac{\partial n_B}{\partial \mu_q}\right)_T\), and \(\left(\frac{\partial n_B}{\partial T}\right)_{\mu_q}\) which can be derived from the free energy. The expressions are provided below:
\begin{eqnarray}
	\FB{\frac{\partial \mathcal{M}}{\partial T}}_{\mu_q}&=&{3\sum_{nfs}^{}\frac{|{e_f}|}{2\pi}\int^{+\infty}_{-\infty}\frac{dp_z}{2\pi}f_\Lambda\frac{\partial\omega_{nfs}}{\partial M}\frac{\partial M}{\partial T}+3\sum_{nfs}^{}\frac{|{e_fB}|}{2\pi}\int^{+\infty}_{-\infty}\frac{dp_z}{2\pi}\frac{\partial f_\Lambda}{\partial  B}}\frac{\partial\omega_{nfs}}{\partial M}\frac{\partial M}{\partial T}\nn\\
	&+&\sum_{nfs}^{}\frac{|e_f|}{2\pi}\int^{+\infty}_{-\infty}\frac{dp_z}{2\pi}\SB{{\text{ln}g^+}+{\text{ln}g^-}}+T\sum_{nfs}^{}\frac{|e_f|}{2\pi}\int^{+\infty}_{-\infty}\frac{dp_z}{2\pi}\SB{\frac{\partial\text{ln}g^++}{\partial T}+\frac{\partial\text{ln}g^-}{\partial T}}\nn\\
	&+&3\sum_{nfs}^{}\frac{|e_fB|}{2\pi}\int^{+\infty}_{-\infty}\frac{dp_z}{2\pi}\frac{-1}{\omega^2_{nfs}}\frac{\partial\omega_{nfs}}{\partial M}\frac{\partial M}{\partial T}\SB{1-\frac{s\kappa_fe_fB}{M_{nfs}}}\SB{\frac{2n+1-s}{2}|e_f|-s\kappa_fe_fM_{nfs}}\SB{f_\Lambda-f^+-f^-}~
	\nn\\
	&+&3\sum_{nfs}^{}\frac{|e_fB|}{2\pi}\int^{+\infty}_{-\infty}\frac{dp_z}{2\pi}\frac{1}{\omega_{nfs}}\SB{-s\kappa_fe_fB\FB{\frac{-M}{M^3_{nfs}}}\frac{\partial M}{\partial T}}\SB{\frac{2n+1-s}{2}|e_f|-s\kappa_fe_fM_{nfs}}\SB{f_\Lambda-f^+-f^-}~
	\nn\\
	&+&3\sum_{nfs}^{}\frac{|e_fB|}{2\pi}\int^{+\infty}_{-\infty}\frac{dp_z}{2\pi}\frac{1}{\omega_{nfs}}\SB{1-\frac{s\kappa_fe_fB}{M_{nfs}}}\FB{-s\kappa_fe_f\FB{\frac{M}{M_{nfs}}}\frac{\partial M}{\partial T}}\SB{f_\Lambda-f^+-f^-}~
	\nn\\
	&+&3\sum_{nfs}^{}\frac{|e_fB|}{2\pi}\int^{+\infty}_{-\infty}\frac{dp_z}{2\pi}\frac{1}{\omega_{nfs}}\SB{1-\frac{s\kappa_fe_fB}{M_{nfs}}}\SB{\frac{2n+1-s}{2}|e_f|-s\kappa_fe_fM_{nfs}}\SB{-\frac{\partial f^+}{\partial T}-\frac{\partial f^-}{\partial T}}
\end{eqnarray}
\begin{eqnarray}
	\FB{\frac{\partial \mathcal{M}}{\partial{\mu_q}}}_T&=&{3\sum_{nfs}^{}\frac{|{e_f}|}{2\pi}\int^{+\infty}_{-\infty}\frac{dp_z}{2\pi}f_\Lambda\frac{\partial\omega_{nfs}}{\partial M}\frac{\partial M}{\partial\mu_q}+3\sum_{nfs}^{}\frac{|{e_fB}|}{2\pi}\int^{+\infty}_{-\infty}\frac{dp_z}{2\pi}\frac{\partial f_\Lambda}{\partial  B}}\frac{\partial\omega_{nfs}}{\partial M}\frac{\partial M}{\partial\mu_q}
	+T\sum_{nfs}^{}\frac{|e_f|}{2\pi}\int^{+\infty}_{-\infty}\frac{dp_z}{2\pi}\SB{\frac{\partial\text{ln}g^++}{\partial \mu_q}+\frac{\partial\text{ln}g^-}{\partial\mu_q}}\nn\\
	&+&3\sum_{nfs}^{}\frac{|e_fB|}{2\pi}\int^{+\infty}_{-\infty}\frac{dp_z}{2\pi}\frac{-1}{\omega^2_{nfs}}\frac{\partial\omega_{nfs}}{\partial M}\frac{\partial M}{\partial\mu_q}\SB{1-\frac{s\kappa_fe_fB}{M_{nfs}}}\SB{\frac{2n+1-s}{2}|e_f|-s\kappa_fe_fM_{nfs}}\SB{f_\Lambda-f^+-f^-}~
	\nn\\
	&+&3\sum_{nfs}^{}\frac{|e_fB|}{2\pi}\int^{+\infty}_{-\infty}\frac{dp_z}{2\pi}\frac{1}{\omega_{nfs}}\SB{-s\kappa_fe_fB\FB{\frac{-M}{M^3_{nfs}}}\frac{\partial M}{\partial\mu_q}}\SB{\frac{2n+1-s}{2}|e_f|-s\kappa_fe_fM_{nfs}}\SB{f_\Lambda-f^+-f^-}~
	\nn\\
	&+&3\sum_{nfs}^{}\frac{|e_fB|}{2\pi}\int^{+\infty}_{-\infty}\frac{dp_z}{2\pi}\frac{1}{\omega_{nfs}}\SB{1-\frac{s\kappa_fe_fB}{M_{nfs}}}\SB{-s\kappa_fe_f\FB{\frac{M}{M_{nfs}}}\frac{\partial M}{\partial\mu_q}}\SB{f_\Lambda-f^+-f^-}~
	\nn\\
	&+&3\sum_{nfs}^{}\frac{|e_fB|}{2\pi}\int^{+\infty}_{-\infty}\frac{dp_z}{2\pi}\frac{1}{\omega_{nfs}}\SB{1-\frac{s\kappa_fe_fB}{M_{nfs}}}\SB{\frac{2n+1-s}{2}|e_f|-s\kappa_fe_fM_{nfs}}\SB{-\frac{\partial f^+}{\partial\mu_q}-\frac{\partial f^-}{\partial\mu_q}}
\end{eqnarray}
\begin{eqnarray}
	\FB{\frac{\partial s}{\partial T}}_{\mu_q}&=&\frac{\partial{s_0}}{\partial T}+\sum_{nfs}^{}\frac{|e_fB|}{2\pi}\int^{+\infty}_{-\infty}\frac{dp_z}{2\pi}\SB{\frac{\partial\text{ln}g^+}{\partial T}+\frac{\partial\text{ln}g^-}{\partial T}}+3\sum_{nfs}^{}\frac{|e_fB|}{2\pi}\int^{+\infty}_{-\infty}\frac{dp_z}{2\pi}\SB{\frac{\omega_{nfs}-\mu_q}{T}\frac{\partial f^+}{\partial T}+\frac{\omega_{nfs}+\mu_q}{T}\frac{\partial f^-}{\partial T}}\nn\\
	&+&3\sum_{nfs}^{}\frac{|e_fB|}{2\pi}\int^{+\infty}_{-\infty}\frac{dp_z}{2\pi}\SB{\FB{\frac{\omega_{nfs}-\mu_q}{-T^2}+\frac{1}{T}\frac{\partial\omega_{nfs}}{\partial M}\frac{\partial M}{\partial T}}f^++\FB{\frac{\omega_{nfs}+\mu_q}{-T^2}+\frac{1}{T}\frac{\partial\omega_{nfs}}{\partial M}\frac{\partial M}{\partial T}}f^-}
\end{eqnarray}
\begin{eqnarray}
	\FB{\frac{\partial s}{\partial\mu_q}}_{T}&=&\frac{\partial{s_0}}{\partial {\mu_q}}+\sum_{nfs}^{}\frac{|e_fB|}{2\pi}\int^{+\infty}_{-\infty}\frac{dp_z}{2\pi}\SB{\frac{\partial\text{ln}g^+}{\partial\mu_q}+\frac{\partial\text{ln}g^-}{\partial\mu_q}}+3\sum_{nfs}^{}\frac{|e_fB|}{2\pi}\int^{+\infty}_{-\infty}\frac{dp_z}{2\pi}\SB{\frac{\omega_{nfs}-\mu_q}{T}\frac{\partial f^+}{\partial\mu_q}+\frac{\omega_{nfs}+\mu_q}{T}\frac{\partial f^-}{\partial\mu_q}}\nn\\
	&+&3\sum_{nfs}^{}\frac{|e_fB|}{2\pi}\int^{+\infty}_{-\infty}\frac{dp_z}{2\pi}\SB{\FB{\frac{\omega_{nfs}-\mu_q}{-T^2}+\frac{1}{T}\frac{\partial\omega_{nfs}}{\partial M}\frac{\partial M}{\partial\mu_q}}f^++\FB{\frac{\omega_{nfs}+\mu_q}{-T^2}+\frac{1}{T}\frac{\partial\omega_{nfs}}{\partial M}\frac{\partial M}{\partial\mu_q}}f^-}
\end{eqnarray}
\begin{eqnarray}
	\FB{\frac{\partial n_q}{\partial T}}_{\mu_q}&=&3\sum_{nfs}^{}\frac{|e_fB|}{2\pi}\int^{+\infty}_{-\infty}\frac{dp_z}{2\pi}\FB{\frac{\partial f^+}{\partial T}-\frac{\partial f^-}{\partial T}}\\
	\FB{\frac{\partial n_q}{\partial\mu_q}}_{T}&=&3\sum_{nfs}^{}\frac{|e_fB|}{2\pi}\int^{+\infty}_{-\infty}\frac{dp_z}{2\pi}\FB{\frac{\partial f^+}{\partial\mu_q}-\frac{\partial f^-}{\partial\mu_q}}\label{Appendix.dnqbydNu}	
\end{eqnarray}

The partial derivaties - $\frac{\partial M}{\partial T},~ \frac{\partial\phi}{\partial T},~\frac{\partial\bar{\phi}}{\partial T}$ and $
\frac{\partial M}{\partial \mu_q},~\frac{\partial\phi}{\partial \mu_q},~\frac{\partial\bar{\phi}}{\partial \mu_q}$ - involved in previous equations can be obtained from the coupled Eqs.~\eqref{SCE1}-\eqref{SCE3}. We define $X,~Y,~Z$ using the gap Eqs.~\eqref{SCE1}-\eqref{SCE3} as
\begin{eqnarray}
	X&=&(M-m)-6G\sum_{nfs}\frac{|e_fB|}{2\pi}\int^{+\infty}_{-\infty}\frac{dp_z}{2\pi}\frac{1}{\omega_{nfs}}\frac{M}{M_{nfs}}\SB{M_{nfs}-s\kappa_fe_fB}\FB{1-f^+-f^-}\label{XX}\\
	Y&=&\frac{\partial U}{\partial\Phi}-3T\sum_{nfs}\frac{|e_fB|}{2\pi}\int^{+\infty}_{-\infty}\frac{dp_z}{2\pi}\SB{\frac{e^{-\frac{\omega_{nfs}-\mu_q}{T}}}{g^+}+\frac{e^{-2\frac{\omega_{nfs}+\mu_q}{T}}}{g^-}}\label{YY}\\
	Z&=&\frac{\partial U}{\partial\bar{\Phi}}-3T\sum_{nfs}\frac{|e_fB|}{2\pi}\int^{+\infty}_{-\infty}\frac{dp_z}{2\pi}\SB{\frac{e^{-2\frac{\omega_{nfs}-\mu_q}{T}}}{g^+}+\frac{e^{-\frac{\omega_{nfs}+\mu_q}{T}}}{g^-}}\label{ZZ}
\end{eqnarray}
Taking derivative with respect to temperature and chemical potential, the Eqs.~\eqref{XX}-\eqref{ZZ} can be expressed in the matrix form as
\begin{eqnarray}
\begin{pmatrix}
	X_M	& X_\phi & X_{\bar{\Phi}}\\
	Y_M	& Y_\phi & Y_{\bar{\Phi}}\\
	Z_M	& Z_\phi & Z_{\bar{\Phi}}
\end{pmatrix}
\begin{pmatrix}
	\frac{\partial M}{\partial T}\\ \frac{\partial\phi}{\partial T}\\ \frac{\partial\bar{\phi}}{\partial T}
\end{pmatrix}
=\begin{pmatrix} X_T\\ Y_T\\ Z_T \end{pmatrix}
\end{eqnarray}
\begin{eqnarray}
	\begin{pmatrix}
		X_M	& X_\phi & X_{\bar{\Phi}}\\
		Y_M	& Y_\phi & Y_{\bar{\Phi}}\\
		Z_M	& Z_\phi & Z_{\bar{\Phi}}
	\end{pmatrix}
	\begin{pmatrix}
		\frac{\partial M}{\partial \mu_q}\\ \frac{\partial\phi}{\partial \mu_q}\\ \frac{\partial\bar{\phi}}{\partial \mu_q}
	\end{pmatrix}
	=\begin{pmatrix} X_{\mu_q}\\ Y_{\mu_q}\\ Z_{\mu_q} \end{pmatrix}
\end{eqnarray}
where
\begin{eqnarray}
	X_M&=&1-6G\sum_{nfs}\frac{|e_fB|}{2\pi}\int^{+\infty}_{-\infty}\frac{dp_z}{2\pi}\left[ \SB{\frac{1}{\omega_{nfs}}\frac{1}{M_{nfs}}-\frac{1}{\omega^2_{nfs}}\frac{M}{M_{nfs}}\frac{\partial\omega_{nfs}}{\partial M}-\frac{1}{\omega_{nfs}}\frac{M^2}{M^3_{nfs}}}\FB{M_{nfs}-sk_fe_fB}\FB{1-f^+-f^-}\right. \nn\\&&
	+\left. \frac{1}{\omega_{nfs}}\FB{\frac{M}{M_{nfs}}}^2\FB{1-f^+-f^-}-\frac{1}{\omega_{nfs}}\frac{M}{M_{nfs}}\FB{M_{nfs}-sk_fe_fB}\FB{-\frac{\partial f^+}{\partial M}-\frac{\partial f^-}{\partial M}}\right] \\
	X_\phi&=&6G\sum_{nfs}\frac{|e_fB|}{2\pi}\int^{+\infty}_{-\infty}\frac{dp_z}{2\pi}\frac{1}{\omega_{nfs}}\frac{M}{M_{nfs}}\FB{M_{nfs}-sk_fe_fB}\FB{\frac{\partial f^+}{\partial\phi}+\frac{\partial f^-}{\partial\phi}}\\
	X_{\bar{\Phi}}&=&6G\sum_{nfs}\frac{|e_fB|}{2\pi}\int^{+\infty}_{-\infty}\frac{dp_z}{2\pi}\frac{1}{\omega_{nfs}}\frac{M}{M_{nfs}}\FB{M_{nfs}-sk_fe_fB}\FB{\frac{\partial f^+}{\partial\bar{\phi}}+\frac{\partial f^-}{\partial\bar{\phi}}}\\
	X_T&=&-6G\sum_{nfs}\frac{|e_fB|}{2\pi}\int^{+\infty}_{-\infty}\frac{dp_z}{2\pi}\frac{1}{\omega_{nfs}}\frac{M}{M_{nfs}}\FB{M_{nfs}-sk_fe_fB}\FB{\frac{\partial f^+}{\partial T}+\frac{\partial f^-}{\partial T}}\\	
	X_{\mu_q}&=&-6G\sum_{nfs}\frac{|e_fB|}{2\pi}\int^{+\infty}_{-\infty}\frac{dp_z}{2\pi}\frac{1}{\omega_{nfs}}\frac{M}{M_{nfs}}\FB{M_{nfs}-sk_fe_fB}\FB{\frac{\partial f^+}{\partial\mu_q}+\frac{\partial f^-}{\partial\mu_q}}\\
\end{eqnarray}

\begin{eqnarray}
	Y_M&=&-3T\sum_{nfs}\frac{|e_fB|}{2\pi}\int^{+\infty}_{-\infty}\frac{dp_z}{2\pi}\left[-\frac{e^{-\frac{{\omega_{nfs}-\mu_q}}{T}}}{(g^+)^2} \frac{\partial g^+}{\partial M}-\frac{e^{-\frac{{\omega_{nfs}-\mu_q}}{T}}}{Tg^+} \frac{\partial\omega_{nfs}}{\partial M} -\frac{e^{-2{\frac{{\omega_{nfs}+\mu_q}}{T}}}}{(g^-)^2} \frac{\partial g^-}{\partial M}-\frac{2e^{-2\frac{{\omega_{nfs}+\mu_q}}{T}}}{Tg^-} \frac{\partial\omega_{nfs}}{\partial M} 
	\right] \\
	Y_\phi&=&\frac{\partial^2U}{\partial\phi^2}-3T\sum_{nfs}\frac{|e_fB|}{2\pi}\int^{+\infty}_{-\infty}\frac{dp_z}{2\pi}\TB{-\frac{e^{-\frac{{\omega_{nfs}-\mu_q}}{T}}}{\FB{g^+}^2}\frac{\partial g^+}{\partial\phi}-\frac{e^{-2\frac{{\omega_{nfs}+\mu_q}}{T}}}{\FB{g^-}^2}\frac{\partial g^-}{\partial\phi}}\\
	Y_{\bar{\Phi}}&=&\frac{\partial^2U}{\partial\Phi\partial\bar{\Phi}}-3T\sum_{nfs}\frac{|e_fB|}{2\pi}\int^{+\infty}_{-\infty}\frac{dp_z}{2\pi}\TB{-\frac{e^{-\frac{{\omega_{nfs}-\mu_q}}{T}}}{\FB{g^+}^2}\frac{\partial g^+}{\partial\bar{\Phi}}-\frac{e^{-2\frac{{\omega_{nfs}+\mu_q}}{T}}}{\FB{g^-}^2}\frac{\partial g^-}{\partial\bar{\Phi}}}\\
	Y_T&=&-\frac{\partial^2U}{\partial T\partial\Phi}+\sum_{nfs}\frac{|e_fB|}{2\pi}\int^{+\infty}_{-\infty}\frac{dp_z}{2\pi}\left[  3\SB{\frac{e^{-\frac{\omega_{nfs}-\mu}{T}}}{g^+}+\frac{e^{-2\frac{\omega_{nfs}+\mu}{T}}}{g^-}}+3T\SB{\frac{\omega_{nfs}-\mu_q}{T^2}\frac{e^{-\frac{\omega_{nfs}-\mu_q}{T}}}{g^+}-\frac{e^{-\frac{\omega_{nfs}-\mu_q}{T}}}{(g^+)^2}\frac{\partial g^+}{\partial T}\right.\right. \nn\\&&\hspace{5cm}\left.\left. +2\frac{\omega_{nfs}+\mu_q}{T^2}\frac{e^{-2\frac{\omega_{nfs}+\mu_q}{T}}}{g^-}-\frac{e^{-2\frac{\omega_{nfs}+\mu_q}{T}}}{(g^+)^2}\frac{\partial g^-}{\partial T}} \right] \\
	Y_\mu&=&3T\sum_{nfs}\frac{|e_fB|}{2\pi}\int^{+\infty}_{-\infty}\frac{dp_z}{2\pi}\SB{\frac{e^{-\frac{\omega_{nfs}-\mu_q}{T}}}{Tg^+}-\frac{e^{-\frac{\omega_{nfs}-\mu_q}{T}}}{\FB{g^+}^2}\frac{\partial g^+}{\partial\mu_q}-\frac{2e^{-2\frac{\omega_{nfs}+\mu_q}{T}}}{Tg^-}-\frac{e^{-2\frac{\omega_{nfs}+\mu_q}{T}}}{\FB{g^-}^2}\frac{\partial g^-}{\partial\mu_q}}
\end{eqnarray}

\begin{eqnarray}
	Z_M&=&-3T\sum_{nfs}\frac{|e_fB|}{2\pi}\int^{+\infty}_{-\infty}\frac{dp_z}{2\pi}\left[-\frac{2e^{-2\frac{{\omega_{nfs}-\mu_q}}{T}}}{Tg^+} \frac{\partial\omega_{nfs}}{\partial M}-\frac{e^{-2\frac{{\omega_{nfs}-\mu_q}}{T}}}{\FB{g^+}^2} \frac{\partial g^+}{\partial M}  -\frac{e^{-\frac{{\omega_{nfs}+\mu_q}}{T}}}{Tg^-} \frac{\partial\omega_{nfs}}{\partial M}-\frac{e^{-\frac{{\omega_{nfs}+\mu_q}}{T}}}{\FB{g^-}^2} \frac{\partial g^-}{\partial M}\right] \\
	Z_\Phi&=&\frac{\partial^2U}{\partial\Phi\bar{\Phi}}-3T\sum_{nfs}\frac{|e_fB|}{2\pi}\int^{+\infty}_{-\infty}\frac{dp_z}{2\pi}\left[-\frac{e^{-2\frac{\omega_{nfs}-\mu_q}{T}}}{\FB{g^+}^2}\frac{\partial g^+}{\partial\Phi}-\frac{e^{-\frac{\omega_{nfs}+\mu_q}{T}}}{\FB{g^-}^2}\frac{\partial g^-}{\partial\Phi} \right] \\
	Z_{\bar{\Phi}}&=&\frac{\partial^2U}{\partial\bar{\Phi}^2}-3T\sum_{nfs}\frac{|e_fB|}{2\pi}\int^{+\infty}_{-\infty}\frac{dp_z}{2\pi}\left[-\frac{e^{-2\frac{\omega_{nfs}-\mu_q}{T}}}{\FB{g^+}^2}\frac{\partial g^+}{\partial\bar{\Phi}}-\frac{e^{-\frac{\omega_{nfs}+\mu_q}{T}}}{\FB{g^-}^2}\frac{\partial g^-}{\partial\bar{\Phi}} \right]\\
	Z_T&=&-\frac{\partial^2U}{\partial T\partial\bar{\Phi}}+\sum_{nfs}\frac{|e_fB|}{2\pi}\int^{+\infty}_{-\infty}\frac{dp_z}{2\pi}\left[ 3\SB{\frac{e^{-2\frac{\omega_{nfs}-\mu_q}{T}}}{g^+}+\frac{e^{-\frac{\omega_{nfs}+\mu_q}{T}}}{g^-}}\right. \\&&\hspace{1cm}\left. +3T\SB{-\frac{e^{-2\frac{\omega_{nfs}-\mu_q}{T}}}{(g^+)^2}\frac{\partial g^+}{\partial T}+\frac{2\FB{\omega_{nfs}-\mu_q}}{T^2}\frac{e^{-2\frac{\omega_{nfs}-\mu_q}{T}}}{g^+}-\frac{e^{-\frac{\omega_{nfs}+\mu_q}{T}}}{(g^-)^2}\frac{\partial g^-}{\partial T}+\frac{\omega_{nfs}+\mu_q}{T^2}\frac{e^{-\frac{\omega_{nfs}+\mu_q}{T}}}{g^-}}\right]\\ 
	Z_{\mu_q}&=&3T\sum_{nfs}\frac{|e_fB|}{2\pi}\int^{+\infty}_{-\infty}\frac{dp_z}{2\pi}\TB{\frac{2e^{-2\frac{\omega_{nfs}-\mu_q}{T}}}{Tg^+}-\frac{e^{-\frac{\omega_{nfs}+\mu_q}{T}}}{Tg^-}-\frac{e^{-2\frac{\omega_{nfs}-\mu_q}{T}}}{(g^+)^2}\frac{\partial g^+}{\partial \mu_q}-\frac{e^{-\frac{\omega_{nfs}+\mu_q}{T}}}{(g^-)^2}\frac{\partial g^-}{\partial \mu_q}}
\end{eqnarray}
\bibliographystyle{apsrev4-1}
\bibliography{Ref.bib}

\begin{thebibliography}{111}%
\makeatletter
\providecommand \@ifxundefined [1]{%
 \@ifx{#1\undefined}
}%
\providecommand \@ifnum [1]{%
 \ifnum #1\expandafter \@firstoftwo
 \else \expandafter \@secondoftwo
 \fi
}%
\providecommand \@ifx [1]{%
 \ifx #1\expandafter \@firstoftwo
 \else \expandafter \@secondoftwo
 \fi
}%
\providecommand \natexlab [1]{#1}%
\providecommand \enquote  [1]{``#1''}%
\providecommand \bibnamefont  [1]{#1}%
\providecommand \bibfnamefont [1]{#1}%
\providecommand \citenamefont [1]{#1}%
\providecommand \href@noop [0]{\@secondoftwo}%
\providecommand \href [0]{\begingroup \@sanitize@url \@href}%
\providecommand \@href[1]{\@@startlink{#1}\@@href}%
\providecommand \@@href[1]{\endgroup#1\@@endlink}%
\providecommand \@sanitize@url [0]{\catcode `\\12\catcode `\$12\catcode
  `\&12\catcode `\#12\catcode `\^12\catcode `\_12\catcode `\%12\relax}%
\providecommand \@@startlink[1]{}%
\providecommand \@@endlink[0]{}%
\providecommand \url  [0]{\begingroup\@sanitize@url \@url }%
\providecommand \@url [1]{\endgroup\@href {#1}{\urlprefix }}%
\providecommand \urlprefix  [0]{URL }%
\providecommand \Eprint [0]{\href }%
\providecommand \doibase [0]{http://dx.doi.org/}%
\providecommand \selectlanguage [0]{\@gobble}%
\providecommand \bibinfo  [0]{\@secondoftwo}%
\providecommand \bibfield  [0]{\@secondoftwo}%
\providecommand \translation [1]{[#1]}%
\providecommand \BibitemOpen [0]{}%
\providecommand \bibitemStop [0]{}%
\providecommand \bibitemNoStop [0]{.\EOS\space}%
\providecommand \EOS [0]{\spacefactor3000\relax}%
\providecommand \BibitemShut  [1]{\csname bibitem#1\endcsname}%
\let\auto@bib@innerbib\@empty
\bibitem [{\citenamefont {Kharzeev}\ \emph {et~al.}(2008)\citenamefont
  {Kharzeev}, \citenamefont {McLerran},\ and\ \citenamefont
  {Warringa}}]{Kharzeev:2007jp}%
  \BibitemOpen
  \bibfield  {author} {\bibinfo {author} {\bibfnamefont {D.~E.}\ \bibnamefont
  {Kharzeev}}, \bibinfo {author} {\bibfnamefont {L.~D.}\ \bibnamefont
  {McLerran}}, \ and\ \bibinfo {author} {\bibfnamefont {H.~J.}\ \bibnamefont
  {Warringa}},\ }\href {\doibase 10.1016/j.nuclphysa.2008.02.298} {\bibfield
  {journal} {\bibinfo  {journal} {Nucl. Phys. A}\ }\textbf {\bibinfo {volume}
  {803}},\ \bibinfo {pages} {227} (\bibinfo {year} {2008})},\ \Eprint
  {http://arxiv.org/abs/0711.0950} {arXiv:0711.0950 [hep-ph]} \BibitemShut
  {NoStop}%
\bibitem [{\citenamefont {Skokov}\ \emph {et~al.}(2009)\citenamefont {Skokov},
  \citenamefont {Illarionov},\ and\ \citenamefont {Toneev}}]{Skokov:2009qp}%
  \BibitemOpen
  \bibfield  {author} {\bibinfo {author} {\bibfnamefont {V.}~\bibnamefont
  {Skokov}}, \bibinfo {author} {\bibfnamefont {A.~Y.}\ \bibnamefont
  {Illarionov}}, \ and\ \bibinfo {author} {\bibfnamefont {V.}~\bibnamefont
  {Toneev}},\ }\href {\doibase 10.1142/S0217751X09047570} {\bibfield  {journal}
  {\bibinfo  {journal} {Int. J. Mod. Phys. A}\ }\textbf {\bibinfo {volume}
  {24}},\ \bibinfo {pages} {5925} (\bibinfo {year} {2009})},\ \Eprint
  {http://arxiv.org/abs/0907.1396} {arXiv:0907.1396 [nucl-th]} \BibitemShut
  {NoStop}%
\bibitem [{\citenamefont {Tuchin}(2013)}]{Tuchin:2013apa}%
  \BibitemOpen
  \bibfield  {author} {\bibinfo {author} {\bibfnamefont {K.}~\bibnamefont
  {Tuchin}},\ }\href {\doibase 10.1103/PhysRevC.88.024911} {\bibfield
  {journal} {\bibinfo  {journal} {Phys. Rev. C}\ }\textbf {\bibinfo {volume}
  {88}},\ \bibinfo {pages} {024911} (\bibinfo {year} {2013})},\ \Eprint
  {http://arxiv.org/abs/1305.5806} {arXiv:1305.5806 [hep-ph]} \BibitemShut
  {NoStop}%
\bibitem [{\citenamefont {Gursoy}\ \emph {et~al.}(2014)\citenamefont {Gursoy},
  \citenamefont {Kharzeev},\ and\ \citenamefont {Rajagopal}}]{Gursoy:2014aka}%
  \BibitemOpen
  \bibfield  {author} {\bibinfo {author} {\bibfnamefont {U.}~\bibnamefont
  {Gursoy}}, \bibinfo {author} {\bibfnamefont {D.}~\bibnamefont {Kharzeev}}, \
  and\ \bibinfo {author} {\bibfnamefont {K.}~\bibnamefont {Rajagopal}},\ }\href
  {\doibase 10.1103/PhysRevC.89.054905} {\bibfield  {journal} {\bibinfo
  {journal} {Phys. Rev.}\ }\textbf {\bibinfo {volume} {C89}},\ \bibinfo {pages}
  {054905} (\bibinfo {year} {2014})},\ \Eprint {http://arxiv.org/abs/1401.3805}
  {arXiv:1401.3805 [hep-ph]} \BibitemShut {NoStop}%
\bibitem [{\citenamefont {Inghirami}\ \emph {et~al.}(2016)\citenamefont
  {Inghirami}, \citenamefont {Del~Zanna}, \citenamefont {Beraudo},
  \citenamefont {Moghaddam}, \citenamefont {Becattini},\ and\ \citenamefont
  {Bleicher}}]{Inghirami:2016iru}%
  \BibitemOpen
  \bibfield  {author} {\bibinfo {author} {\bibfnamefont {G.}~\bibnamefont
  {Inghirami}}, \bibinfo {author} {\bibfnamefont {L.}~\bibnamefont
  {Del~Zanna}}, \bibinfo {author} {\bibfnamefont {A.}~\bibnamefont {Beraudo}},
  \bibinfo {author} {\bibfnamefont {M.~H.}\ \bibnamefont {Moghaddam}}, \bibinfo
  {author} {\bibfnamefont {F.}~\bibnamefont {Becattini}}, \ and\ \bibinfo
  {author} {\bibfnamefont {M.}~\bibnamefont {Bleicher}},\ }\href {\doibase
  10.1140/epjc/s10052-016-4516-8} {\bibfield  {journal} {\bibinfo  {journal}
  {Eur. Phys. J. C}\ }\textbf {\bibinfo {volume} {76}},\ \bibinfo {pages} {659}
  (\bibinfo {year} {2016})},\ \Eprint {http://arxiv.org/abs/1609.03042}
  {arXiv:1609.03042 [hep-ph]} \BibitemShut {NoStop}%
\bibitem [{\citenamefont {Kalikotay}\ \emph {et~al.}(2020)\citenamefont
  {Kalikotay}, \citenamefont {Ghosh}, \citenamefont {Chaudhuri}, \citenamefont
  {Roy},\ and\ \citenamefont {Sarkar}}]{Kalikotay:2020snc}%
  \BibitemOpen
  \bibfield  {author} {\bibinfo {author} {\bibfnamefont {P.}~\bibnamefont
  {Kalikotay}}, \bibinfo {author} {\bibfnamefont {S.}~\bibnamefont {Ghosh}},
  \bibinfo {author} {\bibfnamefont {N.}~\bibnamefont {Chaudhuri}}, \bibinfo
  {author} {\bibfnamefont {P.}~\bibnamefont {Roy}}, \ and\ \bibinfo {author}
  {\bibfnamefont {S.}~\bibnamefont {Sarkar}},\ }\href {\doibase
  10.1103/PhysRevD.102.076007} {\bibfield  {journal} {\bibinfo  {journal}
  {Phys. Rev. D}\ }\textbf {\bibinfo {volume} {102}},\ \bibinfo {pages}
  {076007} (\bibinfo {year} {2020})},\ \Eprint
  {http://arxiv.org/abs/2009.10493} {arXiv:2009.10493 [hep-ph]} \BibitemShut
  {NoStop}%
\bibitem [{\citenamefont {Friman}\ \emph {et~al.}(2011)\citenamefont {Friman},
  \citenamefont {Hohne}, \citenamefont {Knoll}, \citenamefont {Leupold},
  \citenamefont {Randrup}, \citenamefont {Rapp},\ and\ \citenamefont
  {Senger}}]{Friman:2011zz}%
  \BibitemOpen
  \bibinfo {editor} {\bibfnamefont {B.}~\bibnamefont {Friman}}, \bibinfo
  {editor} {\bibfnamefont {C.}~\bibnamefont {Hohne}}, \bibinfo {editor}
  {\bibfnamefont {J.}~\bibnamefont {Knoll}}, \bibinfo {editor} {\bibfnamefont
  {S.}~\bibnamefont {Leupold}}, \bibinfo {editor} {\bibfnamefont
  {J.}~\bibnamefont {Randrup}}, \bibinfo {editor} {\bibfnamefont
  {R.}~\bibnamefont {Rapp}}, \ and\ \bibinfo {editor} {\bibfnamefont
  {P.}~\bibnamefont {Senger}},\ eds.,\ \href {\doibase
  10.1007/978-3-642-13293-3} {\emph {\bibinfo {title} {{The CBM physics book:
  Compressed baryonic matter in laboratory experiments}}}},\ Vol.\ \bibinfo
  {volume} {814}\ (\bibinfo {year} {2011})\BibitemShut {NoStop}%
\bibitem [{\citenamefont {Miransky}\ and\ \citenamefont
  {Shovkovy}(2015)}]{Miransky:2015ava}%
  \BibitemOpen
  \bibfield  {author} {\bibinfo {author} {\bibfnamefont {V.~A.}\ \bibnamefont
  {Miransky}}\ and\ \bibinfo {author} {\bibfnamefont {I.~A.}\ \bibnamefont
  {Shovkovy}},\ }\href {\doibase 10.1016/j.physrep.2015.02.003} {\bibfield
  {journal} {\bibinfo  {journal} {Phys. Rept.}\ }\textbf {\bibinfo {volume}
  {576}},\ \bibinfo {pages} {1} (\bibinfo {year} {2015})},\ \Eprint
  {http://arxiv.org/abs/1503.00732} {arXiv:1503.00732 [hep-ph]} \BibitemShut
  {NoStop}%
\bibitem [{\citenamefont {Kharzeev}\ \emph {et~al.}(2013)\citenamefont
  {Kharzeev}, \citenamefont {Landsteiner}, \citenamefont {Schmitt},\ and\
  \citenamefont {Yee}}]{Kharzeev:2013jha}%
  \BibitemOpen
  \bibinfo {editor} {\bibfnamefont {D.}~\bibnamefont {Kharzeev}}, \bibinfo
  {editor} {\bibfnamefont {K.}~\bibnamefont {Landsteiner}}, \bibinfo {editor}
  {\bibfnamefont {A.}~\bibnamefont {Schmitt}}, \ and\ \bibinfo {editor}
  {\bibfnamefont {H.-U.}\ \bibnamefont {Yee}},\ eds.,\ \href {\doibase
  10.1007/978-3-642-37305-3} {\emph {\bibinfo {title} {{Strongly Interacting
  Matter in Magnetic Fields}}}},\ Vol.\ \bibinfo {volume} {871}\ (\bibinfo
  {year} {2013})\BibitemShut {NoStop}%
\bibitem [{\citenamefont {Chatrchyan}\ \emph {et~al.}(2012)\citenamefont
  {Chatrchyan} \emph {et~al.}}]{CMS:2012sap}%
  \BibitemOpen
  \bibfield  {author} {\bibinfo {author} {\bibfnamefont {S.}~\bibnamefont
  {Chatrchyan}} \emph {et~al.} (\bibinfo {collaboration} {CMS}),\ }\href
  {\doibase 10.1007/JHEP05(2012)055} {\bibfield  {journal} {\bibinfo  {journal}
  {JHEP}\ }\textbf {\bibinfo {volume} {05}},\ \bibinfo {pages} {055} (\bibinfo
  {year} {2012})},\ \Eprint {http://arxiv.org/abs/1202.5535} {arXiv:1202.5535
  [hep-ex]} \BibitemShut {NoStop}%
\bibitem [{\citenamefont {Fukushima}\ \emph {et~al.}(2008)\citenamefont
  {Fukushima}, \citenamefont {Kharzeev},\ and\ \citenamefont
  {Warringa}}]{Fukushima:2008xe}%
  \BibitemOpen
  \bibfield  {author} {\bibinfo {author} {\bibfnamefont {K.}~\bibnamefont
  {Fukushima}}, \bibinfo {author} {\bibfnamefont {D.~E.}\ \bibnamefont
  {Kharzeev}}, \ and\ \bibinfo {author} {\bibfnamefont {H.~J.}\ \bibnamefont
  {Warringa}},\ }\href {\doibase 10.1103/PhysRevD.78.074033} {\bibfield
  {journal} {\bibinfo  {journal} {Phys. Rev. D}\ }\textbf {\bibinfo {volume}
  {78}},\ \bibinfo {pages} {074033} (\bibinfo {year} {2008})},\ \Eprint
  {http://arxiv.org/abs/0808.3382} {arXiv:0808.3382 [hep-ph]} \BibitemShut
  {NoStop}%
\bibitem [{\citenamefont {Kharzeev}\ and\ \citenamefont
  {Warringa}(2009)}]{Kharzeev:2009pj}%
  \BibitemOpen
  \bibfield  {author} {\bibinfo {author} {\bibfnamefont {D.~E.}\ \bibnamefont
  {Kharzeev}}\ and\ \bibinfo {author} {\bibfnamefont {H.~J.}\ \bibnamefont
  {Warringa}},\ }\href {\doibase 10.1103/PhysRevD.80.034028} {\bibfield
  {journal} {\bibinfo  {journal} {Phys. Rev. D}\ }\textbf {\bibinfo {volume}
  {80}},\ \bibinfo {pages} {034028} (\bibinfo {year} {2009})},\ \Eprint
  {http://arxiv.org/abs/0907.5007} {arXiv:0907.5007 [hep-ph]} \BibitemShut
  {NoStop}%
\bibitem [{\citenamefont {Shovkovy}(2013)}]{Shovkovy:2012zn}%
  \BibitemOpen
  \bibfield  {author} {\bibinfo {author} {\bibfnamefont {I.~A.}\ \bibnamefont
  {Shovkovy}},\ }\href {\doibase 10.1007/978-3-642-37305-3_2} {\bibfield
  {journal} {\bibinfo  {journal} {Lect. Notes Phys.}\ }\textbf {\bibinfo
  {volume} {871}},\ \bibinfo {pages} {13} (\bibinfo {year} {2013})},\ \Eprint
  {http://arxiv.org/abs/1207.5081} {arXiv:1207.5081 [hep-ph]} \BibitemShut
  {NoStop}%
\bibitem [{\citenamefont {Gusynin}\ \emph {et~al.}(1994)\citenamefont
  {Gusynin}, \citenamefont {Miransky},\ and\ \citenamefont
  {Shovkovy}}]{Gusynin:1994re}%
  \BibitemOpen
  \bibfield  {author} {\bibinfo {author} {\bibfnamefont {V.~P.}\ \bibnamefont
  {Gusynin}}, \bibinfo {author} {\bibfnamefont {V.~A.}\ \bibnamefont
  {Miransky}}, \ and\ \bibinfo {author} {\bibfnamefont {I.~A.}\ \bibnamefont
  {Shovkovy}},\ }\href {\doibase 10.1103/PhysRevLett.73.3499} {\bibfield
  {journal} {\bibinfo  {journal} {Phys. Rev. Lett.}\ }\textbf {\bibinfo
  {volume} {73}},\ \bibinfo {pages} {3499} (\bibinfo {year} {1994})},\ \bibinfo
  {note} {[Erratum: Phys.Rev.Lett. 76, 1005 (1996)]},\ \Eprint
  {http://arxiv.org/abs/hep-ph/9405262} {arXiv:hep-ph/9405262} \BibitemShut
  {NoStop}%
\bibitem [{\citenamefont {Gusynin}\ \emph {et~al.}(1996)\citenamefont
  {Gusynin}, \citenamefont {Miransky},\ and\ \citenamefont
  {Shovkovy}}]{Gusynin:1995nb}%
  \BibitemOpen
  \bibfield  {author} {\bibinfo {author} {\bibfnamefont {V.~P.}\ \bibnamefont
  {Gusynin}}, \bibinfo {author} {\bibfnamefont {V.~A.}\ \bibnamefont
  {Miransky}}, \ and\ \bibinfo {author} {\bibfnamefont {I.~A.}\ \bibnamefont
  {Shovkovy}},\ }\href {\doibase 10.1016/0550-3213(96)00021-1} {\bibfield
  {journal} {\bibinfo  {journal} {Nucl. Phys. B}\ }\textbf {\bibinfo {volume}
  {462}},\ \bibinfo {pages} {249} (\bibinfo {year} {1996})},\ \Eprint
  {http://arxiv.org/abs/hep-ph/9509320} {arXiv:hep-ph/9509320} \BibitemShut
  {NoStop}%
\bibitem [{\citenamefont {Gusynin}\ \emph {et~al.}(1999)\citenamefont
  {Gusynin}, \citenamefont {Miransky},\ and\ \citenamefont
  {Shovkovy}}]{Gusynin:1999pq}%
  \BibitemOpen
  \bibfield  {author} {\bibinfo {author} {\bibfnamefont {V.~P.}\ \bibnamefont
  {Gusynin}}, \bibinfo {author} {\bibfnamefont {V.~A.}\ \bibnamefont
  {Miransky}}, \ and\ \bibinfo {author} {\bibfnamefont {I.~A.}\ \bibnamefont
  {Shovkovy}},\ }\href {\doibase 10.1016/S0550-3213(99)00573-8} {\bibfield
  {journal} {\bibinfo  {journal} {Nucl. Phys. B}\ }\textbf {\bibinfo {volume}
  {563}},\ \bibinfo {pages} {361} (\bibinfo {year} {1999})},\ \Eprint
  {http://arxiv.org/abs/hep-ph/9908320} {arXiv:hep-ph/9908320} \BibitemShut
  {NoStop}%
\bibitem [{\citenamefont {Preis}\ \emph {et~al.}(2011)\citenamefont {Preis},
  \citenamefont {Rebhan},\ and\ \citenamefont {Schmitt}}]{Preis:2010cq}%
  \BibitemOpen
  \bibfield  {author} {\bibinfo {author} {\bibfnamefont {F.}~\bibnamefont
  {Preis}}, \bibinfo {author} {\bibfnamefont {A.}~\bibnamefont {Rebhan}}, \
  and\ \bibinfo {author} {\bibfnamefont {A.}~\bibnamefont {Schmitt}},\ }\href
  {\doibase 10.1007/JHEP03(2011)033} {\bibfield  {journal} {\bibinfo  {journal}
  {JHEP}\ }\textbf {\bibinfo {volume} {03}},\ \bibinfo {pages} {033} (\bibinfo
  {year} {2011})},\ \Eprint {http://arxiv.org/abs/1012.4785} {arXiv:1012.4785
  [hep-th]} \BibitemShut {NoStop}%
\bibitem [{\citenamefont {Preis}\ \emph {et~al.}(2013)\citenamefont {Preis},
  \citenamefont {Rebhan},\ and\ \citenamefont {Schmitt}}]{Preis:2012fh}%
  \BibitemOpen
  \bibfield  {author} {\bibinfo {author} {\bibfnamefont {F.}~\bibnamefont
  {Preis}}, \bibinfo {author} {\bibfnamefont {A.}~\bibnamefont {Rebhan}}, \
  and\ \bibinfo {author} {\bibfnamefont {A.}~\bibnamefont {Schmitt}},\ }\href
  {\doibase 10.1007/978-3-642-37305-3_3} {\bibfield  {journal} {\bibinfo
  {journal} {Lect. Notes Phys.}\ }\textbf {\bibinfo {volume} {871}},\ \bibinfo
  {pages} {51} (\bibinfo {year} {2013})},\ \Eprint
  {http://arxiv.org/abs/1208.0536} {arXiv:1208.0536 [hep-ph]} \BibitemShut
  {NoStop}%
\bibitem [{\citenamefont {Muroya}\ \emph {et~al.}(2003)\citenamefont {Muroya},
  \citenamefont {Nakamura}, \citenamefont {Nonaka},\ and\ \citenamefont
  {Takaishi}}]{Muroya:2003qs}%
  \BibitemOpen
  \bibfield  {author} {\bibinfo {author} {\bibfnamefont {S.}~\bibnamefont
  {Muroya}}, \bibinfo {author} {\bibfnamefont {A.}~\bibnamefont {Nakamura}},
  \bibinfo {author} {\bibfnamefont {C.}~\bibnamefont {Nonaka}}, \ and\ \bibinfo
  {author} {\bibfnamefont {T.}~\bibnamefont {Takaishi}},\ }\href {\doibase
  10.1143/PTP.110.615} {\bibfield  {journal} {\bibinfo  {journal} {Prog. Theor.
  Phys.}\ }\textbf {\bibinfo {volume} {110}},\ \bibinfo {pages} {615} (\bibinfo
  {year} {2003})},\ \Eprint {http://arxiv.org/abs/hep-lat/0306031}
  {arXiv:hep-lat/0306031} \BibitemShut {NoStop}%
\bibitem [{\citenamefont {Aoki}\ \emph
  {et~al.}(2006{\natexlab{a}})\citenamefont {Aoki}, \citenamefont {Fodor},
  \citenamefont {Katz},\ and\ \citenamefont {Szabo}}]{Aoki:2006br}%
  \BibitemOpen
  \bibfield  {author} {\bibinfo {author} {\bibfnamefont {Y.}~\bibnamefont
  {Aoki}}, \bibinfo {author} {\bibfnamefont {Z.}~\bibnamefont {Fodor}},
  \bibinfo {author} {\bibfnamefont {S.~D.}\ \bibnamefont {Katz}}, \ and\
  \bibinfo {author} {\bibfnamefont {K.~K.}\ \bibnamefont {Szabo}},\ }\href
  {\doibase 10.1016/j.physletb.2006.10.021} {\bibfield  {journal} {\bibinfo
  {journal} {Phys. Lett. B}\ }\textbf {\bibinfo {volume} {643}},\ \bibinfo
  {pages} {46} (\bibinfo {year} {2006}{\natexlab{a}})},\ \Eprint
  {http://arxiv.org/abs/hep-lat/0609068} {arXiv:hep-lat/0609068} \BibitemShut
  {NoStop}%
\bibitem [{\citenamefont {Bazavov}\ \emph {et~al.}(2009)\citenamefont {Bazavov}
  \emph {et~al.}}]{Bazavov:2009zn}%
  \BibitemOpen
  \bibfield  {author} {\bibinfo {author} {\bibfnamefont {A.}~\bibnamefont
  {Bazavov}} \emph {et~al.},\ }\href {\doibase 10.1103/PhysRevD.80.014504}
  {\bibfield  {journal} {\bibinfo  {journal} {Phys. Rev. D}\ }\textbf {\bibinfo
  {volume} {80}},\ \bibinfo {pages} {014504} (\bibinfo {year} {2009})},\
  \Eprint {http://arxiv.org/abs/0903.4379} {arXiv:0903.4379 [hep-lat]}
  \BibitemShut {NoStop}%
\bibitem [{\citenamefont {Cheng}\ \emph {et~al.}(2008)\citenamefont {Cheng}
  \emph {et~al.}}]{Cheng:2007jq}%
  \BibitemOpen
  \bibfield  {author} {\bibinfo {author} {\bibfnamefont {M.}~\bibnamefont
  {Cheng}} \emph {et~al.},\ }\href {\doibase 10.1103/PhysRevD.77.014511}
  {\bibfield  {journal} {\bibinfo  {journal} {Phys. Rev. D}\ }\textbf {\bibinfo
  {volume} {77}},\ \bibinfo {pages} {014511} (\bibinfo {year} {2008})},\
  \Eprint {http://arxiv.org/abs/0710.0354} {arXiv:0710.0354 [hep-lat]}
  \BibitemShut {NoStop}%
\bibitem [{\citenamefont {Klevansky}(1992)}]{Klevansky:1992qe}%
  \BibitemOpen
  \bibfield  {author} {\bibinfo {author} {\bibfnamefont {S.~P.}\ \bibnamefont
  {Klevansky}},\ }\href {\doibase 10.1103/RevModPhys.64.649} {\bibfield
  {journal} {\bibinfo  {journal} {Rev. Mod. Phys.}\ }\textbf {\bibinfo {volume}
  {64}},\ \bibinfo {pages} {649} (\bibinfo {year} {1992})}\BibitemShut
  {NoStop}%
\bibitem [{\citenamefont {Vogl}\ and\ \citenamefont
  {Weise}(1991)}]{Vogl:1991qt}%
  \BibitemOpen
  \bibfield  {author} {\bibinfo {author} {\bibfnamefont {U.}~\bibnamefont
  {Vogl}}\ and\ \bibinfo {author} {\bibfnamefont {W.}~\bibnamefont {Weise}},\
  }\href {\doibase 10.1016/0146-6410(91)90005-9} {\bibfield  {journal}
  {\bibinfo  {journal} {Prog. Part. Nucl. Phys.}\ }\textbf {\bibinfo {volume}
  {27}},\ \bibinfo {pages} {195} (\bibinfo {year} {1991})}\BibitemShut
  {NoStop}%
\bibitem [{\citenamefont {Buballa}(2005)}]{Buballa:2003qv}%
  \BibitemOpen
  \bibfield  {author} {\bibinfo {author} {\bibfnamefont {M.}~\bibnamefont
  {Buballa}},\ }\href {\doibase 10.1016/j.physrep.2004.11.004} {\bibfield
  {journal} {\bibinfo  {journal} {Phys. Rept.}\ }\textbf {\bibinfo {volume}
  {407}},\ \bibinfo {pages} {205} (\bibinfo {year} {2005})},\ \Eprint
  {http://arxiv.org/abs/hep-ph/0402234} {arXiv:hep-ph/0402234} \BibitemShut
  {NoStop}%
\bibitem [{\citenamefont {Nambu}\ and\ \citenamefont
  {Jona-Lasinio}(1961{\natexlab{a}})}]{Nambu:1961fr}%
  \BibitemOpen
  \bibfield  {author} {\bibinfo {author} {\bibfnamefont {Y.}~\bibnamefont
  {Nambu}}\ and\ \bibinfo {author} {\bibfnamefont {G.}~\bibnamefont
  {Jona-Lasinio}},\ }\href {\doibase 10.1103/PhysRev.124.246} {\bibfield
  {journal} {\bibinfo  {journal} {Phys. Rev.}\ }\textbf {\bibinfo {volume}
  {124}},\ \bibinfo {pages} {246} (\bibinfo {year}
  {1961}{\natexlab{a}})}\BibitemShut {NoStop}%
\bibitem [{\citenamefont {Nambu}\ and\ \citenamefont
  {Jona-Lasinio}(1961{\natexlab{b}})}]{Nambu:1961tp}%
  \BibitemOpen
  \bibfield  {author} {\bibinfo {author} {\bibfnamefont {Y.}~\bibnamefont
  {Nambu}}\ and\ \bibinfo {author} {\bibfnamefont {G.}~\bibnamefont
  {Jona-Lasinio}},\ }\href {\doibase 10.1103/PhysRev.122.345} {\bibfield
  {journal} {\bibinfo  {journal} {Phys. Rev.}\ }\textbf {\bibinfo {volume}
  {122}},\ \bibinfo {pages} {345} (\bibinfo {year}
  {1961}{\natexlab{b}})}\BibitemShut {NoStop}%
\bibitem [{\citenamefont {Bijnens}(1996)}]{Bijnens:1995ww}%
  \BibitemOpen
  \bibfield  {author} {\bibinfo {author} {\bibfnamefont {J.}~\bibnamefont
  {Bijnens}},\ }\href {\doibase 10.1016/0370-1573(95)00051-8} {\bibfield
  {journal} {\bibinfo  {journal} {Phys. Rept.}\ }\textbf {\bibinfo {volume}
  {265}},\ \bibinfo {pages} {369} (\bibinfo {year} {1996})},\ \Eprint
  {http://arxiv.org/abs/hep-ph/9502335} {arXiv:hep-ph/9502335} \BibitemShut
  {NoStop}%
\bibitem [{\citenamefont {Ratti}\ \emph {et~al.}(2006)\citenamefont {Ratti},
  \citenamefont {Thaler},\ and\ \citenamefont {Weise}}]{Ratti:2005jh}%
  \BibitemOpen
  \bibfield  {author} {\bibinfo {author} {\bibfnamefont {C.}~\bibnamefont
  {Ratti}}, \bibinfo {author} {\bibfnamefont {M.~A.}\ \bibnamefont {Thaler}}, \
  and\ \bibinfo {author} {\bibfnamefont {W.}~\bibnamefont {Weise}},\ }\href
  {\doibase 10.1103/PhysRevD.73.014019} {\bibfield  {journal} {\bibinfo
  {journal} {Phys. Rev. D}\ }\textbf {\bibinfo {volume} {73}},\ \bibinfo
  {pages} {014019} (\bibinfo {year} {2006})},\ \Eprint
  {http://arxiv.org/abs/hep-ph/0506234} {arXiv:hep-ph/0506234} \BibitemShut
  {NoStop}%
\bibitem [{\citenamefont {Ratti}\ \emph {et~al.}(2007)\citenamefont {Ratti},
  \citenamefont {Roessner}, \citenamefont {Thaler},\ and\ \citenamefont
  {Weise}}]{Ratti:2006wg}%
  \BibitemOpen
  \bibfield  {author} {\bibinfo {author} {\bibfnamefont {C.}~\bibnamefont
  {Ratti}}, \bibinfo {author} {\bibfnamefont {S.}~\bibnamefont {Roessner}},
  \bibinfo {author} {\bibfnamefont {M.~A.}\ \bibnamefont {Thaler}}, \ and\
  \bibinfo {author} {\bibfnamefont {W.}~\bibnamefont {Weise}},\ }\href
  {\doibase 10.1140/epjc/s10052-006-0065-x} {\bibfield  {journal} {\bibinfo
  {journal} {Eur. Phys. J. C}\ }\textbf {\bibinfo {volume} {49}},\ \bibinfo
  {pages} {213} (\bibinfo {year} {2007})},\ \Eprint
  {http://arxiv.org/abs/hep-ph/0609218} {arXiv:hep-ph/0609218} \BibitemShut
  {NoStop}%
\bibitem [{\citenamefont {Andersen}\ \emph {et~al.}(2016)\citenamefont
  {Andersen}, \citenamefont {Naylor},\ and\ \citenamefont
  {Tranberg}}]{Andersen:2014xxa}%
  \BibitemOpen
  \bibfield  {author} {\bibinfo {author} {\bibfnamefont {J.~O.}\ \bibnamefont
  {Andersen}}, \bibinfo {author} {\bibfnamefont {W.~R.}\ \bibnamefont
  {Naylor}}, \ and\ \bibinfo {author} {\bibfnamefont {A.}~\bibnamefont
  {Tranberg}},\ }\href {\doibase 10.1103/RevModPhys.88.025001} {\bibfield
  {journal} {\bibinfo  {journal} {Rev. Mod. Phys.}\ }\textbf {\bibinfo {volume}
  {88}},\ \bibinfo {pages} {025001} (\bibinfo {year} {2016})},\ \Eprint
  {http://arxiv.org/abs/1411.7176} {arXiv:1411.7176 [hep-ph]} \BibitemShut
  {NoStop}%
\bibitem [{\citenamefont {Chaudhuri}\ \emph {et~al.}(2020)\citenamefont
  {Chaudhuri}, \citenamefont {Ghosh}, \citenamefont {Sarkar},\ and\
  \citenamefont {Roy}}]{Chaudhuri:2020lga}%
  \BibitemOpen
  \bibfield  {author} {\bibinfo {author} {\bibfnamefont {N.}~\bibnamefont
  {Chaudhuri}}, \bibinfo {author} {\bibfnamefont {S.}~\bibnamefont {Ghosh}},
  \bibinfo {author} {\bibfnamefont {S.}~\bibnamefont {Sarkar}}, \ and\ \bibinfo
  {author} {\bibfnamefont {P.}~\bibnamefont {Roy}},\ }\href {\doibase
  10.1140/epja/s10050-020-00222-9} {\bibfield  {journal} {\bibinfo  {journal}
  {Eur. Phys. J. A}\ }\textbf {\bibinfo {volume} {56}},\ \bibinfo {pages} {213}
  (\bibinfo {year} {2020})},\ \Eprint {http://arxiv.org/abs/2003.05692}
  {arXiv:2003.05692 [nucl-th]} \BibitemShut {NoStop}%
\bibitem [{\citenamefont {Gatto}\ and\ \citenamefont
  {Ruggieri}(2010)}]{Gatto:2010qs}%
  \BibitemOpen
  \bibfield  {author} {\bibinfo {author} {\bibfnamefont {R.}~\bibnamefont
  {Gatto}}\ and\ \bibinfo {author} {\bibfnamefont {M.}~\bibnamefont
  {Ruggieri}},\ }\href {\doibase 10.1103/PhysRevD.82.054027} {\bibfield
  {journal} {\bibinfo  {journal} {Phys. Rev. D}\ }\textbf {\bibinfo {volume}
  {82}},\ \bibinfo {pages} {054027} (\bibinfo {year} {2010})},\ \Eprint
  {http://arxiv.org/abs/1007.0790} {arXiv:1007.0790 [hep-ph]} \BibitemShut
  {NoStop}%
\bibitem [{\citenamefont {Fukushima}(2008)}]{Fukushima:2008wg}%
  \BibitemOpen
  \bibfield  {author} {\bibinfo {author} {\bibfnamefont {K.}~\bibnamefont
  {Fukushima}},\ }\href {\doibase 10.1103/PhysRevD.77.114028} {\bibfield
  {journal} {\bibinfo  {journal} {Phys. Rev. D}\ }\textbf {\bibinfo {volume}
  {77}},\ \bibinfo {pages} {114028} (\bibinfo {year} {2008})},\ \bibinfo {note}
  {[Erratum: Phys.Rev.D 78, 039902 (2008)]},\ \Eprint
  {http://arxiv.org/abs/0803.3318} {arXiv:0803.3318 [hep-ph]} \BibitemShut
  {NoStop}%
\bibitem [{\citenamefont {Mattos}\ \emph
  {et~al.}(2021{\natexlab{a}})\citenamefont {Mattos}, \citenamefont
  {Frederico},\ and\ \citenamefont {Louren\c{c}o}}]{Mattos:2021alf}%
  \BibitemOpen
  \bibfield  {author} {\bibinfo {author} {\bibfnamefont {O.~A.}\ \bibnamefont
  {Mattos}}, \bibinfo {author} {\bibfnamefont {T.}~\bibnamefont {Frederico}}, \
  and\ \bibinfo {author} {\bibfnamefont {O.}~\bibnamefont {Louren\c{c}o}},\
  }\href {\doibase 10.1140/epjc/s10052-021-08827-0} {\bibfield  {journal}
  {\bibinfo  {journal} {Eur. Phys. J. C}\ }\textbf {\bibinfo {volume} {81}},\
  \bibinfo {pages} {24} (\bibinfo {year} {2021}{\natexlab{a}})},\ \Eprint
  {http://arxiv.org/abs/2101.07340} {arXiv:2101.07340 [hep-ph]} \BibitemShut
  {NoStop}%
\bibitem [{\citenamefont {Mattos}\ \emph
  {et~al.}(2021{\natexlab{b}})\citenamefont {Mattos}, \citenamefont
  {Frederico}, \citenamefont {Lenzi}, \citenamefont {Dutra},\ and\
  \citenamefont {Louren\c{c}o}}]{Mattos:2021tmz}%
  \BibitemOpen
  \bibfield  {author} {\bibinfo {author} {\bibfnamefont {O.~A.}\ \bibnamefont
  {Mattos}}, \bibinfo {author} {\bibfnamefont {T.}~\bibnamefont {Frederico}},
  \bibinfo {author} {\bibfnamefont {C.~H.}\ \bibnamefont {Lenzi}}, \bibinfo
  {author} {\bibfnamefont {M.}~\bibnamefont {Dutra}}, \ and\ \bibinfo {author}
  {\bibfnamefont {O.}~\bibnamefont {Louren\c{c}o}},\ }\href {\doibase
  10.1103/PhysRevD.104.116001} {\bibfield  {journal} {\bibinfo  {journal}
  {Phys. Rev. D}\ }\textbf {\bibinfo {volume} {104}},\ \bibinfo {pages}
  {116001} (\bibinfo {year} {2021}{\natexlab{b}})},\ \Eprint
  {http://arxiv.org/abs/2110.05602} {arXiv:2110.05602 [hep-ph]} \BibitemShut
  {NoStop}%
\bibitem [{\citenamefont {Wang}\ and\ \citenamefont
  {Wen}(2022)}]{Wang:2022xxp}%
  \BibitemOpen
  \bibfield  {author} {\bibinfo {author} {\bibfnamefont {Y.}~\bibnamefont
  {Wang}}\ and\ \bibinfo {author} {\bibfnamefont {X.-J.}\ \bibnamefont {Wen}},\
  }\href {\doibase 10.1103/PhysRevD.105.074034} {\bibfield  {journal} {\bibinfo
   {journal} {Phys. Rev. D}\ }\textbf {\bibinfo {volume} {105}},\ \bibinfo
  {pages} {074034} (\bibinfo {year} {2022})},\ \Eprint
  {http://arxiv.org/abs/2204.06135} {arXiv:2204.06135 [hep-ph]} \BibitemShut
  {NoStop}%
\bibitem [{\citenamefont {Bali}\ \emph
  {et~al.}(2012{\natexlab{a}})\citenamefont {Bali}, \citenamefont {Bruckmann},
  \citenamefont {Endrodi}, \citenamefont {Fodor}, \citenamefont {Katz},
  \citenamefont {Krieg}, \citenamefont {Schafer},\ and\ \citenamefont
  {Szabo}}]{Bali:2011qj}%
  \BibitemOpen
  \bibfield  {author} {\bibinfo {author} {\bibfnamefont {G.~S.}\ \bibnamefont
  {Bali}}, \bibinfo {author} {\bibfnamefont {F.}~\bibnamefont {Bruckmann}},
  \bibinfo {author} {\bibfnamefont {G.}~\bibnamefont {Endrodi}}, \bibinfo
  {author} {\bibfnamefont {Z.}~\bibnamefont {Fodor}}, \bibinfo {author}
  {\bibfnamefont {S.~D.}\ \bibnamefont {Katz}}, \bibinfo {author}
  {\bibfnamefont {S.}~\bibnamefont {Krieg}}, \bibinfo {author} {\bibfnamefont
  {A.}~\bibnamefont {Schafer}}, \ and\ \bibinfo {author} {\bibfnamefont
  {K.~K.}\ \bibnamefont {Szabo}},\ }\href {\doibase 10.1007/JHEP02(2012)044}
  {\bibfield  {journal} {\bibinfo  {journal} {JHEP}\ }\textbf {\bibinfo
  {volume} {02}},\ \bibinfo {pages} {044} (\bibinfo {year}
  {2012}{\natexlab{a}})},\ \Eprint {http://arxiv.org/abs/1111.4956}
  {arXiv:1111.4956 [hep-lat]} \BibitemShut {NoStop}%
\bibitem [{\citenamefont {Bali}\ \emph
  {et~al.}(2012{\natexlab{b}})\citenamefont {Bali}, \citenamefont {Collins},
  \citenamefont {Deka}, \citenamefont {Glassle}, \citenamefont {Gockeler},
  \citenamefont {Najjar}, \citenamefont {Nobile}, \citenamefont {Pleiter},
  \citenamefont {Schafer},\ and\ \citenamefont {Sternbeck}}]{Bali:2012av}%
  \BibitemOpen
  \bibfield  {author} {\bibinfo {author} {\bibfnamefont {G.~S.}\ \bibnamefont
  {Bali}}, \bibinfo {author} {\bibfnamefont {S.}~\bibnamefont {Collins}},
  \bibinfo {author} {\bibfnamefont {M.}~\bibnamefont {Deka}}, \bibinfo {author}
  {\bibfnamefont {B.}~\bibnamefont {Glassle}}, \bibinfo {author} {\bibfnamefont
  {M.}~\bibnamefont {Gockeler}}, \bibinfo {author} {\bibfnamefont
  {J.}~\bibnamefont {Najjar}}, \bibinfo {author} {\bibfnamefont
  {A.}~\bibnamefont {Nobile}}, \bibinfo {author} {\bibfnamefont
  {D.}~\bibnamefont {Pleiter}}, \bibinfo {author} {\bibfnamefont
  {A.}~\bibnamefont {Schafer}}, \ and\ \bibinfo {author} {\bibfnamefont
  {A.}~\bibnamefont {Sternbeck}},\ }\href {\doibase 10.1103/PhysRevD.86.054504}
  {\bibfield  {journal} {\bibinfo  {journal} {Phys. Rev. D}\ }\textbf {\bibinfo
  {volume} {86}},\ \bibinfo {pages} {054504} (\bibinfo {year}
  {2012}{\natexlab{b}})},\ \Eprint {http://arxiv.org/abs/1207.1110}
  {arXiv:1207.1110 [hep-lat]} \BibitemShut {NoStop}%
\bibitem [{\citenamefont {Bandyopadhyay}\ and\ \citenamefont
  {Farias}(2021)}]{Bandyopadhyay:2020zte}%
  \BibitemOpen
  \bibfield  {author} {\bibinfo {author} {\bibfnamefont {A.}~\bibnamefont
  {Bandyopadhyay}}\ and\ \bibinfo {author} {\bibfnamefont {R.~L.~S.}\
  \bibnamefont {Farias}},\ }\href {\doibase 10.1140/epjs/s11734-021-00023-1}
  {\bibfield  {journal} {\bibinfo  {journal} {Eur. Phys. J. ST}\ }\textbf
  {\bibinfo {volume} {230}},\ \bibinfo {pages} {719} (\bibinfo {year}
  {2021})},\ \Eprint {http://arxiv.org/abs/2003.11054} {arXiv:2003.11054
  [hep-ph]} \BibitemShut {NoStop}%
\bibitem [{\citenamefont {Ferreira}\ \emph {et~al.}(2014)\citenamefont
  {Ferreira}, \citenamefont {Costa}, \citenamefont {Louren\c{c}o},
  \citenamefont {Frederico},\ and\ \citenamefont
  {Provid\^encia}}]{Ferreira:2014kpa}%
  \BibitemOpen
  \bibfield  {author} {\bibinfo {author} {\bibfnamefont {M.}~\bibnamefont
  {Ferreira}}, \bibinfo {author} {\bibfnamefont {P.}~\bibnamefont {Costa}},
  \bibinfo {author} {\bibfnamefont {O.}~\bibnamefont {Louren\c{c}o}}, \bibinfo
  {author} {\bibfnamefont {T.}~\bibnamefont {Frederico}}, \ and\ \bibinfo
  {author} {\bibfnamefont {C.}~\bibnamefont {Provid\^encia}},\ }\href {\doibase
  10.1103/PhysRevD.89.116011} {\bibfield  {journal} {\bibinfo  {journal} {Phys.
  Rev. D}\ }\textbf {\bibinfo {volume} {89}},\ \bibinfo {pages} {116011}
  (\bibinfo {year} {2014})},\ \Eprint {http://arxiv.org/abs/1404.5577}
  {arXiv:1404.5577 [hep-ph]} \BibitemShut {NoStop}%
\bibitem [{\citenamefont {Farias}\ \emph {et~al.}(2017)\citenamefont {Farias},
  \citenamefont {Timoteo}, \citenamefont {Avancini}, \citenamefont {Pinto},\
  and\ \citenamefont {Krein}}]{Farias:2016gmy}%
  \BibitemOpen
  \bibfield  {author} {\bibinfo {author} {\bibfnamefont {R.~L.~S.}\
  \bibnamefont {Farias}}, \bibinfo {author} {\bibfnamefont {V.~S.}\
  \bibnamefont {Timoteo}}, \bibinfo {author} {\bibfnamefont {S.~S.}\
  \bibnamefont {Avancini}}, \bibinfo {author} {\bibfnamefont {M.~B.}\
  \bibnamefont {Pinto}}, \ and\ \bibinfo {author} {\bibfnamefont
  {G.}~\bibnamefont {Krein}},\ }\href {\doibase 10.1140/epja/i2017-12320-8}
  {\bibfield  {journal} {\bibinfo  {journal} {Eur. Phys. J. A}\ }\textbf
  {\bibinfo {volume} {53}},\ \bibinfo {pages} {101} (\bibinfo {year} {2017})},\
  \Eprint {http://arxiv.org/abs/1603.03847} {arXiv:1603.03847 [hep-ph]}
  \BibitemShut {NoStop}%
\bibitem [{\citenamefont {Avancini}\ \emph {et~al.}(2019)\citenamefont
  {Avancini}, \citenamefont {Farias},\ and\ \citenamefont
  {Tavares}}]{Avancini:2018svs}%
  \BibitemOpen
  \bibfield  {author} {\bibinfo {author} {\bibfnamefont {S.~S.}\ \bibnamefont
  {Avancini}}, \bibinfo {author} {\bibfnamefont {R.~L.}\ \bibnamefont
  {Farias}}, \ and\ \bibinfo {author} {\bibfnamefont {W.~R.}\ \bibnamefont
  {Tavares}},\ }\href {\doibase 10.1103/PhysRevD.99.056009} {\bibfield
  {journal} {\bibinfo  {journal} {Phys. Rev. D}\ }\textbf {\bibinfo {volume}
  {99}},\ \bibinfo {pages} {056009} (\bibinfo {year} {2019})},\ \Eprint
  {http://arxiv.org/abs/1812.00945} {arXiv:1812.00945 [hep-ph]} \BibitemShut
  {NoStop}%
\bibitem [{\citenamefont {Sheng}\ \emph {et~al.}(2022)\citenamefont {Sheng},
  \citenamefont {Wang},\ and\ \citenamefont {Yu}}]{Sheng:2021evj}%
  \BibitemOpen
  \bibfield  {author} {\bibinfo {author} {\bibfnamefont {B.-k.}\ \bibnamefont
  {Sheng}}, \bibinfo {author} {\bibfnamefont {X.}~\bibnamefont {Wang}}, \ and\
  \bibinfo {author} {\bibfnamefont {L.}~\bibnamefont {Yu}},\ }\href {\doibase
  10.1103/PhysRevD.105.034003} {\bibfield  {journal} {\bibinfo  {journal}
  {Phys. Rev. D}\ }\textbf {\bibinfo {volume} {105}},\ \bibinfo {pages}
  {034003} (\bibinfo {year} {2022})},\ \Eprint
  {http://arxiv.org/abs/2110.12811} {arXiv:2110.12811 [hep-ph]} \BibitemShut
  {NoStop}%
\bibitem [{\citenamefont {Mao}(2016)}]{Mao:2016fha}%
  \BibitemOpen
  \bibfield  {author} {\bibinfo {author} {\bibfnamefont {S.}~\bibnamefont
  {Mao}},\ }\href {\doibase 10.1016/j.physletb.2016.05.018} {\bibfield
  {journal} {\bibinfo  {journal} {Phys. Lett. B}\ }\textbf {\bibinfo {volume}
  {758}},\ \bibinfo {pages} {195} (\bibinfo {year} {2016})},\ \Eprint
  {http://arxiv.org/abs/1602.06503} {arXiv:1602.06503 [hep-ph]} \BibitemShut
  {NoStop}%
\bibitem [{\citenamefont {Fayazbakhsh}\ and\ \citenamefont
  {Sadooghi}(2014)}]{Fayazbakhsh:2014mca}%
  \BibitemOpen
  \bibfield  {author} {\bibinfo {author} {\bibfnamefont {S.}~\bibnamefont
  {Fayazbakhsh}}\ and\ \bibinfo {author} {\bibfnamefont {N.}~\bibnamefont
  {Sadooghi}},\ }\href {\doibase 10.1103/PhysRevD.90.105030} {\bibfield
  {journal} {\bibinfo  {journal} {Phys. Rev. D}\ }\textbf {\bibinfo {volume}
  {90}},\ \bibinfo {pages} {105030} (\bibinfo {year} {2014})},\ \Eprint
  {http://arxiv.org/abs/1408.5457} {arXiv:1408.5457 [hep-ph]} \BibitemShut
  {NoStop}%
\bibitem [{\citenamefont {Chaudhuri}\ \emph {et~al.}(2019)\citenamefont
  {Chaudhuri}, \citenamefont {Ghosh}, \citenamefont {Sarkar},\ and\
  \citenamefont {Roy}}]{Chaudhuri:2019lbw}%
  \BibitemOpen
  \bibfield  {author} {\bibinfo {author} {\bibfnamefont {N.}~\bibnamefont
  {Chaudhuri}}, \bibinfo {author} {\bibfnamefont {S.}~\bibnamefont {Ghosh}},
  \bibinfo {author} {\bibfnamefont {S.}~\bibnamefont {Sarkar}}, \ and\ \bibinfo
  {author} {\bibfnamefont {P.}~\bibnamefont {Roy}},\ }\href {\doibase
  10.1103/PhysRevD.99.116025} {\bibfield  {journal} {\bibinfo  {journal} {Phys.
  Rev. D}\ }\textbf {\bibinfo {volume} {99}},\ \bibinfo {pages} {116025}
  (\bibinfo {year} {2019})},\ \Eprint {http://arxiv.org/abs/1907.03990}
  {arXiv:1907.03990 [nucl-th]} \BibitemShut {NoStop}%
\bibitem [{\citenamefont {Chaudhuri}\ \emph {et~al.}(2021)\citenamefont
  {Chaudhuri}, \citenamefont {Ghosh}, \citenamefont {Sarkar},\ and\
  \citenamefont {Roy}}]{Chaudhuri:2021skc}%
  \BibitemOpen
  \bibfield  {author} {\bibinfo {author} {\bibfnamefont {N.}~\bibnamefont
  {Chaudhuri}}, \bibinfo {author} {\bibfnamefont {S.}~\bibnamefont {Ghosh}},
  \bibinfo {author} {\bibfnamefont {S.}~\bibnamefont {Sarkar}}, \ and\ \bibinfo
  {author} {\bibfnamefont {P.}~\bibnamefont {Roy}},\ }\href {\doibase
  10.1103/PhysRevD.103.096021} {\bibfield  {journal} {\bibinfo  {journal}
  {Phys. Rev. D}\ }\textbf {\bibinfo {volume} {103}},\ \bibinfo {pages}
  {096021} (\bibinfo {year} {2021})},\ \Eprint
  {http://arxiv.org/abs/2104.11425} {arXiv:2104.11425 [hep-ph]} \BibitemShut
  {NoStop}%
\bibitem [{\citenamefont {Chaudhuri}\ \emph
  {et~al.}(2022{\natexlab{a}})\citenamefont {Chaudhuri}, \citenamefont
  {Mukherjee}, \citenamefont {Ghosh}, \citenamefont {Sarkar},\ and\
  \citenamefont {Roy}}]{Chaudhuri:2021lui}%
  \BibitemOpen
  \bibfield  {author} {\bibinfo {author} {\bibfnamefont {N.}~\bibnamefont
  {Chaudhuri}}, \bibinfo {author} {\bibfnamefont {A.}~\bibnamefont
  {Mukherjee}}, \bibinfo {author} {\bibfnamefont {S.}~\bibnamefont {Ghosh}},
  \bibinfo {author} {\bibfnamefont {S.}~\bibnamefont {Sarkar}}, \ and\ \bibinfo
  {author} {\bibfnamefont {P.}~\bibnamefont {Roy}},\ }\href {\doibase
  10.1140/epja/s10050-022-00731-9} {\bibfield  {journal} {\bibinfo  {journal}
  {Eur. Phys. J. A}\ }\textbf {\bibinfo {volume} {58}},\ \bibinfo {pages} {82}
  (\bibinfo {year} {2022}{\natexlab{a}})},\ \Eprint
  {http://arxiv.org/abs/2111.12058} {arXiv:2111.12058 [hep-ph]} \BibitemShut
  {NoStop}%
\bibitem [{\citenamefont {Ghosh}\ \emph {et~al.}(2020)\citenamefont {Ghosh},
  \citenamefont {Chaudhuri}, \citenamefont {Sarkar},\ and\ \citenamefont
  {Roy}}]{Ghosh:2020xwp}%
  \BibitemOpen
  \bibfield  {author} {\bibinfo {author} {\bibfnamefont {S.}~\bibnamefont
  {Ghosh}}, \bibinfo {author} {\bibfnamefont {N.}~\bibnamefont {Chaudhuri}},
  \bibinfo {author} {\bibfnamefont {S.}~\bibnamefont {Sarkar}}, \ and\ \bibinfo
  {author} {\bibfnamefont {P.}~\bibnamefont {Roy}},\ }\href {\doibase
  10.1103/PhysRevD.101.096002} {\bibfield  {journal} {\bibinfo  {journal}
  {Phys. Rev. D}\ }\textbf {\bibinfo {volume} {101}},\ \bibinfo {pages}
  {096002} (\bibinfo {year} {2020})},\ \Eprint
  {http://arxiv.org/abs/2004.09203} {arXiv:2004.09203 [nucl-th]} \BibitemShut
  {NoStop}%
\bibitem [{\citenamefont {Xu}\ \emph {et~al.}(2021)\citenamefont {Xu},
  \citenamefont {Chao},\ and\ \citenamefont {Huang}}]{Xu:2020yag}%
  \BibitemOpen
  \bibfield  {author} {\bibinfo {author} {\bibfnamefont {K.}~\bibnamefont
  {Xu}}, \bibinfo {author} {\bibfnamefont {J.}~\bibnamefont {Chao}}, \ and\
  \bibinfo {author} {\bibfnamefont {M.}~\bibnamefont {Huang}},\ }\href
  {\doibase 10.1103/PhysRevD.103.076015} {\bibfield  {journal} {\bibinfo
  {journal} {Phys. Rev. D}\ }\textbf {\bibinfo {volume} {103}},\ \bibinfo
  {pages} {076015} (\bibinfo {year} {2021})},\ \Eprint
  {http://arxiv.org/abs/2007.13122} {arXiv:2007.13122 [hep-ph]} \BibitemShut
  {NoStop}%
\bibitem [{\citenamefont {Mei}\ and\ \citenamefont {Mao}(2020)}]{Mei:2020jzn}%
  \BibitemOpen
  \bibfield  {author} {\bibinfo {author} {\bibfnamefont {J.}~\bibnamefont
  {Mei}}\ and\ \bibinfo {author} {\bibfnamefont {S.}~\bibnamefont {Mao}},\
  }\href {\doibase 10.1103/PhysRevD.102.114035} {\bibfield  {journal} {\bibinfo
   {journal} {Phys. Rev. D}\ }\textbf {\bibinfo {volume} {102}},\ \bibinfo
  {pages} {114035} (\bibinfo {year} {2020})},\ \Eprint
  {http://arxiv.org/abs/2008.12123} {arXiv:2008.12123 [hep-ph]} \BibitemShut
  {NoStop}%
\bibitem [{\citenamefont {Aguirre}(2021)}]{Aguirre:2021ljk}%
  \BibitemOpen
  \bibfield  {author} {\bibinfo {author} {\bibfnamefont {R.~M.}\ \bibnamefont
  {Aguirre}},\ }\href {\doibase 10.1140/epja/s10050-021-00480-1} {\bibfield
  {journal} {\bibinfo  {journal} {Eur. Phys. J. A}\ }\textbf {\bibinfo {volume}
  {57}},\ \bibinfo {pages} {166} (\bibinfo {year} {2021})},\ \Eprint
  {http://arxiv.org/abs/2312.13882} {arXiv:2312.13882 [hep-ph]} \BibitemShut
  {NoStop}%
\bibitem [{\citenamefont {Ghosh}\ \emph {et~al.}(2021)\citenamefont {Ghosh},
  \citenamefont {Chaudhuri}, \citenamefont {Roy},\ and\ \citenamefont
  {Sarkar}}]{Ghosh:2021dlo}%
  \BibitemOpen
  \bibfield  {author} {\bibinfo {author} {\bibfnamefont {S.}~\bibnamefont
  {Ghosh}}, \bibinfo {author} {\bibfnamefont {N.}~\bibnamefont {Chaudhuri}},
  \bibinfo {author} {\bibfnamefont {P.}~\bibnamefont {Roy}}, \ and\ \bibinfo
  {author} {\bibfnamefont {S.}~\bibnamefont {Sarkar}},\ }\href {\doibase
  10.1103/PhysRevD.103.116008} {\bibfield  {journal} {\bibinfo  {journal}
  {Phys. Rev. D}\ }\textbf {\bibinfo {volume} {103}},\ \bibinfo {pages}
  {116008} (\bibinfo {year} {2021})},\ \Eprint
  {http://arxiv.org/abs/2104.14112} {arXiv:2104.14112 [hep-ph]} \BibitemShut
  {NoStop}%
\bibitem [{\citenamefont {Farias}\ \emph {et~al.}(2022)\citenamefont {Farias},
  \citenamefont {Tavares}, \citenamefont {Nunes},\ and\ \citenamefont
  {Avancini}}]{Farias:2021fci}%
  \BibitemOpen
  \bibfield  {author} {\bibinfo {author} {\bibfnamefont {R.~L.~S.}\
  \bibnamefont {Farias}}, \bibinfo {author} {\bibfnamefont {W.~R.}\
  \bibnamefont {Tavares}}, \bibinfo {author} {\bibfnamefont {R.~M.}\
  \bibnamefont {Nunes}}, \ and\ \bibinfo {author} {\bibfnamefont {S.~S.}\
  \bibnamefont {Avancini}},\ }\href {\doibase 10.1140/epjc/s10052-022-10640-2}
  {\bibfield  {journal} {\bibinfo  {journal} {Eur. Phys. J. C}\ }\textbf
  {\bibinfo {volume} {82}},\ \bibinfo {pages} {674} (\bibinfo {year} {2022})},\
  \Eprint {http://arxiv.org/abs/2109.11112} {arXiv:2109.11112 [hep-ph]}
  \BibitemShut {NoStop}%
\bibitem [{\citenamefont {Ferrer}\ \emph {et~al.}(2010)\citenamefont {Ferrer},
  \citenamefont {de~la Incera}, \citenamefont {Keith}, \citenamefont
  {Portillo},\ and\ \citenamefont {Springsteen}}]{Ferrer:2010wz}%
  \BibitemOpen
  \bibfield  {author} {\bibinfo {author} {\bibfnamefont {E.~J.}\ \bibnamefont
  {Ferrer}}, \bibinfo {author} {\bibfnamefont {V.}~\bibnamefont {de~la
  Incera}}, \bibinfo {author} {\bibfnamefont {J.~P.}\ \bibnamefont {Keith}},
  \bibinfo {author} {\bibfnamefont {I.}~\bibnamefont {Portillo}}, \ and\
  \bibinfo {author} {\bibfnamefont {P.~L.}\ \bibnamefont {Springsteen}},\
  }\href {\doibase 10.1103/PhysRevC.82.065802} {\bibfield  {journal} {\bibinfo
  {journal} {Phys. Rev. C}\ }\textbf {\bibinfo {volume} {82}},\ \bibinfo
  {pages} {065802} (\bibinfo {year} {2010})},\ \Eprint
  {http://arxiv.org/abs/1009.3521} {arXiv:1009.3521 [hep-ph]} \BibitemShut
  {NoStop}%
\bibitem [{\citenamefont {Ferrer}\ and\ \citenamefont
  {Hackebill}(2022)}]{Ferrer:2020tlz}%
  \BibitemOpen
  \bibfield  {author} {\bibinfo {author} {\bibfnamefont {E.~J.}\ \bibnamefont
  {Ferrer}}\ and\ \bibinfo {author} {\bibfnamefont {A.}~\bibnamefont
  {Hackebill}},\ }\href {\doibase 10.1142/S0217751X22500488} {\bibfield
  {journal} {\bibinfo  {journal} {Int. J. Mod. Phys. A}\ }\textbf {\bibinfo
  {volume} {37}},\ \bibinfo {pages} {2250048} (\bibinfo {year} {2022})},\
  \Eprint {http://arxiv.org/abs/2010.10574} {arXiv:2010.10574 [nucl-th]}
  \BibitemShut {NoStop}%
\bibitem [{\citenamefont {Ferrer}\ and\ \citenamefont
  {Hackebill}(2023)}]{Ferrer:2022afu}%
  \BibitemOpen
  \bibfield  {author} {\bibinfo {author} {\bibfnamefont {E.~J.}\ \bibnamefont
  {Ferrer}}\ and\ \bibinfo {author} {\bibfnamefont {A.}~\bibnamefont
  {Hackebill}},\ }\href {\doibase 10.1016/j.nuclphysa.2023.122608} {\bibfield
  {journal} {\bibinfo  {journal} {Nucl. Phys. A}\ }\textbf {\bibinfo {volume}
  {1031}},\ \bibinfo {pages} {122608} (\bibinfo {year} {2023})},\ \Eprint
  {http://arxiv.org/abs/2203.16576} {arXiv:2203.16576 [hep-ph]} \BibitemShut
  {NoStop}%
\bibitem [{\citenamefont {Chaudhuri}\ \emph
  {et~al.}(2022{\natexlab{b}})\citenamefont {Chaudhuri}, \citenamefont {Ghosh},
  \citenamefont {Roy},\ and\ \citenamefont {Sarkar}}]{Chaudhuri:2022oru}%
  \BibitemOpen
  \bibfield  {author} {\bibinfo {author} {\bibfnamefont {N.}~\bibnamefont
  {Chaudhuri}}, \bibinfo {author} {\bibfnamefont {S.}~\bibnamefont {Ghosh}},
  \bibinfo {author} {\bibfnamefont {P.}~\bibnamefont {Roy}}, \ and\ \bibinfo
  {author} {\bibfnamefont {S.}~\bibnamefont {Sarkar}},\ }\href {\doibase
  10.1103/PhysRevD.106.056020} {\bibfield  {journal} {\bibinfo  {journal}
  {Phys. Rev. D}\ }\textbf {\bibinfo {volume} {106}},\ \bibinfo {pages}
  {056020} (\bibinfo {year} {2022}{\natexlab{b}})},\ \Eprint
  {http://arxiv.org/abs/2209.02248} {arXiv:2209.02248 [hep-ph]} \BibitemShut
  {NoStop}%
\bibitem [{\citenamefont {Chatterjee}\ \emph {et~al.}(2015)\citenamefont
  {Chatterjee}, \citenamefont {Elghozi}, \citenamefont {Novak},\ and\
  \citenamefont {Oertel}}]{Chatterjee:2014qsa}%
  \BibitemOpen
  \bibfield  {author} {\bibinfo {author} {\bibfnamefont {D.}~\bibnamefont
  {Chatterjee}}, \bibinfo {author} {\bibfnamefont {T.}~\bibnamefont {Elghozi}},
  \bibinfo {author} {\bibfnamefont {J.}~\bibnamefont {Novak}}, \ and\ \bibinfo
  {author} {\bibfnamefont {M.}~\bibnamefont {Oertel}},\ }\href {\doibase
  10.1093/mnras/stu2706} {\bibfield  {journal} {\bibinfo  {journal} {Mon. Not.
  Roy. Astron. Soc.}\ }\textbf {\bibinfo {volume} {447}},\ \bibinfo {pages}
  {3785} (\bibinfo {year} {2015})},\ \Eprint {http://arxiv.org/abs/1410.6332}
  {arXiv:1410.6332 [astro-ph.HE]} \BibitemShut {NoStop}%
\bibitem [{\citenamefont {Canuto}\ and\ \citenamefont
  {Chiu}(1968)}]{Canuto:1968apg}%
  \BibitemOpen
  \bibfield  {author} {\bibinfo {author} {\bibfnamefont {V.}~\bibnamefont
  {Canuto}}\ and\ \bibinfo {author} {\bibfnamefont {H.~Y.}\ \bibnamefont
  {Chiu}},\ }\href {\doibase 10.1103/PhysRev.173.1210} {\bibfield  {journal}
  {\bibinfo  {journal} {Phys. Rev.}\ }\textbf {\bibinfo {volume} {173}},\
  \bibinfo {pages} {1210} (\bibinfo {year} {1968})}\BibitemShut {NoStop}%
\bibitem [{\citenamefont {Martinez}\ \emph {et~al.}(2003)\citenamefont
  {Martinez}, \citenamefont {Rojas},\ and\ \citenamefont
  {Mosquera~Cuesta}}]{Martinez:2003dz}%
  \BibitemOpen
  \bibfield  {author} {\bibinfo {author} {\bibfnamefont {A.~P.}\ \bibnamefont
  {Martinez}}, \bibinfo {author} {\bibfnamefont {H.~P.}\ \bibnamefont {Rojas}},
  \ and\ \bibinfo {author} {\bibfnamefont {H.~J.}\ \bibnamefont
  {Mosquera~Cuesta}},\ }\href {\doibase 10.1140/epjc/s2003-01192-6} {\bibfield
  {journal} {\bibinfo  {journal} {Eur. Phys. J. C}\ }\textbf {\bibinfo {volume}
  {29}},\ \bibinfo {pages} {111} (\bibinfo {year} {2003})},\ \Eprint
  {http://arxiv.org/abs/astro-ph/0303213} {arXiv:astro-ph/0303213} \BibitemShut
  {NoStop}%
\bibitem [{\citenamefont {Noronha}\ and\ \citenamefont
  {Shovkovy}(2007)}]{Noronha:2007wg}%
  \BibitemOpen
  \bibfield  {author} {\bibinfo {author} {\bibfnamefont {J.~L.}\ \bibnamefont
  {Noronha}}\ and\ \bibinfo {author} {\bibfnamefont {I.~A.}\ \bibnamefont
  {Shovkovy}},\ }\href {\doibase 10.1103/PhysRevD.76.105030} {\bibfield
  {journal} {\bibinfo  {journal} {Phys. Rev. D}\ }\textbf {\bibinfo {volume}
  {76}},\ \bibinfo {pages} {105030} (\bibinfo {year} {2007})},\ \bibinfo {note}
  {[Erratum: Phys.Rev.D 86, 049901 (2012)]},\ \Eprint
  {http://arxiv.org/abs/0708.0307} {arXiv:0708.0307 [hep-ph]} \BibitemShut
  {NoStop}%
\bibitem [{\citenamefont {Huang}\ \emph {et~al.}(2010)\citenamefont {Huang},
  \citenamefont {Huang}, \citenamefont {Rischke},\ and\ \citenamefont
  {Sedrakian}}]{Huang:2009ue}%
  \BibitemOpen
  \bibfield  {author} {\bibinfo {author} {\bibfnamefont {X.-G.}\ \bibnamefont
  {Huang}}, \bibinfo {author} {\bibfnamefont {M.}~\bibnamefont {Huang}},
  \bibinfo {author} {\bibfnamefont {D.~H.}\ \bibnamefont {Rischke}}, \ and\
  \bibinfo {author} {\bibfnamefont {A.}~\bibnamefont {Sedrakian}},\ }\href
  {\doibase 10.1103/PhysRevD.81.045015} {\bibfield  {journal} {\bibinfo
  {journal} {Phys. Rev. D}\ }\textbf {\bibinfo {volume} {81}},\ \bibinfo
  {pages} {045015} (\bibinfo {year} {2010})},\ \Eprint
  {http://arxiv.org/abs/0910.3633} {arXiv:0910.3633 [astro-ph.HE]} \BibitemShut
  {NoStop}%
\bibitem [{\citenamefont {Strickland}\ \emph {et~al.}(2012)\citenamefont
  {Strickland}, \citenamefont {Dexheimer},\ and\ \citenamefont
  {Menezes}}]{Strickland:2012vu}%
  \BibitemOpen
  \bibfield  {author} {\bibinfo {author} {\bibfnamefont {M.}~\bibnamefont
  {Strickland}}, \bibinfo {author} {\bibfnamefont {V.}~\bibnamefont
  {Dexheimer}}, \ and\ \bibinfo {author} {\bibfnamefont {D.~P.}\ \bibnamefont
  {Menezes}},\ }\href {\doibase 10.1103/PhysRevD.86.125032} {\bibfield
  {journal} {\bibinfo  {journal} {Phys. Rev. D}\ }\textbf {\bibinfo {volume}
  {86}},\ \bibinfo {pages} {125032} (\bibinfo {year} {2012})},\ \Eprint
  {http://arxiv.org/abs/1209.3276} {arXiv:1209.3276 [nucl-th]} \BibitemShut
  {NoStop}%
\bibitem [{\citenamefont {Dexheimer}\ \emph {et~al.}(2014)\citenamefont
  {Dexheimer}, \citenamefont {Menezes},\ and\ \citenamefont
  {Strickland}}]{Dexheimer:2012mk}%
  \BibitemOpen
  \bibfield  {author} {\bibinfo {author} {\bibfnamefont {V.}~\bibnamefont
  {Dexheimer}}, \bibinfo {author} {\bibfnamefont {D.~P.}\ \bibnamefont
  {Menezes}}, \ and\ \bibinfo {author} {\bibfnamefont {M.}~\bibnamefont
  {Strickland}},\ }\href {\doibase 10.1088/0954-3899/41/1/015203} {\bibfield
  {journal} {\bibinfo  {journal} {J. Phys. G}\ }\textbf {\bibinfo {volume}
  {41}},\ \bibinfo {pages} {015203} (\bibinfo {year} {2014})},\ \Eprint
  {http://arxiv.org/abs/1210.4526} {arXiv:1210.4526 [nucl-th]} \BibitemShut
  {NoStop}%
\bibitem [{\citenamefont {Sinha}\ \emph {et~al.}(2013)\citenamefont {Sinha},
  \citenamefont {Huang},\ and\ \citenamefont {Sedrakian}}]{Sinha:2013dfa}%
  \BibitemOpen
  \bibfield  {author} {\bibinfo {author} {\bibfnamefont {M.}~\bibnamefont
  {Sinha}}, \bibinfo {author} {\bibfnamefont {X.-G.}\ \bibnamefont {Huang}}, \
  and\ \bibinfo {author} {\bibfnamefont {A.}~\bibnamefont {Sedrakian}},\ }\href
  {\doibase 10.1103/PhysRevD.88.025008} {\bibfield  {journal} {\bibinfo
  {journal} {Phys. Rev. D}\ }\textbf {\bibinfo {volume} {88}},\ \bibinfo
  {pages} {025008} (\bibinfo {year} {2013})},\ \Eprint
  {http://arxiv.org/abs/1306.3300} {arXiv:1306.3300 [astro-ph.HE]} \BibitemShut
  {NoStop}%
\bibitem [{\citenamefont {Peres~Menezes}\ and\ \citenamefont
  {La\'ercio~Lopes}(2016)}]{PeresMenezes:2015ukv}%
  \BibitemOpen
  \bibfield  {author} {\bibinfo {author} {\bibfnamefont {D.}~\bibnamefont
  {Peres~Menezes}}\ and\ \bibinfo {author} {\bibfnamefont {L.}~\bibnamefont
  {La\'ercio~Lopes}},\ }\href {\doibase 10.1140/epja/i2016-16017-2} {\bibfield
  {journal} {\bibinfo  {journal} {Eur. Phys. J. A}\ }\textbf {\bibinfo {volume}
  {52}},\ \bibinfo {pages} {17} (\bibinfo {year} {2016})},\ \Eprint
  {http://arxiv.org/abs/1505.06714} {arXiv:1505.06714 [nucl-th]} \BibitemShut
  {NoStop}%
\bibitem [{\citenamefont {Menezes}\ \emph {et~al.}(2015)\citenamefont
  {Menezes}, \citenamefont {Pinto},\ and\ \citenamefont
  {Provid\^encia}}]{Menezes:2015fla}%
  \BibitemOpen
  \bibfield  {author} {\bibinfo {author} {\bibfnamefont {D.~P.}\ \bibnamefont
  {Menezes}}, \bibinfo {author} {\bibfnamefont {M.~B.}\ \bibnamefont {Pinto}},
  \ and\ \bibinfo {author} {\bibfnamefont {C.}~\bibnamefont {Provid\^encia}},\
  }\href {\doibase 10.1103/PhysRevC.91.065205} {\bibfield  {journal} {\bibinfo
  {journal} {Phys. Rev. C}\ }\textbf {\bibinfo {volume} {91}},\ \bibinfo
  {pages} {065205} (\bibinfo {year} {2015})},\ \Eprint
  {http://arxiv.org/abs/1503.08666} {arXiv:1503.08666 [hep-ph]} \BibitemShut
  {NoStop}%
\bibitem [{\citenamefont {Ferrer}\ \emph {et~al.}(2015)\citenamefont {Ferrer},
  \citenamefont {de~la Incera}, \citenamefont {Manreza~Paret}, \citenamefont
  {P\'erez~Mart\'\i{}nez},\ and\ \citenamefont {Sanchez}}]{Ferrer:2015wca}%
  \BibitemOpen
  \bibfield  {author} {\bibinfo {author} {\bibfnamefont {E.~J.}\ \bibnamefont
  {Ferrer}}, \bibinfo {author} {\bibfnamefont {V.}~\bibnamefont {de~la
  Incera}}, \bibinfo {author} {\bibfnamefont {D.}~\bibnamefont
  {Manreza~Paret}}, \bibinfo {author} {\bibfnamefont {A.}~\bibnamefont
  {P\'erez~Mart\'\i{}nez}}, \ and\ \bibinfo {author} {\bibfnamefont
  {A.}~\bibnamefont {Sanchez}},\ }\href {\doibase 10.1103/PhysRevD.91.085041}
  {\bibfield  {journal} {\bibinfo  {journal} {Phys. Rev. D}\ }\textbf {\bibinfo
  {volume} {91}},\ \bibinfo {pages} {085041} (\bibinfo {year} {2015})},\
  \Eprint {http://arxiv.org/abs/1501.06616} {arXiv:1501.06616 [hep-ph]}
  \BibitemShut {NoStop}%
\bibitem [{\citenamefont {Bazavov}\ \emph {et~al.}(2019)\citenamefont {Bazavov}
  \emph {et~al.}}]{HotQCD:2018pds}%
  \BibitemOpen
  \bibfield  {author} {\bibinfo {author} {\bibfnamefont {A.}~\bibnamefont
  {Bazavov}} \emph {et~al.} (\bibinfo {collaboration} {HotQCD}),\ }\href
  {\doibase 10.1016/j.physletb.2019.05.013} {\bibfield  {journal} {\bibinfo
  {journal} {Phys. Lett. B}\ }\textbf {\bibinfo {volume} {795}},\ \bibinfo
  {pages} {15} (\bibinfo {year} {2019})},\ \Eprint
  {http://arxiv.org/abs/1812.08235} {arXiv:1812.08235 [hep-lat]} \BibitemShut
  {NoStop}%
\bibitem [{\citenamefont {Borsanyi}\ \emph {et~al.}(2020)\citenamefont
  {Borsanyi}, \citenamefont {Fodor}, \citenamefont {Guenther}, \citenamefont
  {Kara}, \citenamefont {Katz}, \citenamefont {Parotto}, \citenamefont
  {Pasztor}, \citenamefont {Ratti},\ and\ \citenamefont
  {Szabo}}]{Borsanyi:2020fev}%
  \BibitemOpen
  \bibfield  {author} {\bibinfo {author} {\bibfnamefont {S.}~\bibnamefont
  {Borsanyi}}, \bibinfo {author} {\bibfnamefont {Z.}~\bibnamefont {Fodor}},
  \bibinfo {author} {\bibfnamefont {J.~N.}\ \bibnamefont {Guenther}}, \bibinfo
  {author} {\bibfnamefont {R.}~\bibnamefont {Kara}}, \bibinfo {author}
  {\bibfnamefont {S.~D.}\ \bibnamefont {Katz}}, \bibinfo {author}
  {\bibfnamefont {P.}~\bibnamefont {Parotto}}, \bibinfo {author} {\bibfnamefont
  {A.}~\bibnamefont {Pasztor}}, \bibinfo {author} {\bibfnamefont
  {C.}~\bibnamefont {Ratti}}, \ and\ \bibinfo {author} {\bibfnamefont {K.~K.}\
  \bibnamefont {Szabo}},\ }\href {\doibase 10.1103/PhysRevLett.125.052001}
  {\bibfield  {journal} {\bibinfo  {journal} {Phys. Rev. Lett.}\ }\textbf
  {\bibinfo {volume} {125}},\ \bibinfo {pages} {052001} (\bibinfo {year}
  {2020})},\ \Eprint {http://arxiv.org/abs/2002.02821} {arXiv:2002.02821
  [hep-lat]} \BibitemShut {NoStop}%
\bibitem [{\citenamefont {Philipsen}(2013)}]{Philipsen:2012nu}%
  \BibitemOpen
  \bibfield  {author} {\bibinfo {author} {\bibfnamefont {O.}~\bibnamefont
  {Philipsen}},\ }\href {\doibase 10.1016/j.ppnp.2012.09.003} {\bibfield
  {journal} {\bibinfo  {journal} {Prog. Part. Nucl. Phys.}\ }\textbf {\bibinfo
  {volume} {70}},\ \bibinfo {pages} {55} (\bibinfo {year} {2013})},\ \Eprint
  {http://arxiv.org/abs/1207.5999} {arXiv:1207.5999 [hep-lat]} \BibitemShut
  {NoStop}%
\bibitem [{\citenamefont {Aoki}\ \emph
  {et~al.}(2006{\natexlab{b}})\citenamefont {Aoki}, \citenamefont {Endrodi},
  \citenamefont {Fodor}, \citenamefont {Katz},\ and\ \citenamefont
  {Szabo}}]{Aoki:2006we}%
  \BibitemOpen
  \bibfield  {author} {\bibinfo {author} {\bibfnamefont {Y.}~\bibnamefont
  {Aoki}}, \bibinfo {author} {\bibfnamefont {G.}~\bibnamefont {Endrodi}},
  \bibinfo {author} {\bibfnamefont {Z.}~\bibnamefont {Fodor}}, \bibinfo
  {author} {\bibfnamefont {S.~D.}\ \bibnamefont {Katz}}, \ and\ \bibinfo
  {author} {\bibfnamefont {K.~K.}\ \bibnamefont {Szabo}},\ }\href {\doibase
  10.1038/nature05120} {\bibfield  {journal} {\bibinfo  {journal} {Nature}\
  }\textbf {\bibinfo {volume} {443}},\ \bibinfo {pages} {675} (\bibinfo {year}
  {2006}{\natexlab{b}})},\ \Eprint {http://arxiv.org/abs/hep-lat/0611014}
  {arXiv:hep-lat/0611014} \BibitemShut {NoStop}%
\bibitem [{\citenamefont {Bazavov}\ \emph {et~al.}(2014)\citenamefont {Bazavov}
  \emph {et~al.}}]{HotQCD:2014kol}%
  \BibitemOpen
  \bibfield  {author} {\bibinfo {author} {\bibfnamefont {A.}~\bibnamefont
  {Bazavov}} \emph {et~al.} (\bibinfo {collaboration} {HotQCD}),\ }\href
  {\doibase 10.1103/PhysRevD.90.094503} {\bibfield  {journal} {\bibinfo
  {journal} {Phys. Rev. D}\ }\textbf {\bibinfo {volume} {90}},\ \bibinfo
  {pages} {094503} (\bibinfo {year} {2014})},\ \Eprint
  {http://arxiv.org/abs/1407.6387} {arXiv:1407.6387 [hep-lat]} \BibitemShut
  {NoStop}%
\bibitem [{\citenamefont {Goswami}\ \emph {et~al.}(2023)\citenamefont
  {Goswami}, \citenamefont {Sahu}, \citenamefont {Dey}, \citenamefont {Sahoo},\
  and\ \citenamefont {Stock}}]{Goswami:2023eol}%
  \BibitemOpen
  \bibfield  {author} {\bibinfo {author} {\bibfnamefont {K.}~\bibnamefont
  {Goswami}}, \bibinfo {author} {\bibfnamefont {D.}~\bibnamefont {Sahu}},
  \bibinfo {author} {\bibfnamefont {J.}~\bibnamefont {Dey}}, \bibinfo {author}
  {\bibfnamefont {R.}~\bibnamefont {Sahoo}}, \ and\ \bibinfo {author}
  {\bibfnamefont {R.}~\bibnamefont {Stock}},\ }\href@noop {} {\  (\bibinfo
  {year} {2023})},\ \Eprint {http://arxiv.org/abs/2310.02711} {arXiv:2310.02711
  [hep-ph]} \BibitemShut {NoStop}%
\bibitem [{\citenamefont {He}\ \emph {et~al.}(2022)\citenamefont {He},
  \citenamefont {Shao}, \citenamefont {Gao}, \citenamefont {Yang},\ and\
  \citenamefont {Xie}}]{He:2022kbc}%
  \BibitemOpen
  \bibfield  {author} {\bibinfo {author} {\bibfnamefont {W.-b.}\ \bibnamefont
  {He}}, \bibinfo {author} {\bibfnamefont {G.-y.}\ \bibnamefont {Shao}},
  \bibinfo {author} {\bibfnamefont {X.-y.}\ \bibnamefont {Gao}}, \bibinfo
  {author} {\bibfnamefont {X.-r.}\ \bibnamefont {Yang}}, \ and\ \bibinfo
  {author} {\bibfnamefont {C.-l.}\ \bibnamefont {Xie}},\ }\href {\doibase
  10.1103/PhysRevD.105.094024} {\bibfield  {journal} {\bibinfo  {journal}
  {Phys. Rev. D}\ }\textbf {\bibinfo {volume} {105}},\ \bibinfo {pages}
  {094024} (\bibinfo {year} {2022})},\ \Eprint
  {http://arxiv.org/abs/2205.04614} {arXiv:2205.04614 [hep-ph]} \BibitemShut
  {NoStop}%
\bibitem [{\citenamefont {Marty}\ \emph {et~al.}(2013)\citenamefont {Marty},
  \citenamefont {Bratkovskaya}, \citenamefont {Cassing}, \citenamefont
  {Aichelin},\ and\ \citenamefont {Berrehrah}}]{Marty:2013ita}%
  \BibitemOpen
  \bibfield  {author} {\bibinfo {author} {\bibfnamefont {R.}~\bibnamefont
  {Marty}}, \bibinfo {author} {\bibfnamefont {E.}~\bibnamefont {Bratkovskaya}},
  \bibinfo {author} {\bibfnamefont {W.}~\bibnamefont {Cassing}}, \bibinfo
  {author} {\bibfnamefont {J.}~\bibnamefont {Aichelin}}, \ and\ \bibinfo
  {author} {\bibfnamefont {H.}~\bibnamefont {Berrehrah}},\ }\href {\doibase
  10.1103/PhysRevC.88.045204} {\bibfield  {journal} {\bibinfo  {journal} {Phys.
  Rev. C}\ }\textbf {\bibinfo {volume} {88}},\ \bibinfo {pages} {045204}
  (\bibinfo {year} {2013})},\ \Eprint {http://arxiv.org/abs/1305.7180}
  {arXiv:1305.7180 [hep-ph]} \BibitemShut {NoStop}%
\bibitem [{\citenamefont {Peterson}\ \emph {et~al.}(2023)\citenamefont
  {Peterson}, \citenamefont {Costa}, \citenamefont {Kumar}, \citenamefont
  {Dexheimer}, \citenamefont {Negreiros},\ and\ \citenamefont
  {Providencia}}]{Peterson:2023bmr}%
  \BibitemOpen
  \bibfield  {author} {\bibinfo {author} {\bibfnamefont {J.}~\bibnamefont
  {Peterson}}, \bibinfo {author} {\bibfnamefont {P.}~\bibnamefont {Costa}},
  \bibinfo {author} {\bibfnamefont {R.}~\bibnamefont {Kumar}}, \bibinfo
  {author} {\bibfnamefont {V.}~\bibnamefont {Dexheimer}}, \bibinfo {author}
  {\bibfnamefont {R.}~\bibnamefont {Negreiros}}, \ and\ \bibinfo {author}
  {\bibfnamefont {C.}~\bibnamefont {Providencia}},\ }\href {\doibase
  10.1103/PhysRevD.108.063011} {\bibfield  {journal} {\bibinfo  {journal}
  {Phys. Rev. D}\ }\textbf {\bibinfo {volume} {108}},\ \bibinfo {pages}
  {063011} (\bibinfo {year} {2023})},\ \Eprint
  {http://arxiv.org/abs/2304.02454} {arXiv:2304.02454 [nucl-th]} \BibitemShut
  {NoStop}%
\bibitem [{\citenamefont {Abhishek}\ \emph {et~al.}(2018)\citenamefont
  {Abhishek}, \citenamefont {Mishra},\ and\ \citenamefont
  {Ghosh}}]{Abhishek:2017pkp}%
  \BibitemOpen
  \bibfield  {author} {\bibinfo {author} {\bibfnamefont {A.}~\bibnamefont
  {Abhishek}}, \bibinfo {author} {\bibfnamefont {H.}~\bibnamefont {Mishra}}, \
  and\ \bibinfo {author} {\bibfnamefont {S.}~\bibnamefont {Ghosh}},\ }\href
  {\doibase 10.1103/PhysRevD.97.014005} {\bibfield  {journal} {\bibinfo
  {journal} {Phys. Rev. D}\ }\textbf {\bibinfo {volume} {97}},\ \bibinfo
  {pages} {014005} (\bibinfo {year} {2018})},\ \Eprint
  {http://arxiv.org/abs/1709.08013} {arXiv:1709.08013 [hep-ph]} \BibitemShut
  {NoStop}%
\bibitem [{\citenamefont {Schaefer}\ \emph {et~al.}(2010)\citenamefont
  {Schaefer}, \citenamefont {Wagner},\ and\ \citenamefont
  {Wambach}}]{Schaefer:2009ui}%
  \BibitemOpen
  \bibfield  {author} {\bibinfo {author} {\bibfnamefont {B.-J.}\ \bibnamefont
  {Schaefer}}, \bibinfo {author} {\bibfnamefont {M.}~\bibnamefont {Wagner}}, \
  and\ \bibinfo {author} {\bibfnamefont {J.}~\bibnamefont {Wambach}},\ }\href
  {\doibase 10.1103/PhysRevD.81.074013} {\bibfield  {journal} {\bibinfo
  {journal} {Phys. Rev. D}\ }\textbf {\bibinfo {volume} {81}},\ \bibinfo
  {pages} {074013} (\bibinfo {year} {2010})},\ \Eprint
  {http://arxiv.org/abs/0910.5628} {arXiv:0910.5628 [hep-ph]} \BibitemShut
  {NoStop}%
\bibitem [{\citenamefont {Venugopalan}\ and\ \citenamefont
  {Prakash}(1992)}]{Venugopalan:1992hy}%
  \BibitemOpen
  \bibfield  {author} {\bibinfo {author} {\bibfnamefont {R.}~\bibnamefont
  {Venugopalan}}\ and\ \bibinfo {author} {\bibfnamefont {M.}~\bibnamefont
  {Prakash}},\ }\href {\doibase 10.1016/0375-9474(92)90005-5} {\bibfield
  {journal} {\bibinfo  {journal} {Nucl. Phys. A}\ }\textbf {\bibinfo {volume}
  {546}},\ \bibinfo {pages} {718} (\bibinfo {year} {1992})}\BibitemShut
  {NoStop}%
\bibitem [{\citenamefont {Bluhm}\ \emph {et~al.}(2014)\citenamefont {Bluhm},
  \citenamefont {Alba}, \citenamefont {Alberico}, \citenamefont {Beraudo},\
  and\ \citenamefont {Ratti}}]{Bluhm:2013yga}%
  \BibitemOpen
  \bibfield  {author} {\bibinfo {author} {\bibfnamefont {M.}~\bibnamefont
  {Bluhm}}, \bibinfo {author} {\bibfnamefont {P.}~\bibnamefont {Alba}},
  \bibinfo {author} {\bibfnamefont {W.}~\bibnamefont {Alberico}}, \bibinfo
  {author} {\bibfnamefont {A.}~\bibnamefont {Beraudo}}, \ and\ \bibinfo
  {author} {\bibfnamefont {C.}~\bibnamefont {Ratti}},\ }\href {\doibase
  10.1016/j.nuclphysa.2014.06.013} {\bibfield  {journal} {\bibinfo  {journal}
  {Nucl. Phys. A}\ }\textbf {\bibinfo {volume} {929}},\ \bibinfo {pages} {157}
  (\bibinfo {year} {2014})},\ \Eprint {http://arxiv.org/abs/1306.6188}
  {arXiv:1306.6188 [hep-ph]} \BibitemShut {NoStop}%
\bibitem [{\citenamefont {Khaidukov}\ \emph {et~al.}(2018)\citenamefont
  {Khaidukov}, \citenamefont {Lukashov},\ and\ \citenamefont
  {Simonov}}]{Khaidukov:2018lor}%
  \BibitemOpen
  \bibfield  {author} {\bibinfo {author} {\bibfnamefont {Z.~V.}\ \bibnamefont
  {Khaidukov}}, \bibinfo {author} {\bibfnamefont {M.~S.}\ \bibnamefont
  {Lukashov}}, \ and\ \bibinfo {author} {\bibfnamefont {Y.~A.}\ \bibnamefont
  {Simonov}},\ }\href {\doibase 10.1103/PhysRevD.98.074031} {\bibfield
  {journal} {\bibinfo  {journal} {Phys. Rev. D}\ }\textbf {\bibinfo {volume}
  {98}},\ \bibinfo {pages} {074031} (\bibinfo {year} {2018})},\ \Eprint
  {http://arxiv.org/abs/1806.09407} {arXiv:1806.09407 [hep-ph]} \BibitemShut
  {NoStop}%
\bibitem [{\citenamefont {Khaidukov}\ and\ \citenamefont
  {Simonov}(2019)}]{Khaidukov:2019icg}%
  \BibitemOpen
  \bibfield  {author} {\bibinfo {author} {\bibfnamefont {Z.~V.}\ \bibnamefont
  {Khaidukov}}\ and\ \bibinfo {author} {\bibfnamefont {Y.~A.}\ \bibnamefont
  {Simonov}},\ }\href {\doibase 10.1103/PhysRevD.100.076009} {\bibfield
  {journal} {\bibinfo  {journal} {Phys. Rev. D}\ }\textbf {\bibinfo {volume}
  {100}},\ \bibinfo {pages} {076009} (\bibinfo {year} {2019})},\ \Eprint
  {http://arxiv.org/abs/1906.08677} {arXiv:1906.08677 [hep-ph]} \BibitemShut
  {NoStop}%
\bibitem [{\citenamefont {Pal}\ and\ \citenamefont
  {Chaudhuri}(2023)}]{Pal:2023dlv}%
  \BibitemOpen
  \bibfield  {author} {\bibinfo {author} {\bibfnamefont {S.}~\bibnamefont
  {Pal}}\ and\ \bibinfo {author} {\bibfnamefont {G.}~\bibnamefont
  {Chaudhuri}},\ }\href {\doibase 10.1103/PhysRevD.108.103028} {\bibfield
  {journal} {\bibinfo  {journal} {Phys. Rev. D}\ }\textbf {\bibinfo {volume}
  {108}},\ \bibinfo {pages} {103028} (\bibinfo {year} {2023})}\BibitemShut
  {NoStop}%
\bibitem [{\citenamefont {Pal}\ \emph {et~al.}(2023)\citenamefont {Pal},
  \citenamefont {Podder}, \citenamefont {Sen},\ and\ \citenamefont
  {Chaudhuri}}]{Pal:2023quk}%
  \BibitemOpen
  \bibfield  {author} {\bibinfo {author} {\bibfnamefont {S.}~\bibnamefont
  {Pal}}, \bibinfo {author} {\bibfnamefont {S.}~\bibnamefont {Podder}},
  \bibinfo {author} {\bibfnamefont {D.}~\bibnamefont {Sen}}, \ and\ \bibinfo
  {author} {\bibfnamefont {G.}~\bibnamefont {Chaudhuri}},\ }\href {\doibase
  10.1103/PhysRevD.107.063019} {\bibfield  {journal} {\bibinfo  {journal}
  {Phys. Rev. D}\ }\textbf {\bibinfo {volume} {107}},\ \bibinfo {pages}
  {063019} (\bibinfo {year} {2023})},\ \Eprint
  {http://arxiv.org/abs/2303.04653} {arXiv:2303.04653 [nucl-th]} \BibitemShut
  {NoStop}%
\bibitem [{\citenamefont {Mykhaylova}\ and\ \citenamefont
  {Sasaki}(2021)}]{Mykhaylova:2020pfk}%
  \BibitemOpen
  \bibfield  {author} {\bibinfo {author} {\bibfnamefont {V.}~\bibnamefont
  {Mykhaylova}}\ and\ \bibinfo {author} {\bibfnamefont {C.}~\bibnamefont
  {Sasaki}},\ }\href {\doibase 10.1103/PhysRevD.103.014007} {\bibfield
  {journal} {\bibinfo  {journal} {Phys. Rev. D}\ }\textbf {\bibinfo {volume}
  {103}},\ \bibinfo {pages} {014007} (\bibinfo {year} {2021})},\ \Eprint
  {http://arxiv.org/abs/2007.06846} {arXiv:2007.06846 [hep-ph]} \BibitemShut
  {NoStop}%
\bibitem [{\citenamefont {\"Ozel}\ and\ \citenamefont
  {Freire}(2016)}]{Ozel:2016oaf}%
  \BibitemOpen
  \bibfield  {author} {\bibinfo {author} {\bibfnamefont {F.}~\bibnamefont
  {\"Ozel}}\ and\ \bibinfo {author} {\bibfnamefont {P.}~\bibnamefont
  {Freire}},\ }\href {\doibase 10.1146/annurev-astro-081915-023322} {\bibfield
  {journal} {\bibinfo  {journal} {Ann. Rev. Astron. Astrophys.}\ }\textbf
  {\bibinfo {volume} {54}},\ \bibinfo {pages} {401} (\bibinfo {year} {2016})},\
  \Eprint {http://arxiv.org/abs/1603.02698} {arXiv:1603.02698 [astro-ph.HE]}
  \BibitemShut {NoStop}%
\bibitem [{\citenamefont {Bedaque}\ and\ \citenamefont
  {Steiner}(2015)}]{Bedaque:2014sqa}%
  \BibitemOpen
  \bibfield  {author} {\bibinfo {author} {\bibfnamefont {P.}~\bibnamefont
  {Bedaque}}\ and\ \bibinfo {author} {\bibfnamefont {A.~W.}\ \bibnamefont
  {Steiner}},\ }\href {\doibase 10.1103/PhysRevLett.114.031103} {\bibfield
  {journal} {\bibinfo  {journal} {Phys. Rev. Lett.}\ }\textbf {\bibinfo
  {volume} {114}},\ \bibinfo {pages} {031103} (\bibinfo {year} {2015})},\
  \Eprint {http://arxiv.org/abs/1408.5116} {arXiv:1408.5116 [nucl-th]}
  \BibitemShut {NoStop}%
\bibitem [{\citenamefont {Tews}\ \emph {et~al.}(2018)\citenamefont {Tews},
  \citenamefont {Carlson}, \citenamefont {Gandolfi},\ and\ \citenamefont
  {Reddy}}]{Tews:2018kmu}%
  \BibitemOpen
  \bibfield  {author} {\bibinfo {author} {\bibfnamefont {I.}~\bibnamefont
  {Tews}}, \bibinfo {author} {\bibfnamefont {J.}~\bibnamefont {Carlson}},
  \bibinfo {author} {\bibfnamefont {S.}~\bibnamefont {Gandolfi}}, \ and\
  \bibinfo {author} {\bibfnamefont {S.}~\bibnamefont {Reddy}},\ }\href
  {\doibase 10.3847/1538-4357/aac267} {\bibfield  {journal} {\bibinfo
  {journal} {Astrophys. J.}\ }\textbf {\bibinfo {volume} {860}},\ \bibinfo
  {pages} {149} (\bibinfo {year} {2018})},\ \Eprint
  {http://arxiv.org/abs/1801.01923} {arXiv:1801.01923 [nucl-th]} \BibitemShut
  {NoStop}%
\bibitem [{\citenamefont {McLerran}\ and\ \citenamefont
  {Reddy}(2019)}]{McLerran:2018hbz}%
  \BibitemOpen
  \bibfield  {author} {\bibinfo {author} {\bibfnamefont {L.}~\bibnamefont
  {McLerran}}\ and\ \bibinfo {author} {\bibfnamefont {S.}~\bibnamefont
  {Reddy}},\ }\href {\doibase 10.1103/PhysRevLett.122.122701} {\bibfield
  {journal} {\bibinfo  {journal} {Phys. Rev. Lett.}\ }\textbf {\bibinfo
  {volume} {122}},\ \bibinfo {pages} {122701} (\bibinfo {year} {2019})},\
  \Eprint {http://arxiv.org/abs/1811.12503} {arXiv:1811.12503 [nucl-th]}
  \BibitemShut {NoStop}%
\bibitem [{\citenamefont {Fujimoto}\ \emph {et~al.}(2020)\citenamefont
  {Fujimoto}, \citenamefont {Fukushima},\ and\ \citenamefont
  {Murase}}]{Fujimoto:2019hxv}%
  \BibitemOpen
  \bibfield  {author} {\bibinfo {author} {\bibfnamefont {Y.}~\bibnamefont
  {Fujimoto}}, \bibinfo {author} {\bibfnamefont {K.}~\bibnamefont {Fukushima}},
  \ and\ \bibinfo {author} {\bibfnamefont {K.}~\bibnamefont {Murase}},\ }\href
  {\doibase 10.1103/PhysRevD.101.054016} {\bibfield  {journal} {\bibinfo
  {journal} {Phys. Rev. D}\ }\textbf {\bibinfo {volume} {101}},\ \bibinfo
  {pages} {054016} (\bibinfo {year} {2020})},\ \Eprint
  {http://arxiv.org/abs/1903.03400} {arXiv:1903.03400 [nucl-th]} \BibitemShut
  {NoStop}%
\bibitem [{\citenamefont {Jaikumar}\ \emph {et~al.}(2021)\citenamefont
  {Jaikumar}, \citenamefont {Semposki}, \citenamefont {Prakash},\ and\
  \citenamefont {Constantinou}}]{Jaikumar:2021jbw}%
  \BibitemOpen
  \bibfield  {author} {\bibinfo {author} {\bibfnamefont {P.}~\bibnamefont
  {Jaikumar}}, \bibinfo {author} {\bibfnamefont {A.}~\bibnamefont {Semposki}},
  \bibinfo {author} {\bibfnamefont {M.}~\bibnamefont {Prakash}}, \ and\
  \bibinfo {author} {\bibfnamefont {C.}~\bibnamefont {Constantinou}},\ }\href
  {\doibase 10.1103/PhysRevD.103.123009} {\bibfield  {journal} {\bibinfo
  {journal} {Phys. Rev. D}\ }\textbf {\bibinfo {volume} {103}},\ \bibinfo
  {pages} {123009} (\bibinfo {year} {2021})},\ \Eprint
  {http://arxiv.org/abs/2101.06349} {arXiv:2101.06349 [nucl-th]} \BibitemShut
  {NoStop}%
\bibitem [{\citenamefont {Dolan}(2011)}]{Dolan:2011jm}%
  \BibitemOpen
  \bibfield  {author} {\bibinfo {author} {\bibfnamefont {B.~P.}\ \bibnamefont
  {Dolan}},\ }\href {\doibase 10.1103/PhysRevD.84.127503} {\bibfield  {journal}
  {\bibinfo  {journal} {Phys. Rev. D}\ }\textbf {\bibinfo {volume} {84}},\
  \bibinfo {pages} {127503} (\bibinfo {year} {2011})},\ \Eprint
  {http://arxiv.org/abs/1109.0198} {arXiv:1109.0198 [gr-qc]} \BibitemShut
  {NoStop}%
\bibitem [{\citenamefont {Bhattacharyya}\ \emph {et~al.}(2012)\citenamefont
  {Bhattacharyya}, \citenamefont {Ghosh}, \citenamefont {Majumder},\ and\
  \citenamefont {Ray}}]{Bhattacharyya:2011na}%
  \BibitemOpen
  \bibfield  {author} {\bibinfo {author} {\bibfnamefont {A.}~\bibnamefont
  {Bhattacharyya}}, \bibinfo {author} {\bibfnamefont {S.~K.}\ \bibnamefont
  {Ghosh}}, \bibinfo {author} {\bibfnamefont {S.}~\bibnamefont {Majumder}}, \
  and\ \bibinfo {author} {\bibfnamefont {R.}~\bibnamefont {Ray}},\ }\href
  {\doibase 10.1103/PhysRevD.86.096006} {\bibfield  {journal} {\bibinfo
  {journal} {Phys. Rev. D}\ }\textbf {\bibinfo {volume} {86}},\ \bibinfo
  {pages} {096006} (\bibinfo {year} {2012})},\ \Eprint
  {http://arxiv.org/abs/1107.5941} {arXiv:1107.5941 [hep-ph]} \BibitemShut
  {NoStop}%
\bibitem [{\citenamefont {Iwasaki}(2004)}]{Iwasaki:2004nz}%
  \BibitemOpen
  \bibfield  {author} {\bibinfo {author} {\bibfnamefont {M.}~\bibnamefont
  {Iwasaki}},\ }\href {\doibase 10.1103/PhysRevD.70.114031} {\bibfield
  {journal} {\bibinfo  {journal} {Phys. Rev. D}\ }\textbf {\bibinfo {volume}
  {70}},\ \bibinfo {pages} {114031} (\bibinfo {year} {2004})},\ \Eprint
  {http://arxiv.org/abs/hep-ph/0411199} {arXiv:hep-ph/0411199} \BibitemShut
  {NoStop}%
\bibitem [{\citenamefont {Yang}\ and\ \citenamefont
  {Wen}(2021)}]{Yang:2021rdo}%
  \BibitemOpen
  \bibfield  {author} {\bibinfo {author} {\bibfnamefont {L.}~\bibnamefont
  {Yang}}\ and\ \bibinfo {author} {\bibfnamefont {X.-J.}\ \bibnamefont {Wen}},\
  }\href {\doibase 10.1103/PhysRevD.104.114010} {\bibfield  {journal} {\bibinfo
   {journal} {Phys. Rev. D}\ }\textbf {\bibinfo {volume} {104}},\ \bibinfo
  {pages} {114010} (\bibinfo {year} {2021})},\ \Eprint
  {http://arxiv.org/abs/2111.10008} {arXiv:2111.10008 [hep-ph]} \BibitemShut
  {NoStop}%
\bibitem [{\citenamefont {Ding}\ \emph {et~al.}(2023)\citenamefont {Ding},
  \citenamefont {Li}, \citenamefont {Liu},\ and\ \citenamefont
  {Wang}}]{Ding:2022uwj}%
  \BibitemOpen
  \bibfield  {author} {\bibinfo {author} {\bibfnamefont {H.-T.}\ \bibnamefont
  {Ding}}, \bibinfo {author} {\bibfnamefont {S.-T.}\ \bibnamefont {Li}},
  \bibinfo {author} {\bibfnamefont {J.-H.}\ \bibnamefont {Liu}}, \ and\
  \bibinfo {author} {\bibfnamefont {X.-D.}\ \bibnamefont {Wang}},\ }\href
  {\doibase 10.5506/APhysPolBSupp.16.1-A134} {\bibfield  {journal} {\bibinfo
  {journal} {Acta Phys. Polon. Supp.}\ }\textbf {\bibinfo {volume} {16}},\
  \bibinfo {pages} {1} (\bibinfo {year} {2023})},\ \Eprint
  {http://arxiv.org/abs/2208.07285} {arXiv:2208.07285 [hep-lat]} \BibitemShut
  {NoStop}%
\bibitem [{\citenamefont {Ding}\ \emph {et~al.}(2024)\citenamefont {Ding},
  \citenamefont {Gu}, \citenamefont {Kumar}, \citenamefont {Li},\ and\
  \citenamefont {Liu}}]{Ding:2023bft}%
  \BibitemOpen
  \bibfield  {author} {\bibinfo {author} {\bibfnamefont {H.-T.}\ \bibnamefont
  {Ding}}, \bibinfo {author} {\bibfnamefont {J.-B.}\ \bibnamefont {Gu}},
  \bibinfo {author} {\bibfnamefont {A.}~\bibnamefont {Kumar}}, \bibinfo
  {author} {\bibfnamefont {S.-T.}\ \bibnamefont {Li}}, \ and\ \bibinfo {author}
  {\bibfnamefont {J.-H.}\ \bibnamefont {Liu}},\ }\href {\doibase
  10.1103/PhysRevLett.132.201903} {\bibfield  {journal} {\bibinfo  {journal}
  {Phys. Rev. Lett.}\ }\textbf {\bibinfo {volume} {132}},\ \bibinfo {pages}
  {201903} (\bibinfo {year} {2024})},\ \Eprint
  {http://arxiv.org/abs/2312.08860} {arXiv:2312.08860 [hep-lat]} \BibitemShut
  {NoStop}%
\bibitem [{\citenamefont {Marczenko}\ \emph {et~al.}(2024)\citenamefont
  {Marczenko}, \citenamefont {Szyma\'nski}, \citenamefont {Lo}, \citenamefont
  {Karmakar}, \citenamefont {Huovinen}, \citenamefont {Sasaki},\ and\
  \citenamefont {Redlich}}]{Marczenko:2024kko}%
  \BibitemOpen
  \bibfield  {author} {\bibinfo {author} {\bibfnamefont {M.}~\bibnamefont
  {Marczenko}}, \bibinfo {author} {\bibfnamefont {M.}~\bibnamefont
  {Szyma\'nski}}, \bibinfo {author} {\bibfnamefont {P.~M.}\ \bibnamefont {Lo}},
  \bibinfo {author} {\bibfnamefont {B.}~\bibnamefont {Karmakar}}, \bibinfo
  {author} {\bibfnamefont {P.}~\bibnamefont {Huovinen}}, \bibinfo {author}
  {\bibfnamefont {C.}~\bibnamefont {Sasaki}}, \ and\ \bibinfo {author}
  {\bibfnamefont {K.}~\bibnamefont {Redlich}},\ }\href@noop {} {\  (\bibinfo
  {year} {2024})},\ \Eprint {http://arxiv.org/abs/2405.15745} {arXiv:2405.15745
  [hep-ph]} \BibitemShut {NoStop}%
\bibitem [{\citenamefont {Vovchenko}(2024)}]{Vovchenko:2024wbg}%
  \BibitemOpen
  \bibfield  {author} {\bibinfo {author} {\bibfnamefont {V.}~\bibnamefont
  {Vovchenko}},\ }\href@noop {} {\  (\bibinfo {year} {2024})},\ \Eprint
  {http://arxiv.org/abs/2405.16306} {arXiv:2405.16306 [hep-ph]} \BibitemShut
  {NoStop}%
\bibitem [{\citenamefont {Mondal}\ \emph {et~al.}(2024)\citenamefont {Mondal},
  \citenamefont {Chaudhuri}, \citenamefont {Roy},\ and\ \citenamefont
  {Sarkar}}]{Mondal:2023baz}%
  \BibitemOpen
  \bibfield  {author} {\bibinfo {author} {\bibfnamefont {R.}~\bibnamefont
  {Mondal}}, \bibinfo {author} {\bibfnamefont {N.}~\bibnamefont {Chaudhuri}},
  \bibinfo {author} {\bibfnamefont {P.}~\bibnamefont {Roy}}, \ and\ \bibinfo
  {author} {\bibfnamefont {S.}~\bibnamefont {Sarkar}},\ }\href {\doibase
  10.1103/PhysRevC.109.054911} {\bibfield  {journal} {\bibinfo  {journal}
  {Phys. Rev. C}\ }\textbf {\bibinfo {volume} {109}},\ \bibinfo {pages}
  {054911} (\bibinfo {year} {2024})},\ \Eprint
  {http://arxiv.org/abs/2312.03310} {arXiv:2312.03310 [nucl-th]} \BibitemShut
  {NoStop}%
\bibitem [{\citenamefont {Roessner}\ \emph {et~al.}(2007)\citenamefont
  {Roessner}, \citenamefont {Ratti},\ and\ \citenamefont
  {Weise}}]{Roessner:2006xn}%
  \BibitemOpen
  \bibfield  {author} {\bibinfo {author} {\bibfnamefont {S.}~\bibnamefont
  {Roessner}}, \bibinfo {author} {\bibfnamefont {C.}~\bibnamefont {Ratti}}, \
  and\ \bibinfo {author} {\bibfnamefont {W.}~\bibnamefont {Weise}},\ }\href
  {\doibase 10.1103/PhysRevD.75.034007} {\bibfield  {journal} {\bibinfo
  {journal} {Phys. Rev. D}\ }\textbf {\bibinfo {volume} {75}},\ \bibinfo
  {pages} {034007} (\bibinfo {year} {2007})},\ \Eprint
  {http://arxiv.org/abs/hep-ph/0609281} {arXiv:hep-ph/0609281} \BibitemShut
  {NoStop}%
\bibitem [{\citenamefont {Fukushima}\ \emph {et~al.}(2010)\citenamefont
  {Fukushima}, \citenamefont {Ruggieri},\ and\ \citenamefont
  {Gatto}}]{Fukushima:2010fe}%
  \BibitemOpen
  \bibfield  {author} {\bibinfo {author} {\bibfnamefont {K.}~\bibnamefont
  {Fukushima}}, \bibinfo {author} {\bibfnamefont {M.}~\bibnamefont {Ruggieri}},
  \ and\ \bibinfo {author} {\bibfnamefont {R.}~\bibnamefont {Gatto}},\ }\href
  {\doibase 10.1103/PhysRevD.81.114031} {\bibfield  {journal} {\bibinfo
  {journal} {Phys. Rev. D}\ }\textbf {\bibinfo {volume} {81}},\ \bibinfo
  {pages} {114031} (\bibinfo {year} {2010})},\ \Eprint
  {http://arxiv.org/abs/1003.0047} {arXiv:1003.0047 [hep-ph]} \BibitemShut
  {NoStop}%
\bibitem [{\citenamefont {Kapusta}\ and\ \citenamefont
  {Gale}(2011)}]{Kapusta:2006pm}%
  \BibitemOpen
  \bibfield  {author} {\bibinfo {author} {\bibfnamefont {J.~I.}\ \bibnamefont
  {Kapusta}}\ and\ \bibinfo {author} {\bibfnamefont {C.}~\bibnamefont {Gale}},\
  }\href {\doibase 10.1017/CBO9780511535130} {\emph {\bibinfo {title}
  {{Finite-temperature field theory: Principles and applications}}}},\
  Cambridge Monographs on Mathematical Physics\ (\bibinfo  {publisher}
  {Cambridge University Press},\ \bibinfo {year} {2011})\BibitemShut {NoStop}%
\bibitem [{\citenamefont {Gatto}\ and\ \citenamefont
  {Ruggieri}(2011)}]{Gatto:2010pt}%
  \BibitemOpen
  \bibfield  {author} {\bibinfo {author} {\bibfnamefont {R.}~\bibnamefont
  {Gatto}}\ and\ \bibinfo {author} {\bibfnamefont {M.}~\bibnamefont
  {Ruggieri}},\ }\href {\doibase 10.1103/PhysRevD.83.034016} {\bibfield
  {journal} {\bibinfo  {journal} {Phys. Rev. D}\ }\textbf {\bibinfo {volume}
  {83}},\ \bibinfo {pages} {034016} (\bibinfo {year} {2011})},\ \Eprint
  {http://arxiv.org/abs/1012.1291} {arXiv:1012.1291 [hep-ph]} \BibitemShut
  {NoStop}%
\bibitem [{\citenamefont {Sasaki}\ \emph {et~al.}(2007)\citenamefont {Sasaki},
  \citenamefont {Friman},\ and\ \citenamefont {Redlich}}]{Sasaki:2006ww}%
  \BibitemOpen
  \bibfield  {author} {\bibinfo {author} {\bibfnamefont {C.}~\bibnamefont
  {Sasaki}}, \bibinfo {author} {\bibfnamefont {B.}~\bibnamefont {Friman}}, \
  and\ \bibinfo {author} {\bibfnamefont {K.}~\bibnamefont {Redlich}},\ }\href
  {\doibase 10.1103/PhysRevD.75.074013} {\bibfield  {journal} {\bibinfo
  {journal} {Phys. Rev. D}\ }\textbf {\bibinfo {volume} {75}},\ \bibinfo
  {pages} {074013} (\bibinfo {year} {2007})},\ \Eprint
  {http://arxiv.org/abs/hep-ph/0611147} {arXiv:hep-ph/0611147} \BibitemShut
  {NoStop}%
\bibitem [{\citenamefont {Yao}\ \emph {et~al.}(2024)\citenamefont {Yao},
  \citenamefont {Sorensen}, \citenamefont {Dexheimer},\ and\ \citenamefont
  {Noronha-Hostler}}]{Yao:2023yda}%
  \BibitemOpen
  \bibfield  {author} {\bibinfo {author} {\bibfnamefont {N.}~\bibnamefont
  {Yao}}, \bibinfo {author} {\bibfnamefont {A.}~\bibnamefont {Sorensen}},
  \bibinfo {author} {\bibfnamefont {V.}~\bibnamefont {Dexheimer}}, \ and\
  \bibinfo {author} {\bibfnamefont {J.}~\bibnamefont {Noronha-Hostler}},\
  }\href {\doibase 10.1103/PhysRevC.109.065803} {\bibfield  {journal} {\bibinfo
   {journal} {Phys. Rev. C}\ }\textbf {\bibinfo {volume} {109}},\ \bibinfo
  {pages} {065803} (\bibinfo {year} {2024})},\ \Eprint
  {http://arxiv.org/abs/2311.18819} {arXiv:2311.18819 [nucl-th]} \BibitemShut
  {NoStop}%
\bibitem [{\citenamefont {Sorensen}\ \emph {et~al.}(2021)\citenamefont
  {Sorensen}, \citenamefont {Oliinychenko}, \citenamefont {Koch},\ and\
  \citenamefont {McLerran}}]{Sorensen:2021zme}%
  \BibitemOpen
  \bibfield  {author} {\bibinfo {author} {\bibfnamefont {A.}~\bibnamefont
  {Sorensen}}, \bibinfo {author} {\bibfnamefont {D.}~\bibnamefont
  {Oliinychenko}}, \bibinfo {author} {\bibfnamefont {V.}~\bibnamefont {Koch}},
  \ and\ \bibinfo {author} {\bibfnamefont {L.}~\bibnamefont {McLerran}},\
  }\href {\doibase 10.1103/PhysRevLett.127.042303} {\bibfield  {journal}
  {\bibinfo  {journal} {Phys. Rev. Lett.}\ }\textbf {\bibinfo {volume} {127}},\
  \bibinfo {pages} {042303} (\bibinfo {year} {2021})},\ \Eprint
  {http://arxiv.org/abs/2103.07365} {arXiv:2103.07365 [nucl-th]} \BibitemShut
  {NoStop}%
\bibitem [{\citenamefont {Hung}\ and\ \citenamefont
  {Shuryak}(1995)}]{Hung:1994eq}%
  \BibitemOpen
  \bibfield  {author} {\bibinfo {author} {\bibfnamefont {C.~M.}\ \bibnamefont
  {Hung}}\ and\ \bibinfo {author} {\bibfnamefont {E.~V.}\ \bibnamefont
  {Shuryak}},\ }\href {\doibase 10.1103/PhysRevLett.75.4003} {\bibfield
  {journal} {\bibinfo  {journal} {Phys. Rev. Lett.}\ }\textbf {\bibinfo
  {volume} {75}},\ \bibinfo {pages} {4003} (\bibinfo {year} {1995})},\ \Eprint
  {http://arxiv.org/abs/hep-ph/9412360} {arXiv:hep-ph/9412360} \BibitemShut
  {NoStop}%
\end{thebibliography}%

\end{document}